\documentclass[reviewcopy]{elsarticle}
\usepackage{graphicx}
\usepackage[reviewcopy]{adndt}
\usepackage{longtable}
\usepackage{bm}
\usepackage{CJK,upgreek,fancyhdr}
\usepackage{multicol}
\usepackage{graphicx}
\usepackage{booktabs}
\usepackage{amssymb,bm,mathrsfs,bbm,amscd}
\usepackage[tbtags]{amsmath}
\usepackage{lastpage}
\usepackage[usenames, dvipsnames]{xcolor}

\usepackage[normalem]{ulem}
\usepackage{lipsum}

\usepackage{amsmath}

\usepackage{amssymb}

\biboptions{square,sort&compress}
\bibpunct[]{[}{]}{,}{n}{}{;}
\newcommand{\journaltitle}{Atomic Data and Nuclear Data Tables}
\newcommand{\journalhome}{\texttt{http://www.elsevier.com/locate/jnlabr/yadndt}}
\newcommand{\journalmail}{\texttt{adndt@elsevier.com}}
\newcommand{\cmd}[1]{\texttt{$\backslash${#1}}}
\newcommand{\templatename}{\texttt{tmpadndt.tex}}
\newcommand{\adndtstyle}{\texttt{adndt}}
\newcommand{\adndtbst}{\texttt{adndt.bst}}
\newcommand{\adndtguide}{\texttt{../doc/ADNDdoc.pdf}}
\begin{document}

\begin{frontmatter}

\journal{Atomic Data and Nuclear Data Tables}

\title{The limits of the nuclear landscape explored by the relativistic continuum Hatree-Bogoliubov theory}

  \author[One]{X. W. Xia}
  \author[Two,Three]{Y. Lim}
  \author[Four,Five]{P. W. Zhao}
  \author[Six]{H. Z. Liang}
  \author[One,Seven]{X. Y. Qu}
  \author[Four,Eight]{Y. Chen}
  \author[Four]{H. Liu}
  \author[Four]{L. F. Zhang}
  \author[Four]{S. Q. Zhang}
  \ead{E-mail: sqzhang@pku.edu.cn}
  \author[Three]{Y. Kim}
  \ead{E-mail: ykim@ibs.re.kr}
 \author[Four,One,Nine]{J. Meng\corref{cor1}}
  \ead{E-mail: mengj@pku.edu.cn}

  \cortext[cor1]{Corresponding author.}

\address[One]{School of Physics and Nuclear Energy Engineering, Beihang University, Beijing 100191, China}
\address[Two]{Cyclotron Institute, Texas $A\&M$ University, College Station, Texas 77843, USA}
\address[Three]{Rare Isotope Science Project, Institute for Basic Science, Daejeon 305-811, Republic of Korea}
\address[Four]{State Key Laboratory of Nuclear Physics and Technology, School of Physics,
 Peking University, Beijing 100871, China}
\address[Five]{Physics Division, Argonne National Laboratory, Argonne, Illinois 60439, USA}
\address[Six]{RIKEN Nishina Center, Wako 351-0198, Japan}
\address[Seven]{School of Mechatronics Engineering, Guizhou Minzu University}
\address[Eight]{Institute of materials, China Academy of Engineering Physics, Sichuan, 621907, China}
\address[Nine]{Department of Physics, University of Stellenbosch, Stellenbosch, South Africa}

\date{16.04.2017} 

\begin{abstract}
 The ground-state properties of nuclei with $8\leqslant Z\leqslant 120$ from the proton drip line to the neutron drip line have been investigated using the spherical relativistic continuum Hartree-Bogoliubov (RCHB) theory with the relativistic density functional PC-PK1. With the effects of the continuum included, there are totally 9035 nuclei predicted to be bound, which largely extends the existing nuclear landscapes predicted with other methods. The calculated binding energies,  separation energies, neutron and proton Fermi surfaces, root-mean-square (rms) radii of neutron, proton, matter, and charge distributions, ground-state spins and parities are tabulated. The extension of the nuclear landscape obtained with RCHB is discussed in detail, in particular for the neutron-rich side, in comparison with the relativistic mean field calculations without pairing correlations and also other predicted landscapes. It is found that the coupling between the bound states and the continuum due to the pairing correlations plays an essential role in extending the nuclear landscape.
 The systematics of the separation energies, radii, densities, potentials and pairing energies of the RCHB calculations are also discussed. In addition, the $\alpha$-decay energies and proton emitters based on the RCHB calculations are investigated.
\end{abstract}

\end{frontmatter}

\newpage

\tableofcontents
\listofDtables
\listofDfigures
\vskip5pc

\newpage
\section{Introduction}

Nuclear landscape or nuclear chart is a diagrammatic representation of atomic nuclei consisting of a bound collection of $Z$ protons and $N$ neutrons using $Z$ and $N$ as axes in the plot. Along or near to the diagonal $N=Z$ line, the stable nuclei inhabit, forming the valley of stability. Away from this valley by adding or removing nucleons, there emerge the vast territory of short-lived radioactive nuclei, which disintegrate by emitting particles or other transmutation. Sequentially adding or removing more nucleons, the boundary of nuclear landscape, i.e., the drip line, is reached. Exploring the limit of nuclear binding has always been a priority in nuclear physics and investigating the properties of all nuclei in the nuclear landscape is crucial for understanding the nuclear force, and the formation of elements in astrophysics, etc.~\cite{http://www.nndc.bnl.gov/, Thoennessen2004RPP}.

To date, the existence of about 3000 nuclides has been confirmed experimentally~\cite{Thoennessen2004RPP, http://www.nndc.bnl.gov/}. The proton-rich boundary of the nuclear territory has been experimentally determined up to protactinium~\cite{http://www.nndc.bnl.gov/, Thoennessen2004RPP}. However, the neutron-rich boundary is known only up to oxygen~\cite{Thoennessen2004RPP, http://www.nndc.bnl.gov/}, and most of the neutron-rich nuclei far from the valley of stability seem still beyond the experimental capability in the foreseeable future.

The nuclear mass or binding energy is of crucial importance not only to nuclear structure, but also to the weak interactions~\cite{Lunney2003RevModPhys.75.1021, Blaum2006Phys.Rep.} and the origin of elements~\cite{Schatz2006Europhy.news}. Despite of the considerable achievements in the precise
mass measurements~\cite{Wang2012CPC}, the vast territory of neutron-rich nuclei are still beyond the experimental capability. Therefore, the development of a reliable theoretical nuclear mass model is very important to grasp a complete understanding of the nature.

Many theoretical efforts have been made to predict nuclear binding energies and to explore the great unknowns of the nuclear landscape~\cite{Moller1995ADNDT, Moller2012PRL.108.052501, Moller2016ADNDT, Aboussir1995ADNDT, Wang2010PRC.82.044304, Liu2011PRC.84.014333, Wang2014PLB, HFZhang2014NPA, Goriely2002PhysRevC.66.024326, Goriely2007PRC.75.064312, Goriely2009PRL.102.242501, Goriely2010PRC.82.035804, Goriely2013PRC.88.024308, Erler2012Nature, Hirata1997NPA, Lalazissis1999ADNDT, Geng2005PTP, Afanasjev2013PLB, Agbemava2014PRC.89.054320, Afanasjev2015PhysRevC.91.014324, Zhang2014Frontier, Lu2015PhysRevC.91.027304}. With macroscopic-microscopic models, precise descriptions of nuclear masses have been achieved~\cite{Moller1995ADNDT, Moller2012PRL.108.052501, Moller2016ADNDT, Aboussir1995ADNDT, Wang2010PRC.82.044304, Liu2011PRC.84.014333, Wang2014PLB, HFZhang2014NPA}. Based on the non-relativistic density functional theory (DFT), a series of Hartree-Fock-Bogoliubov mass models has been developed~\cite{Goriely2002PhysRevC.66.024326, Goriely2007PRC.75.064312, Goriely2009PRL.102.242501, Goriely2010PRC.82.035804, Goriely2013PRC.88.024308}. Recently, to explore the limit of the nuclear landscape and to estimate theoretical uncertainties in drip lines, the Skyrme-DFT calculations with six different parametrizations have been performed, and
approximately 7000 bound nuclei with $Z\leq$120 were predicted~\cite{Erler2012Nature}. However, the continuum effects are absent in these investigations.

The covariant density functional theory (CDFT) has attracted wide attentions for its successful description of many nuclear phenomena~\cite{Ring1996PPNP, Vretenar2005Phys.Rep., Meng2006PPNP, Nikvsic2011PPNP, Meng2013Fron.Phy., Meng2015JPG}. It includes naturally the nucleonic spin degree of freedom and automatically results in the nuclear spin-orbit potential with the empirical strength in a covariant way. It can reproduce well the isotopic shifts in the Pb region~\cite{Sharma1993PLB}, and give naturally the origin of the pseudospin symmetry~\cite{Liang2015report} and the spin symmetry in the anti-nucleon spectrum~\cite{Zhou2003PRL.91.262501, Liang2010EPJA}.
Furthermore, it can include the nuclear magnetism~\cite{Koepf1989NPA}, that is, a consistent description of currents and time-odd fields, which plays an important role in nuclear magnetic moments~\cite{Yao2006PRC.74.024307, Arima2011Sci.china, JLi2011Sci.China, JLi2011PTP, Wei2012PTP} and nuclear rotations~\cite{Meng2013Fron.Phy., Konig1993PhysRevLett.71.3079, Afanasjev2000NPA.676.196, Afanasjev2000PhysRevC.62.031302, Afanasjev2010PhysRevC.82.034329, Zhao2011PhysRevLett.107.122501, Zhao2011PLB, Zhao2012PRC.85.054310}.

Therefore, it is natural to investigate the nuclear landscape based on the CDFT.
The masses of about 2000 even-even nuclei with $8 \leq Z \leq 120$ up to the proton and neutron drip lines were investigated using relativistic mean field (RMF) theory without pairing correlations in 1997~\cite{Hirata1997NPA}. Later on, by including pairing correlations with Bardeen-Cooper-Schrieffer (BCS) method, the ground-state properties of 1315 even-even nuclei with $10 \leq Z \leq 98$ were calculated~\cite{Lalazissis1999ADNDT}. By employing the state-dependent BCS method with a zero-range pairing force, the first systematic study of the ground-state properties for over 7000 nuclei ranging from the proton drip line to the neutron drip line was performed~\cite{Geng2005PTP}. To explore the location of the proton and neutron drip lines, a systematic investigation has been performed for even-even nuclei by employing relativistic Hartree-Bogoliubov (RHB) theory with four sets of parametrizations~\cite{Afanasjev2013PLB, Agbemava2014PRC.89.054320, Afanasjev2015PhysRevC.91.014324}. Very recently, the binding energies for even-even nuclei with proton number ranging from $Z=8$ to $Z=108$ were calculated and the dynamical correlation energies which include the rotational correction energies and quadrupole vibrational correction energies were investigated~\cite{Zhang2014Frontier,Lu2015PhysRevC.91.027304}.

It is well known that continuum and pairing correlations play a critical role in exotic nuclei. For the exotic nuclei close to the nucleon drip lines, where the Fermi levels are very close to the continuum threshold, pairing correlations can scatter the valence nucleons between the bound states and continuum, and therefore provide a significant coupling between them. As a result, some unbound nuclei predicted without pairing correlations can
exist as a bound state.
For example, it is found that after taking into account pairing correlations and the contribution from continuum, the neutron-rich nuclei $^{62-72}$Ca predicted unbound without pairing correlations are found to be bound~\cite{Meng2002PhysRevC.65.041302}.
Therefore, the couplings between the bound states and the continuum due to the pairing correlations can strongly influence the position of the drip line.

A consistent description of neutron-rich nuclei requires a unified and self-consistent treatment of continuum, mean-field potentials and pairing correlations. By extending the relativistic mean field theory with the Bogoliubov transformation in the coordinate representation, the relativistic continuum Hartree-Bogoliubov (RCHB) theory was developed~\cite{Meng1996PhysRevLett.77.3963,Meng1998NPA} and it provides a proper treatment of pairing correlations and mean-field potentials in the presence of the continuum. With the RCHB theory, the first microscopic self-consistent description of halo in $^{11}$Li has been
provided~\cite{Meng1996PhysRevLett.77.3963} and the giant halos in light and medium-heavy nuclei have been predicted~\cite{Meng1998PhysRevLett.80.460, Meng2002PhysRevC.65.041302, Zhang2003Sci.inChina}. The RCHB theory has been generalized to treat the odd nucleon system~\cite{Meng1998PLB}, and together with the Glauber model, the charge-changing cross sections from carbon (C) to fluorine (F) isotopes on a carbon target have been well reproduced~\cite{Meng2002PLB}.
For deformed nuclei, much effort has been made to develop a deformed RHB theory in continuum and an interesting shape decoupling between the core and the halo was predicted~\cite{Zhou2010PhysRevC.82.011301,Li2012PhysRevC.85.024312}. Later, the deformed RHB theory in continuum has been extended to incorporate the density-dependent meson-nucleon couplings~\cite{Chen2012PhysRevC.85.067301}, and the blocking effect which is required for the description of odd-nucleon systems~\cite{Li2012CPL}.

As one of the most successful relativistic energy density functionals, PC-PK1~\cite{Zhao2010PhysRevC.82.054319}, which is fitted to the binding energies, charge radii, and empirical pairing gaps of 60 selected spherical nuclei, has been used successfully in describing not only nuclear ground-state properties~\cite{Zhao2012PhysRevC.86.064324, Zhang2014Frontier, Lu2015PhysRevC.91.027304} but also various excited state properties~\cite{Yao2013PLB, Yao2014PhysRevC.89.054306, Wu2014PhysRevC.89.017304,Li2012PLB,Fu2013PhysRevC.87.054305,Li2013PLB,Xiang2013PhysRevC.88.057301,Wang2015JPG}. In particular, PC-PK1 provides a good description for the isospin dependence of binding energy along either the isotopic or the isotonic chain, which makes it more reliable for describing exotic nuclei~\cite{Zhao2010PhysRevC.82.054319, Zhao2012PhysRevC.86.064324}.

Focusing on the continuum effects on the limits of the nuclear landscape, systematic spherical calculations over the nuclear landscape in the framework of RCHB theory with PC-PK1 will be performed. As a first attempt, the results from O to Ti isotopes were reported in Ref.~\cite{Qu2013scichia}, and the nuclear landscape from O to Ti is remarkably extended. Here, the results for all isotopes from $Z=8$ and $Z=120$ are reported. The paper is organized as follows:
In Section~\ref{theo-framew}, the theoretical framework is introduced briefly. The numerical details are given in Section~\ref{numeric}.
Extensive results are compiled in Section~\ref{res-dis}, including the binding energies, one-nucleon and two-nucleon separation energies, neutron and proton Fermi surfaces, root-mean-square (rms) radii of neutron, proton, matter, and charge distributions, ground-state spins and parities. Finally, a brief summary is presented.

\section{Theoretical framework}\label{theo-framew}
The covariant density functional theory is constructed with
either the finite-range meson-exchange interation or the contact interaction in the point-coupling representation between nucleons ~\cite{Ring1996PPNP, Meng2006PPNP, Meng2015JPG, Nikolaus1992PRC, Burvenich2002PhysRevC.65.044308,Zhao2010PhysRevC.82.054319}.  For the former, the nucleus is described as a system of Dirac nucleons that interact with each other via the exchange of mesons (e.g., scalar meson $\sigma$, vector meson $\omega$, and isovector meson $\rho$).
For the latter, the meson exchange in each channel (scalar-isoscalar, vector-isoscalar, scalar-isovector, and vector-isovector) is replaced by the corresponding local four-point (contact) interaction between nucleons.

Following the point-coupling representation in Ref.~\cite{Zhao2010PhysRevC.82.054319}, one starts with the Lagrangian density as
\begin{eqnarray}\label{EQ:LAG}
  {\cal L} &=& \bar\psi(i\gamma_\mu\partial^\mu-M)\psi-\frac{1}{2}\alpha_S(\bar\psi\psi)(\bar\psi\psi)-\frac{1}{2}\alpha_V(\bar\psi\gamma_\mu\psi)(\bar\psi\gamma^\mu\psi)
  -\frac{1}{2}\alpha_{TV}(\bar\psi\vec{\tau}\gamma_\mu\psi)(\bar\psi\vec{\tau}\gamma^\mu\psi)\nonumber\\
  & &-\frac{1}{2}\alpha_{TS}(\bar\psi\vec{\tau}\psi)(\bar\psi\vec{\tau}\psi)
     -\frac{1}{3}\beta_S(\bar\psi\psi)^3\nonumber-\frac{1}{4}\gamma_S(\bar\psi\psi)^4-\frac{1}{4}\gamma_V[(\bar\psi\gamma_\mu\psi)(\bar\psi\gamma^\mu\psi)]^2\nonumber\\
           & &-\frac{1}{2}\delta_S\partial_\nu(\bar\psi\psi)\partial^\nu(\bar\psi\psi)
              -\frac{1}{2}\delta_V\partial_\nu(\bar\psi\gamma_\mu\psi)\partial^\nu(\bar\psi\gamma^\mu\psi)
              -\frac{1}{2}\delta_{TV}\partial_\nu(\bar\psi\vec\tau\gamma_\mu\psi)\partial^\nu(\bar\psi\vec\tau\gamma_\mu\psi)\nonumber\\
           & &-\frac{1}{2}\delta_{TS}\partial_\nu(\bar\psi\vec\tau\psi)\partial^\nu(\bar\psi\vec\tau\psi)
           -\frac{1}{4}F^{\mu\nu}F_{\mu\nu}-e\frac{1-\tau_3}{2}\bar\psi\gamma^\mu\psi A_\mu,
\end{eqnarray}
where $M$ is the nucleon mass; $A_{\mu}$ and  $F_{\mu\nu}$ are respectively the four-vector potential and field strength tensor of the
electromagnetic field.
Here $\alpha_S,~\alpha_V, ~\alpha_{TS}$ and $\alpha_{TV}$ represent the
coupling constants for four-fermion point-coulping terms, $\beta_{S},
~\gamma_S$ and $\gamma_{V}$ are those for the higher-order terms which are responsible for the effects of medium dependence, and $\delta_{S}, ~\delta_{V}, ~\delta_{TS}$ and $\delta_{TV}$ refer to those for the gradient terms which are included to simulate the finite range effects. The subscripts $S$, $V$ and $T$ stand for scalar, vector and isovector, respectively.

The Hamiltonian density can be obtained by the Legendre transformation,
\begin{eqnarray}
\mathcal{H}=\frac{\partial\mathcal{L}}{\partial(\partial_0\phi)}\partial_0\phi-\mathcal{L},
\end{eqnarray}
where $\phi$ represents the nucleon or photon field. Then, the total Hamiltonian reads,
\begin{align}\nonumber
H=&~\int d^3r\mathcal{H}\\ \nonumber
=&~\int d^3r\{\bar{\psi}(-i\gamma^{i}\partial_{i}+M)\psi\\ \nonumber
&+\frac{1}{2}\alpha_{S}(\bar{\psi}\psi)(\bar{\psi}\psi)+\frac{1}{2}\alpha_{V}(\bar{\psi}\gamma_{\mu}\psi)(\bar{\psi}\gamma^{\mu}\psi)
+\frac{1}{2}\alpha_{TV}(\bar{\psi}\vec{\tau}\gamma_{\mu}\psi)(\bar{\psi}\vec{\tau}\gamma^{\mu}\psi)\\ \nonumber
&+\frac{1}{3}\beta_{S}(\bar{\psi}\psi)^3+\frac{1}{4}\gamma_{S}(\bar{\psi}\psi)^4
+\frac{1}{4}\gamma_{V}[(\bar{\psi}\gamma_{\mu}\psi)(\bar{\psi}\gamma^{\mu}\psi)]^2\\ \nonumber
&-\frac{1}{2}\delta_{S}[\partial_0(\bar{\psi}\psi)\partial^0(\bar{\psi}\psi)+\nabla(\bar{\psi}\psi)\cdot\nabla(\bar{\psi}\psi)]\\ \nonumber
&-\frac{1}{2}\delta_{V}[\partial_0(\bar{\psi}\gamma_{\mu}\psi)\partial^0(\bar{\psi}\gamma^{\mu}\psi)
+\nabla(\bar{\psi}\gamma_{\mu}\psi)\cdot\nabla(\bar{\psi}\gamma^{\mu}\psi)]\\ \nonumber
&-\frac{1}{2}\delta_{TV}[\partial_0(\bar{\psi}\vec{\tau}\gamma_{\mu}\psi)\partial^0(\bar{\psi}\vec{\tau}\gamma^{\mu}\psi)
+\nabla(\bar{\psi}\vec{\tau}\gamma_{\mu}\psi)\cdot\nabla(\bar{\psi}\vec{\tau}\gamma^{\mu}\psi)]\\
&+\frac{1}{4}F^{\mu\nu}F_{\mu\nu}+e\bar{\psi}\gamma^{\mu}\frac{1-\tau_3}{2}A_{\mu}\psi-F^{0\mu}\partial^{0}A_{\mu}\}.
\end{align}
With the mean-field and the no-sea approximations,
the nucleon field operators can be written as
\begin{align}\nonumber
\psi(x)=\sum_{k}\psi_k(x)c_k,\\
\psi^{\dag}(x)=\sum_{k}\psi^{\dag}_k(x)c^{\dag}_k,
\end{align}
where $c_k$ is the annihilation operator for a nucleon in the state $k$ and $\psi_k$ is the corresponding single-particle wave function. The operator $c_k$ and its conjugate $c_k^{\dag}$ satisfy the anticommutation relation for fermions,
\begin{eqnarray}\nonumber
&&\{c_k, c^\dag_{k'}\}=\delta_{kk'},\\
&&\{c_k, c_{k'}\}=\{c^\dag_k, c^\dag_{k'}\}=0.
\end{eqnarray}

The ground state of a nucleus can be written as
\begin{eqnarray}
|\Phi\rangle=\prod_i^{A}c_i^\dag|0\rangle~~{\rm with}~~\langle\Phi|\Phi\rangle=1.\label{RMF.G.S.}
\end{eqnarray}
The energy density functional for the nucleus system can be represented as
\begin{align}\nonumber
E_{DF}&=\langle\Phi|H|\Phi\rangle\\ \nonumber
&=\int d^3r\{\sum_{k}\psi^{\dag}_k(\bm{\alpha}\cdot\bm{p}+\beta M)\psi_k+\frac{1}{2}\alpha_{S}\rho_S^2+\frac{1}{2}\alpha_{V}j_{\mu}j^{\mu}+
\frac{1}{2}\alpha_{TV}(\vec{j}_{TV})_{\mu}\vec{j}^{\mu}_{TV}\\ \nonumber
&~~+\frac{1}{3}\beta_{S}\rho_{S}^3+\frac{1}{4}\gamma_{S}\rho_{S}^4+\frac{1}{4}\gamma_{V}(j_{\mu}j^{\mu})^2
-\frac{1}{2}\delta_{S}[\partial_0\rho_S\partial^0\rho_S+\nabla\rho_S\cdot\nabla\rho_S]\\ \nonumber
&~~-\frac{1}{2}\delta_{V}[\partial_0j_{\mu}\partial^0j^{\mu}+\nabla j_{\mu}\cdot\nabla j^{\mu}]
-\frac{1}{2}\delta_{TV}[\partial_0(\vec{j}_{TV})_{\mu}\partial^0\vec{j}^{\mu}_{TV}+\nabla(\vec{j}_{TV})_{\mu}\cdot\nabla\vec{j}^{\mu}_{TV}]\\
&~~+eA_{\mu}j_{p}^{\mu}-F^{0\mu}\partial^{0}A_{\mu}+\frac{1}{4}F_{\mu\nu}F^{\mu\nu}\},\label{RMF.D.F.}
\end{align}
where the local densities and currents are given by
\begin{subequations}
\begin{align}\label{currents}
\rho_S&=\langle\Phi|:\bar{\psi}\psi:|\Phi\rangle=\sum_{k}^{A}\bar{\psi}_k(x)\psi_k(x),\\
j^{\mu}&=\langle\Phi|:\bar{\psi}\gamma^{\mu}\psi:|\Phi\rangle=\sum_{k}^{A}\bar{\psi}_k(x)\gamma^{\mu}\psi_k(x),\\
\vec{j}^{\mu}_{TV}&=\langle\Phi|:\bar{\psi}\gamma^{\mu}\vec{\tau}\psi:|\Phi\rangle=\sum_{k}^{A}\bar{\psi}_k(x)\gamma^{\mu}\vec{\tau}\psi_k(x),\\
j^{\mu}_p&=\langle\Phi|:\bar{\psi}\gamma^{\mu}\frac{1-\tau_3}{2}\psi:|\Phi\rangle=\sum_{k}^{A}\bar{\psi}_k(x)\gamma^{\mu}\frac{1-\tau_3}{2}\psi_k(x).
\end{align}
\end{subequations}

By minimizing the energy density functional with respect to the densities, one obtains the Dirac equation for the nucleons:
\begin{eqnarray}
[\bm{\alpha}\cdot(\bm{p-V})+V^0+\beta(M+S)]\psi_k=\varepsilon\psi_k \label{Dirac},
\end{eqnarray}
with the local scalar $S(\bm{r})$ and vector $V^{\mu}(\bm{r})$ potentials
\begin{eqnarray}
  S(\mathbf{r})&=&\alpha_S \rho_S + \beta_S \rho_S^2 +\gamma_S\rho_S^3 +\delta_S\triangle \rho_S,\\
  \label{eq:vaspot}
V^{\mu}(\mathbf{r})&=&\alpha_V j^{\mu} + \gamma_V(j_{\mu}j^{\mu})j^{\mu} +\delta_V \triangle j^{\mu}+e A^{\mu}
           +\alpha_{TV}\tau_3 \vec{j}^{\mu}_{TV}+\delta_{TV} \tau_3 \triangle \vec{j}^{\mu}_{TV}.
   \label{eq:vavpot}
\end{eqnarray}

For a system with time reversal invariance, the space-like components of the current and the vector potential vanish. Furthermore, one can assume that the nucleon single-particle states do not mix isospin, i.e. the single-particle states are eigenstates of $\tau_3$. Therefore, the energy density functional can be written as
\begin{align}\nonumber
E_{\rm DF}&=\langle\Phi|H|\Phi\rangle\\ \nonumber
&=\int d^3r\{\sum_{k}\psi^{\dag}_k(\bm{\alpha}\cdot\bm{p}+\beta M)\psi_k+\frac{1}{2}\alpha_{S}\rho_S^2+\frac{1}{2}\alpha_{V}\rho_{V}^2+
\frac{1}{2}\alpha_{TV}(\rho_{TV})^2\\ \nonumber
&~~+\frac{1}{3}\beta_{S}\rho_{S}^3+\frac{1}{4}\gamma_{S}\rho_{S}^4+\frac{1}{4}\gamma_{V}(\rho_{V})^4
+\frac{1}{2}\delta_{S}\rho_S\Delta\rho_S\\
&~~+\frac{1}{2}\delta_{V}\rho_{V}\Delta\rho_{V}
+\frac{1}{2}\delta_{TV}\rho_{TV}\Delta\rho_{TV}+\frac{1}{2}eA_{0}\rho_{p}\}. \label{RMF.D.F.-1}
\end{align}

Pairing correlations are crucial in the description of open-shell nuclei. The relativistic Hartree-Bogoliubov model provides a unified description of particle-hole (ph) and particle-particle (pp) correlations on a mean field level by using two average potentials: the self-consistent mean field $\hat{h}$
and a pairing field $\hat{\Delta}$. In analogy to Eq. (\ref{RMF.G.S.}) the ground state of a nucleus is described by a generalized Slater determinant $|\Phi\rangle$ that represents the vacuum with respect to quasiparticles. The quasiparticle operators are defined by the unitary Bogoliubov transformation of the single-nucleon creation and annihilation operators
\begin{eqnarray}
\alpha_{k}^{\dag}=\sum_{n}U_{nk}c^{\dag}_n+V_{nk}c_n,
\end{eqnarray}
and ground state wave function can be written as
\begin{eqnarray}
|\Phi\rangle=\prod_{k}\alpha_k|-\rangle,
\end{eqnarray}
where the index $n$ refers to the original basis, e.g. an oscillator basis or the coordinates in space, spin and isospin $(\bm{r},s,t)$. The matrix $U$ and $V$ are the Hartree-Bogoliubov wave functions determined by the variational principle. In the presence of pairing, the single-particle density matrix is generalized to two different densities: the normal density $\hat{\rho}$ and the pairing tensor $\hat{\kappa}$,
\begin{eqnarray}\nonumber
\hat{\rho}_{nn'}=\langle\Phi|c^{\dag}_{n'}c_{n}|\Phi\rangle,\\
\hat{\kappa}_{nn'}=\langle\Phi|c_{n'}c_{n}|\Phi\rangle.
\end{eqnarray}
The RHB energy density functional thus depends on both $\hat{\rho}$ and $\hat{\kappa}$
\begin{eqnarray}
E_{\rm RHB}[\hat{\rho},\hat{\kappa}]=E_{\rm DF}[\hat{\rho}]+E_{\rm pair}[\hat{\kappa}],
\end{eqnarray}
where $E_{DF}[\hat{\rho}]$ is the usual relativistic mean-field functional defined in Eq. (\ref{RMF.D.F.}) or (\ref{RMF.D.F.-1}), and the pairing part of the RHB functional reads
\begin{eqnarray}
E_{\rm pair}[\hat{\kappa}]=\frac{1}{4}\sum_{n_1n_1'}\sum_{n_2n_2'}\hat{\kappa}^*_{n_1n_1'}\langle n_1n_1'|V^{pp}|n_2n_2'\rangle\hat{\kappa}_{n_2n_2'},
\end{eqnarray}
where $\langle n_1n_1'|V^{pp}|n_2n_2'\rangle$ represent the matrix elements of the pairing interaction. The RHB equation is obtained by the variational principle,

\begin{eqnarray}
 \left(
  \begin{array}{cc}
   \hat{h}_D
   - \lambda &
   \hat{\Delta}
   \\
  -\hat{\Delta}^*
   & -\hat{h}_D
   + \lambda \\
  \end{array}
 \right)
 \left(
  { U_{k}
  \atop V_{k}
   }
 \right)
 & = &
 E_{k}
  \left(
   { U_{k}
   \atop V_{k}
    }
  \right)
 ,
 \label{eq:RHB0}
\end{eqnarray}
where $E_{k}$ is the quasiparticle energy,
$\lambda$ the chemical potential or the Fermi surface,
and $h_D$ the Dirac Hamiltonian in Eq. (\ref{Dirac}), in which the densities can be constructed by quasiparticle wave functions,
  \begin{eqnarray}\nonumber
      \rho_S(\mathbf{r})     &=&\sum_{k>0 }\bar V_k(\mathbf{r})V_k(\mathbf{r}),\\
      \rho_{V}(\mathbf{r})   &=&\sum_{k>0 } V_k^{\dagger}(\mathbf{r})V_k(\mathbf{r}),\\ \nonumber
      \rho_{T V}(\mathbf{r}) &=&\sum_{k>0 } V_k^{\dagger}(\mathbf{r})\tau_3 V_k(\mathbf{r}).
      \label{eq:mesonsource}
  \end{eqnarray}

The pairing potential is determined by
\begin{eqnarray}
\hat{\Delta}_{n_1n_1'}
=\frac{1}{2}\sum_{n_2n_2'}\langle n_1n_1'|V^{pp}|n_2n_2'\rangle\hat{\kappa}_{n_2n_2'},
\end{eqnarray}
which depends on the pairing tensor $ \kappa=U^{*}V^T$ and pairing interaction $V^{pp}$ in the particle-particle channel.
There have been several types of effective pairing forces $V^{pp}$ in the literatures. The finite-range Gogny force can provide a good treatment of the coupling to the highly excited states but requires more numerical costs~\cite{Berger1984NPA}. The separable pairing force is also a finite-range force but has a separable form, which reduces significantly the computational costs~\cite{Tian2009PLB}. The density-dependent zero-range force
has a simple form and a realistic density-dependent behavior
 \begin{eqnarray}
    V^{pp}(\mathbf{r_1},\mathbf{r_2})=V_0\delta(\mathbf{r_1}-\mathbf{r_2})\frac{1}{2}(1-P^{\sigma})
                                        (1-\eta\frac{\rho(\mathbf{r_1})}{\rho_0}).\label{pairingforce}
\end{eqnarray}
Here, $V_0$ is the interaction strength and $\rho_0$ is the saturation density of nuclear matter.
Note that $\eta$ here can be either 0 or 1. For $\eta=1$, it is called as a surface pairing force, because the force is dramatically suppressed in the nuclear interior and has its largest contribution in the surface. Conversely, $\eta=0$ corresponds to a volume pairing force. For a zero-range pairing force, one has to introduce an energy cut-off to avoid the divergence of the pairing energy and, thus,
the interaction strength $V_0$ has to be properly justified
in a given cut-off.

With spherical symmetry, the quasiparticle wave function in the coordinate space can be written as
\begin{eqnarray}
U_k=\frac{1}{r}\left(
\begin{array}{c}
iG_U^{k}(r)Y_{jm}^l(\theta,\phi)\\
F_U^{k}(r)(\pmb{\sigma\cdot \hat{r}})Y_{jm}^l(\theta,\phi)
\end{array}\right)\chi_t(t), ~~V_k=\frac{1}{r}\left(
\begin{array}{c}
iG_V^{k}(r)Y_{jm}^l(\theta,\phi)\\
F_V^{k}(r)(\pmb{\sigma\cdot \hat{r}})Y_{jm}^l(\theta,\phi)
\end{array}\right)\chi_t(t).
\end{eqnarray}
The corresponding RHB equation can be expressed as the following radial integral-differential equations in coordinate space~\cite{Meng1998NPA}:
\begin{eqnarray}\nonumber
\frac{dG_U}{dr}+\frac{\kappa}{r}G_U(r)-(E+\lambda-V(r)+S(r))F_U(r)
+r\int r'dr'\Delta_F(r,r')F_V(r')=0,\\ \nonumber
\frac{dF_U}{dr}-\frac{\kappa}{r}F_U(r)
+(E+\lambda-V(r)-S(r))G_U(r)+r\int r'dr'\Delta_G(r,r')G_V(r')=0,\\ \nonumber
\frac{dG_V}{r}+\frac{\kappa}{r}G_V (r)+(E-\lambda+V(r)-S(r))F_V (r)
+r\int r'dr'\Delta_F(r,r')F_U(r')=0,\\
\frac{dF_V}{r}-\frac{\kappa}{r}F_V (r)-(E-\lambda+V(r)+S(r))G_V (r)
+r\int r'dr'\Delta_G(r,r')G_U(r')=0.\label{rchb-eq}
\end{eqnarray}
If the zero-range pairing force is applied, the above coupled integral-differential equations can be reduced to  differential ones, which can be solved in coordinate space using the shooting method with Runge-Kutta algorithms~\cite{Meng1998NPA}.
After the solution, new densities and fields are obtained, which are iterated in the differential equations until convergence is achieved.

Finally, one can calculate the total energy of a nucleus by~\cite{Meng1998NPA}
\begin{align}\nonumber
E_{\rm RHB}&=\sum_{k}(\lambda-E_{k})V^2_k-E_{\rm pair}-\int d^3r\{\frac{1}{2}\alpha_{S}\rho_S^2+\frac{1}{2}\alpha_{V}\rho_{V}^2+
\frac{1}{2}\alpha_{TV}(\rho_{TV})^2\\ \nonumber
&~~+\frac{2}{3}\beta_{S}\rho_{S}^3+\frac{3}{4}\gamma_{S}\rho_{S}^4+\frac{3}{4}\gamma_{V}(\rho_{V})^4
+\frac{1}{2}\delta_{S}\rho_S\Delta\rho_S\\
&~~+\frac{1}{2}\delta_{V}\rho_{V}\Delta\rho_{V}
+\frac{1}{2}\delta_{TV}\rho_{TV}\Delta\rho_{TV}+\frac{1}{2}eA_{0}\rho_{p}\}-\frac{1}{2MA}\langle \hat{P}^2_{cm}\rangle,
\end{align}
and the root-mean-square (rms) radius by
\begin{align}
R_{\rm{rms}}=\langle r^2\rangle^{1/2}=\biggl\{\int d^3\bm{r}[r^2\rho_V(\bm{r})]\biggr\}^{1/2}.
\end{align}

Note that for odd-$A$ and odd-odd nuclei, the blocking effects of the unpaired nucleon(s) have to be taken into account.
Generally, the ground state of an odd system can be written as an one-quasiparticle state,
\begin{align}
|\Phi_1\rangle=\alpha^{\dag}_{1}\prod_{k}\alpha_{k}|-\rangle,
\end{align}
where $\alpha^{\dag}_{1}$ corresponds to the quasiparticle state which is blocked.
As explained in great detail in Ref.~\cite{Ring2004manybody}, this state $|\Phi_1\rangle$ can be regarded as the vacuum with respect to a new set of quasiparticle operators $(\alpha'_1,...\alpha'_N)$ with
\begin{align}
\alpha'_1=\alpha^{\dag}_1,~\alpha'_2=\alpha_1,~...,~\alpha'_N=\alpha'_N.
\end{align}
In such a way, the odd-nucleon systems can be simply treated just like an even-even system. The only difference is the exchange of the quasiparticle operator $\alpha_{1}^{\dag}\leftrightarrow\alpha_{1}$, which corresponds to the exchange of the column $(U_{n1}, V_{n1})\leftrightarrow(V_{n1}^{*},U_{n1}^{*})$ in the Bogoliubov transform matrix.

\section{Numerical details}\label{numeric}

To explore the nuclear landscape extended by the continuum effects, the systematic calculations for nuclei with $8\leqslant Z\leqslant 120$ from proton drip line to neutron drip line are performed by using the relativistic continuum Hartree-Bogoliubov theory with PC-PK1. In order to describe the continuum and its coupling to the bound states properly, the RCHB equation~(\ref{rchb-eq}) is solved in coordinate space using the shooting method. Each atomic nucleus is described within a box by imposing the boundary condition $U_k(r=R_{\rm box})=V_k(r=R_{\rm box})=0$.

\begin{figure}[ht!]
\centering
 \includegraphics[scale=0.4]{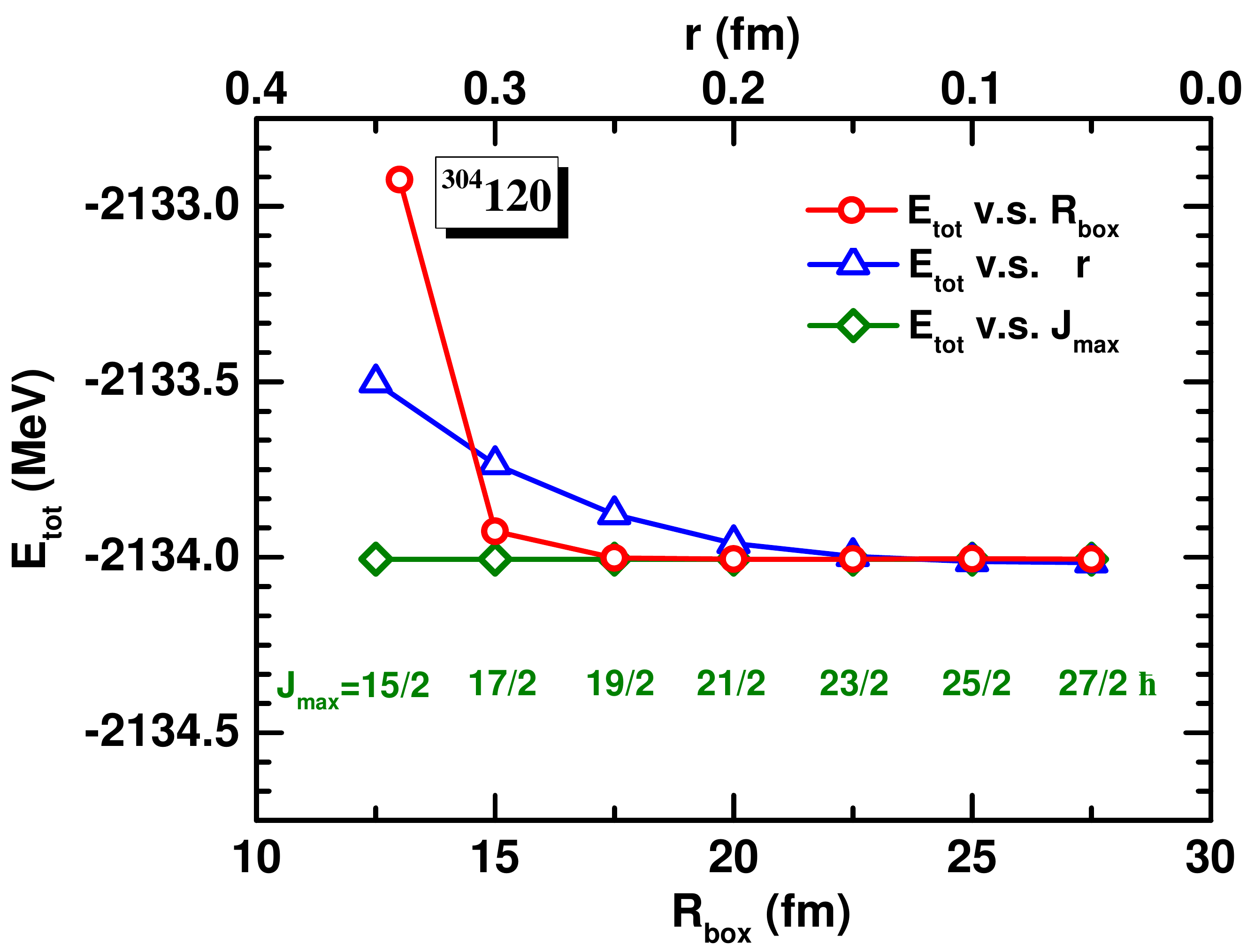}
   \caption{\label{converge-check}(Color online) Total energies of the nuclei $^{304}$120 as a function of box size $R_{\rm box}$ (red line), mesh size $\Delta r$ (blue line) and angular momentum cutoff $J_{\rm max}$ (olive line).}
\end{figure}

As shown in Fig.~\ref{converge-check}, the convergence of the RCHB solutions with respect to the box size, the mesh size and the angular momentum cutoff has been examined for nuclei $^{304}$120.
In Fig.~\ref{converge-check}, the box size $R_{\rm box} = 20$ fm leads to a relative accuracy of 0.002$\%$ in total energy for $^{304}$120 in comparison with $R_{\rm box} = 27.5$ fm.
The energy difference between the calculations with $\Delta r = 0.1$ fm and $\Delta r = 0.05$ fm is smaller than 0.0035 MeV for $^{304}$120, which is about 0.0002$\%$ of the total energy.
Similar convergence has also been confirmed for the angular momentum cutoff $J_{\rm max}=19/2~\hbar$, as shown in Fig.~\ref{converge-check}. Note that the pairing correlation is neglected in the convergence check for angular momentum cutoff in order to avoid renormalizing the strength of the zero-range pairing force  to the corresponding model space. In the following calculations, we fix the box size $R_{\rm box}= 20$ fm, the mesh size $\Delta r =0.1$ fm, and the angular momentum cutoff $J_{\rm max}= 19/2~\hbar$.

\subsection{Relativistic density functional PC-PK1}
In the present systematic calculations,
the relativistic density functional PC-PK1~\cite{Zhao2010PhysRevC.82.054319} listed in Table~\ref{pset-PC-PK1} is adopted for the particle-hole channel. PC-PK1 has turned out to be very successful in providing good descriptions for the isospin dependence of the binding energy along either the isotopic or the isotonic chain. In Ref.~\cite{Zhao2012PhysRevC.86.064324}, a crucial test of the predictive power of PC-PK1
was performed by the comparison with the new and accurate experimental masses in the element range from 50 to 91~\cite{Chen2012NPA}, and it was found that the experimental data can be reproduced quite well with a rms deviation 0.859 MeV.
PC-PK1 has also been applied successfully to investigate nuclear low-lying excited states~\cite{Yao2013PLB, Yao2014PhysRevC.89.054306, Wu2014PhysRevC.89.017304,Li2012PLB,Fu2013PhysRevC.87.054305,Li2013PLB,Xiang2013PhysRevC.88.057301,Wang2015JPG},  magnetic rotation~\cite{Zhao2011PLB, Steppenbeck2012PhysRevC.85.044316, Yu2012PhysRevC.85.024318, LiJ2013PhysRevC.88.014317, Zhao2015PhysRevC.92.034319}, antimagnetic rotation~\cite{Zhao2011PhysRevLett.107.122501, Zhao2012PRC.85.054310, Li2012PhysRevC.86.057305, Zhang2014PhysRevC.89.047302, Peng2015PhysRevC.91.044329}, chiral rotation modes~\cite{Kuti2014PhysRevLett.113.032501}, fission barrier~\cite{Lu2012PhysRevC.85.011301},
and superheavy nuclei~\cite{Agbemave2015PhysRevC.92.054310}, etc.
\begin{table}[ht!]
\caption{\label{pset-PC-PK1} The parameter set of the relativistic density functional PC-PK1~\cite{Zhao2010PhysRevC.82.054319}.}
\begin{center}
\begin{tabular}{ccc}
\hline
\hline
\multicolumn{1}{c}{Coupling constant}
&\multicolumn{1}{c}{Value}
&\multicolumn{1}{c}{Dimension}
\\
\hline
$\alpha_{S}$	&$-3.96291\times 10^{-4}$	&  MeV$^{-2}$ \\
$\beta_{S}$     &$8.6653\times 10^{-11}$	&  MeV$^{-5}$ \\
$\gamma_{S}$	&$-3.80724\times 10^{-17}$	&  MeV$^{-8}$ \\
$\delta_{S}$    &$-1.09108\times 10^{-10}$	&  MeV$^{-4}$ \\
$\alpha_{V}$	&$2.6904\times 10^{-4}$	    &  MeV$^{-2}$ \\
$\gamma_{V}$    &$-3.64219\times 10^{-18}$	&  MeV$^{-8}$ \\
$\delta_{V}$	&$-4.32619\times 10^{-10}$	&  MeV$^{-4}$ \\
$\alpha_{TV}$   &$2.95018\times 10^{-5}$	&  MeV$^{-2}$ \\
$\delta_{TV}$	&$-4.11112\times 10^{-10}$	&  MeV$^{-4}$ \\
\hline
\hline
\end{tabular}
\end{center}
\end{table}

\subsection{Pairing strength}
\begin{figure}[ht!]
\centering
 \includegraphics[scale=0.5]{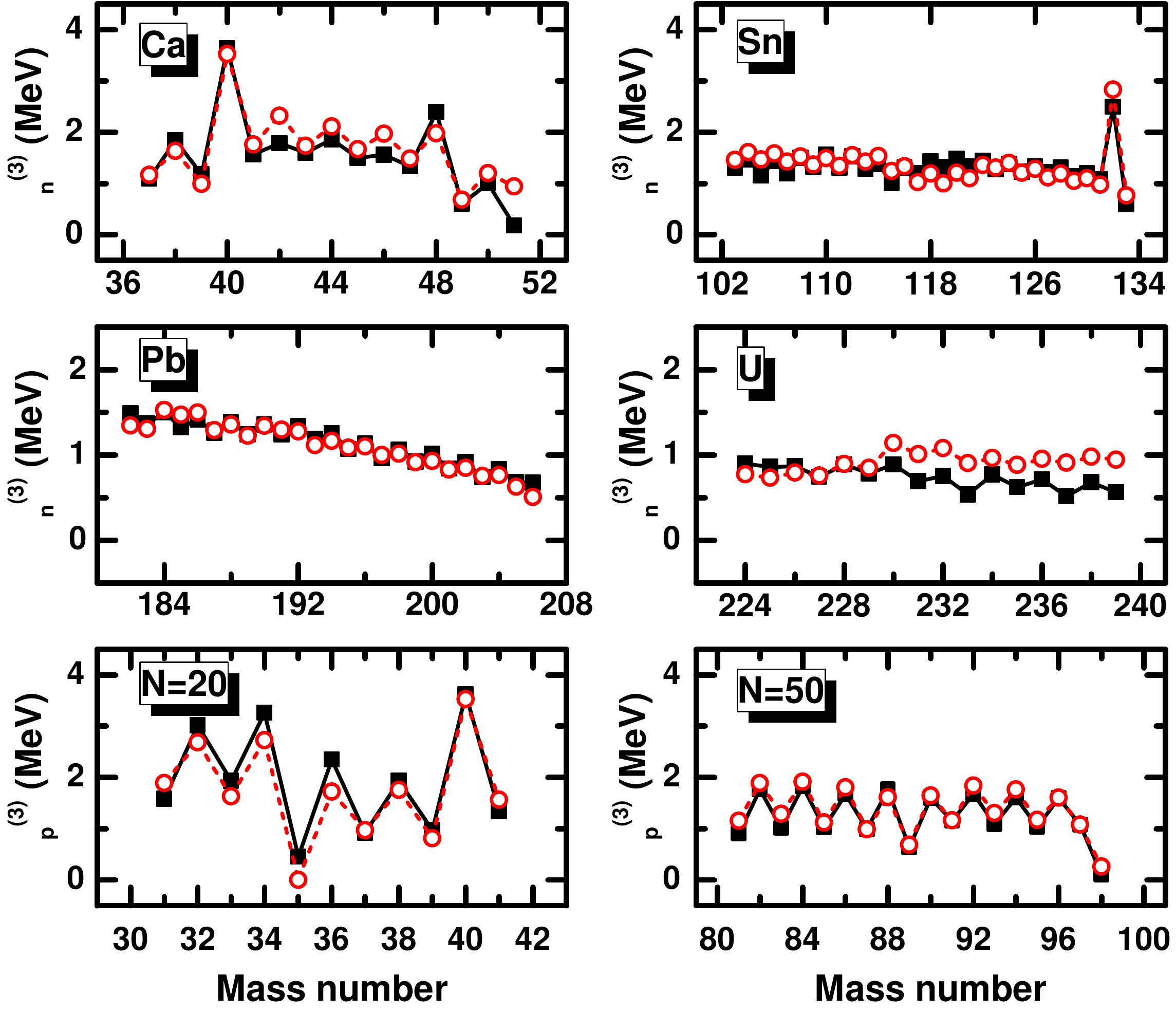}
   \caption{\label{odd-even}(Color online) Odd-even mass differences of Ca, Sn, Pb and U isotopic chains, as well as $N = 20, 50$ isotonic chains in RCHB theory as a function of the mass number, with the pairing strength $V_0 = -342.5$ MeV fm$^3$ and saturation density $\rho_{0} = $ 0.152 fm$^{-3}$. The corresponding experimental data~\cite{Wang2012CPC} are shown for comparison.}
\end{figure}

For the particle-particle channel, the surface pairing force with $\eta =1$ in Eq.~(\ref{pairingforce})
is used. The parameters in the pairing force include the saturation density $\rho_{0}$ and pairing strength $V_0$.
The empirical saturation density of nuclear matter $\rho_{0}$ is 0.152 fm$^{-3}$. The pairing strength $V_0$ is fitted to the odd-even mass differences for given energy cutoff $E_{\rm cut}$.
In the present RCHB calculations, the pairing strength $V_0 =-342.5$ MeV fm$^3$ for both neutron and proton is fixed by reproducing the experimental odd-even mass differences in the Ca, Sn, Pb and U isotopic chains, as well as the $N$  = 20, 50 isotonic chains with a cutoff energy of 100 MeV for the pairing window.
For the odd-even mass difference, a three-point formula is utilized,
\begin{equation}
\Delta = \frac{(-1)^{N}}{2}
\left[\ E_b(Z,N+1) - 2E_b(Z,N) + E_b(Z,N-1)\right]\,.
\end{equation}
The corresponding comparison between the theoretical and experimental odd-even mass differences is
shown in Fig.~\ref{odd-even}.

\subsection{Blocking effects for odd-$A$ and odd-odd nuclei}
For the odd-$A$ or odd-odd nuclei, one has to take into account the blocking effect of the odd nucleon(s)~\cite{Ring2004manybody}. The procedure to determine the ground state of an odd-$A$ nucleus with even $Z$ and odd $N$ in RCHB calculations is as follows:

 (i) Perform the calculation for its neighboring even-even nucleus ($Z$, $N-1$).

 (ii) From the single particle levels in  the even-even nucleus ($Z$, $N-1$), choose two neutron levels just below the Fermi surface (labeled as $E_1$ and $E_2$) and two neutron levels just above the Fermi surface (labeled as $E_3$ and $E_4$).

 (iii) Perform the calculations for the odd-$A$ nucleus ($Z$, $N$) with blocking levels $E_{1}$, $E_{2}$, $E_{3}$
 and $E_{4}$, respectively, and choose the one with the lowest total energy  as the ground state of the odd-$A$ nucleus ($Z$, $N$).

The blocking procedure for odd-odd nucleus ($Z$, $N$) is similar to the odd-$A$ nuclei, but requires blocking for both the proton and neutron levels at the same time. From the single particle levels in the neighboring even-even nucleus ($Z-1$, $N-1$), one chooses the blocking proton and neutron levels (two just below and two just above the Fermi surfaces). By performing the calculations for the odd-odd nucleus ($Z$, $N$) with the blocking proton and neutron levels, the one with the lowest total energy is chosen as the ground state of the odd-odd nucleus ($Z$, $N$).

\section{Results and discussions}\label{res-dis}

\subsection{Nuclear mass}
We perform systematic calculations for all nuclei from $Z=8$ to $Z=120$ by varying the neutron number from the proton drip line to neutron one.
In Table~\ref{masstable}, the ground-state properties of these nuclei are summarized. The mass number $A$, neutron number $N$, binding energy $E_{\rm b}$,  neutron Fermi surface $\lambda_{n}$, proton Fermi surface $\lambda_{p}$, matter root-mean-square (rms) radius $R_m$, neutron rms radius $R_n$, proton rms radius $R_p$, charge radius $R_{\mathrm c}$, ground-state spin and the parity of proton $j^{\pi}(P)$  and neutron $j^{\pi}(N)$ are listed.
The binding energy per nucleon $E_{\rm b}/A$, two-neutron $S_{2n}$, two-proton $S_{2p}$, one-neutron $S_{n}$ and one-proton separation energy $S_{p}$ for each nucleus are also provided.
In addition, the available experimental binding energies~\cite{Wang2012CPC} and charge radii~\cite{Angeli2013ADNDT} are shown for comparison. There are totally 9035 nuclei from O($Z=8$) to $Z=120$ which are predicted to be bound by the RCHB theory with the relativistic density functional PC-PK1.
For guidance, one or two unbound nuclei just outside the drip lines are listed and underlined in the table as well.

\begin{figure}[t]
  \includegraphics[width=16cm]{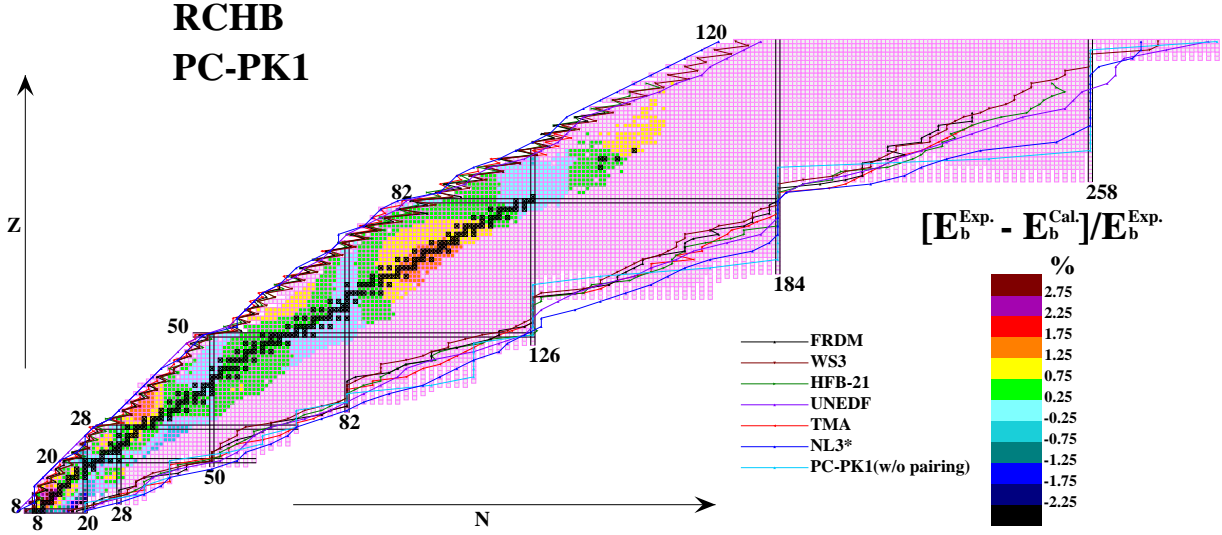}
  \caption{\label{relativediff}(Color online) 9035 bound nuclei from O ($Z=8$) to $Z=120$ predicted by RCHB theory with  PC-PK1~\cite{Zhao2010PhysRevC.82.054319}. For the 2284 nuclei with mass measured, the relative binding energy differences between the data and RCHB calculations are scaled by colors.  Furthermore, the nucleon drip lines predicted by mass table TMA~\cite{Geng2005PTP} , HFB-21~\cite{Goriely2010PRC.82.035804}, FRDM~\cite{Moller1995ADNDT}, WS3~\cite{Wang2010PRC.82.044304}, UNEDF~\cite{Erler2012Nature}, NL3*~\cite{Afanasjev2013PLB,Agbemava2014PRC.89.054320}, and also PC-PK1 without pairing correlations are plotted for comparison.}
\end{figure}

In Fig.~\ref{relativediff}, the nuclear landscape from O ($Z=8$) to $Z=120$ explored by the RCHB theory with PC-PK1 is shown, where the squares represent the bound nuclei. The squares with cross denote the 288 stable or practically stable (that is, have half-lives longer than the expected life of the Solar System) nuclei existing in nature.  Among these 9035 bound nuclei, the masses of 2284 nuclei have been measured experimentally~\cite{Wang2012CPC}. The relative binding energy differences $(E^{\mathrm{Exp.}}_{\mathrm{b}}-E^{\mathrm{Cal.}}_{\mathrm{b}})/E^{\mathrm{Exp.}}_{\mathrm{b}}$ for these measured nuclei scaled by colors are plotted. It is found that most of the deviations are between -0.75\% and  0.75\%, and the root of the relative square (rrs) deviation $\sigma_r=\sqrt{\sum_{i}^{n} \frac{(E^{\rm Exp.}_{\rm b}-E^{\rm Cal.}_{\rm b})^2/(E^{\rm Exp.}_{\rm b})^2}{n}}$ for these 2284 nuclei is around $0.710\%$.  It is noted that the relative differences increase when the neutron (proton) number goes away from the magic number, due to the spherical symmetry adopted here.

Therefore, it is interesting to focus the comparisons on the nuclei with either neutron or proton magic number.  For the nuclei considered here ($Z = $8, 20, 28, 50, 82 or $N = $8, 20, 28, 50, 82, 126), the corresponding rrs deviation is reduced to $0.583\%$.
The root-mean-square (rms) deviation of the binding energy $\sigma=\sqrt{\sum_i^n(E_i^{\mathrm{Exp.}}-E_i^{\mathrm{Cal.}})^2/n}$ for these either neutron or proton magic nuclei is 2.157 MeV, which is remarkably smaller than the rms deviation of 7.960 MeV for the 2284 nuclei with masses measured.
Therefore, we can infer that the deviation between the present RCHB calculations and experimental data comes mainly from the absence of the deformation effects.

\begin{figure}[ht!]
  \centering
  \includegraphics[width=14cm]{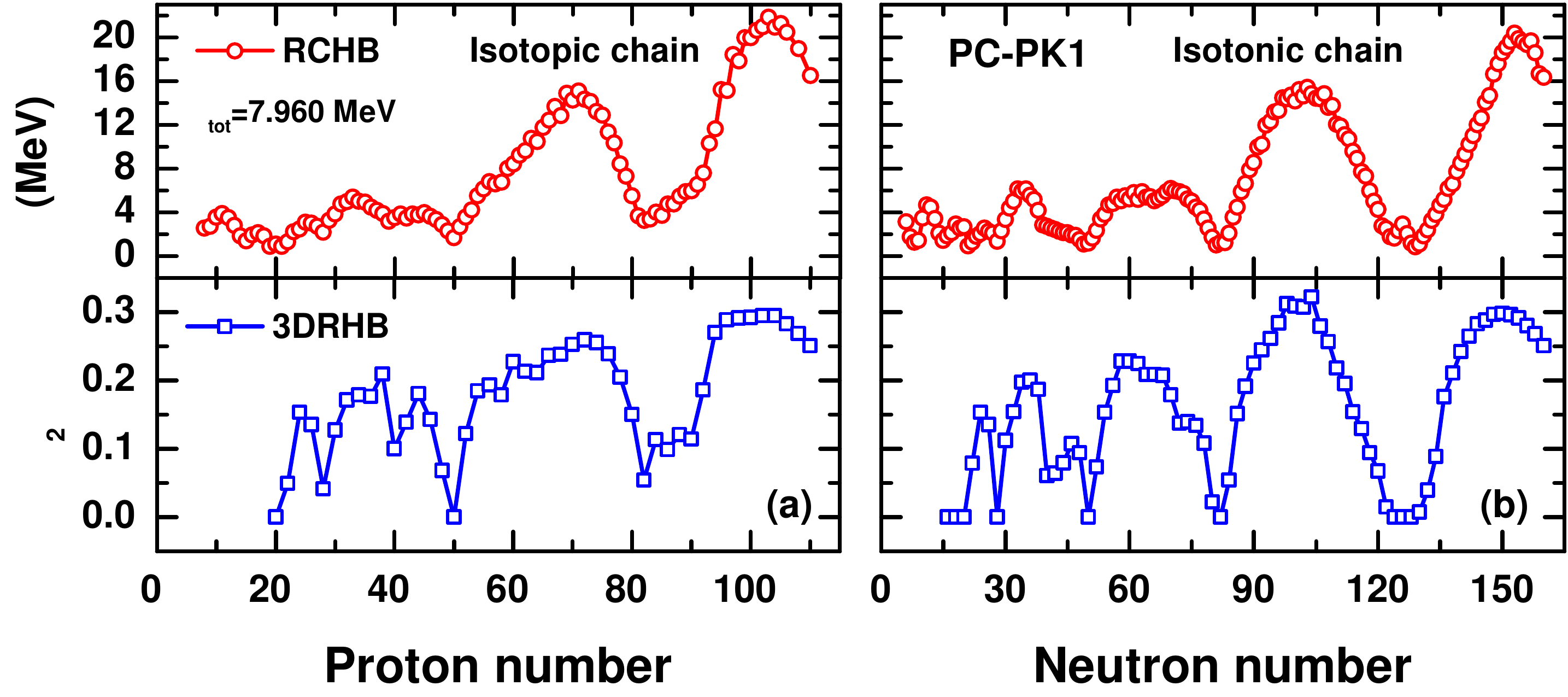}
  \caption{\label{rms-deformation}(Color online) The rms deviation of binding energy $\sigma$ in RCHB calculations compared with the average quadrupole deformation parameter $\langle\beta_2 \rangle= (\sum^{n'}_i |\beta_2^i|)/n'$ for each isotopic chain as a function of the proton number (a), and for each isotonic chain as a function of the neutron number (b). The deformation parameters for even-even nuclei are from 3DRHB~\cite{Niksic2010PhysRevC.81.054318} calculations with PC-PK1.}
\end{figure}

To investigate the deformation effects, the rms deviation of binding energy $\sigma$ in RCHB calculations is compared with the average quadrupole deformation parameter $\langle\beta_2 \rangle= (\sum^{n'}_i |\beta_2^i|)/n'$ for each isotopic chain in Fig.~\ref{rms-deformation}, where the deformation parameters for even-even nuclei are from 3DRHB~\cite{Niksic2010PhysRevC.81.054318} calculations with PC-PK1.
It can be seen that, for isotopic chains, the rms deviations are small at the closed shell $Z = $20, 28, 50, 82, and large around the middle of two closed shells.
Furthermore, a similar tendency for the average deformations is also found in the lower panel of Fig.~\ref{rms-deformation}(a).
Analogously, for isotonic chains, the variation of $\sigma$ is similar to that of $\langle\beta_2 \rangle$ as well.
These correspondences between the $\sigma$ and $\langle\beta_2\rangle$ indicate that taking the deformation effects into account in the future will improve the agreement with the experimental data.

It should be emphasized that it is not the aim of the present RCHB calculations to reproduce the experimental binding energy to a high level of accuracy, but rather to investigate how the continuum states influence the positions of the drip lines, i.e., to which extent the nuclear landscape will be extended by considering the coupling to the continuum.
In Fig.~\ref{relativediff}, the drip lines determined from the mass table TMA~\cite{Geng2005PTP} , HFB-21~\cite{Goriely2010PRC.82.035804}, FRDM~\cite{Moller1995ADNDT}, WS3~\cite{Wang2010PRC.82.044304}, UNEDF~\cite{Erler2012Nature}, NL3*~\cite{Afanasjev2013PLB, Agbemava2014PRC.89.054320}, and also PC-PK1 without pairing correlations are shown.
The comparisons of the nuclear landscape predicted by different models will be discussed
in the following sections in detail.

\subsection{The limits of the nuclear landscape}
\subsubsection{One-nucleon and two-nucleon separation energies}
The one-neutron separation energies $S_{n}$ and one-proton separation energies $S_{p}$ are respectively defined as
\begin{eqnarray}\nonumber
S_{n}(Z,N)=E_{\rm b}(Z,N)-E_{\rm b}(Z,N-1),\\
S_{p}(Z,N)=E_{\rm b}(Z,N)-E_{\rm b}(Z-1,N).
\end{eqnarray}
The two-neutron separation energies $S_{2n}$ and two-proton separation energies $S_{2p}$ are respectively defined as
\begin{eqnarray}\nonumber
S_{2n}(Z,N)=E_{\rm b}(Z,N)-E_{\rm b}(Z,N-2),\\
S_{2p}(Z,N)=E_{\rm b}(Z,N)-E_{\rm b}(Z-2,N).
\end{eqnarray}
These quantities provide information on whether a nucleus is stable  against one or two nucleon emissions, and thus define the nucleon drip lines.
In the present work, we consider a nucleus is bound only if both the one- and two-nucleon separation energies of this nucleus are positive.
For each isotopic chain, the last bound nucleus in the neutron-rich side
is the neutron drip-line nucleus and that in the neutron-deficient side is
the proton drip-line nucleus.

\begin{figure}[h]
  \centering
     \includegraphics[width=16cm]{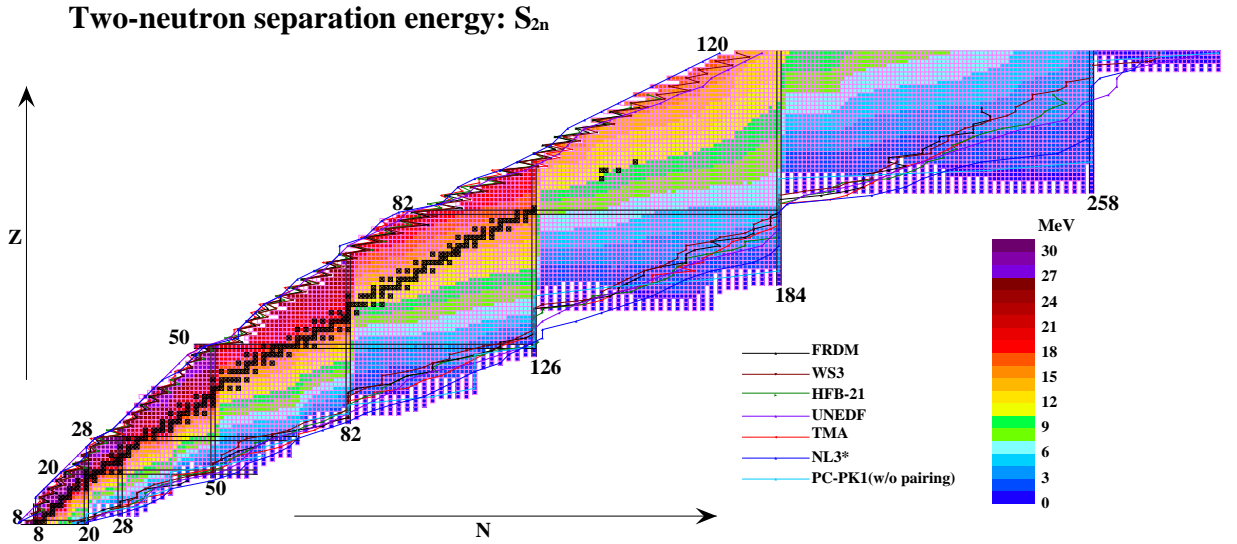}
   \caption{(Color online) Two-neutron separation energies of bound nuclei in the RCHB calculations with PC-PK1 scaled by colors. Furthermore, the nucleon drip lines predicted by mass table TMA~\cite{Geng2005PTP} , HFB-21~\cite{Goriely2010PRC.82.035804}, FRDM~\cite{Moller1995ADNDT}, WS3~\cite{Wang2010PRC.82.044304}, UNEDF~\cite{Erler2012Nature}, NL3*~\cite{Afanasjev2013PLB,Agbemava2014PRC.89.054320}, and also PC-PK1 without pairing correlations are plotted for comparison.}\label{s2n}
\end{figure}

In Fig.~\ref{s2n}, the two-neutron separation energies $S_{2n}$ of
the bound nuclei predicted by the RCHB theory with PC-PK1 are shown.
From a global view, $S_{2n}$ is large at proton drip-line side and
close to zero at neutron drip-line side. For a given isotopic (isotonic) chain, $S_{2n}$ decreases with the increasing neutron number, while increases with the proton number.
It is found that there are 87 nuclei with the predicted two-neutron separation energies larger than 30 MeV, and most of them are located at the proton-rich side of nuclear landscape with $Z \leqslant 50$.
There are 388 nuclei with $S_{2n}$ in the range 21--30 MeV, mainly located in the proton-rich region of the nuclear chart; 1967 nuclei with $S_{2n}$ in the range of 12--21 MeV, most of which are near the valley of stability; 4715 nuclei with $S_{2n}$ in the range 3-12 MeV which lie on the neutron-rich region mostly. In addition, there are 1828 nuclei with their $S_{2n}$ less than 3 MeV, and most of them are located in the region far from the stability line, and are even approaching to the neutron drip line.

It should be noted that, there are 586 weakly bound nuclei with $S_{2n}\leqslant 1$ MeV in the RCHB calculations. They are the extremely neutron-rich nuclei predicted by RCHB, and many of them  lie even beyond the  neutron drip lines predicted by the other
nuclear mass models.
For these weakly bound nuclei, as the neutron Fermi surface is close to the continuum threshold,  pairing correlations could scatter the nucleons from bound states to the continuum, thus provide a significant coupling between the continuum and bound states~\cite{Qu2013scichia}.
The RCHB theory allows a proper treatment of the continuum and the coupling
to the bound states, so it predicts a more extended neutron drip line than the other models.
In addition, the nearly vanishing $S_{2n}$ around the neutron drip line might be regarded as a sign of the neutron giant halo~\cite{Meng1998PhysRevLett.80.460, Meng2002PhysRevC.65.041302}, which invokes the further analysis of neutron radii and single particle levels.

$S_{2n}$ is a widely used probe of the neutron shell structure. Generally, for a given isotopic chain $S_{2n}$ decreases smoothly with the neutron number $N$, except at a magic number where $S_{2n}$ drops significantly.
An abrupt decline of $S_{2n}$ indicates the occurrence of neutron shell closure. It can be seen in Fig.~\ref{s2n} that, the significant drops exist at the traditional magic numbers $N = 20,~28,~50,~82$ and 126, which demonstrates that these shell closures are well reproduced in the RCHB theory.
Apart from this, a dramatic decline of $S_{2n}$ can be found at $N = 184$, which indicates that neutron number $N = 184$ may be a new magic number~\cite{Zhang2005NPA}.

\begin{figure}[ht!]
  \centering
     \includegraphics[width=16cm]{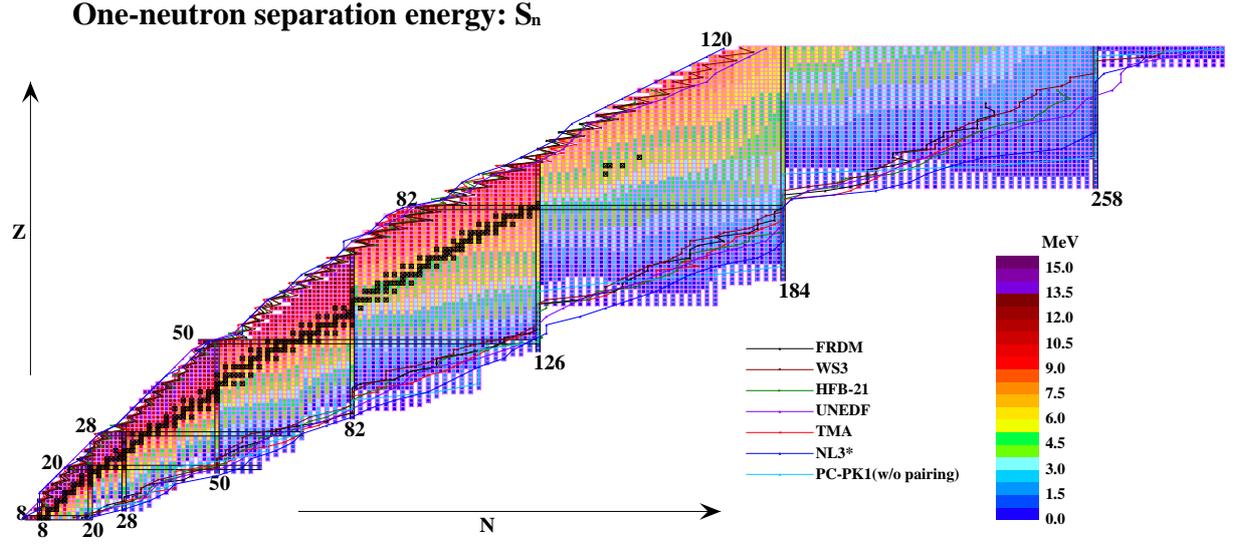}
   \caption{(Color online) One-neutron separation energies of bound nuclei in the RCHB calculations with PC-PK1 scaled by colors. Furthermore, the nucleon drip lines predicted by mass table TMA~\cite{Geng2005PTP} , HFB-21~\cite{Goriely2010PRC.82.035804}, FRDM~\cite{Moller1995ADNDT}, WS3~\cite{Wang2010PRC.82.044304}, UNEDF~\cite{Erler2012Nature}, NL3*~\cite{Afanasjev2013PLB,Agbemava2014PRC.89.054320}, and also PC-PK1 without pairing correlations are plotted for comparison.}\label{sn}
\end{figure}

In Fig.~\ref{sn}, the one-neutron separation energies $S_{n}$ of bound nuclei predicted by the RCHB theory are shown with different colors. Analogous to $S_{2n}$, $S_{n}$ decreases (increases) with the increase of the neutron (proton) number for a given isotopic (isotonic) chain. It is found that 95 nuclei have their $S_{n}$ larger than 15 MeV, 214 nuclei within 12--15 MeV, 610 nuclei within 9--12 MeV, 1516 nuclei within 6--9 MeV, 2583 nuclei within 3--6 MeV, and 4006 nuclei less than 3 MeV.
Note that there are 1096 nuclei with $S_{n} \leqslant 1$ MeV and 466 nuclei with $S_{n} \leqslant 0.5$ MeV. Due to the pairing correlations, which make the nuclei with even nucleon number  more bound than their neighbors with odd nucleon number, $S_{n}$ has a ragged evolution pattern with the variation of neutron number that zigzags between odd- and even-particle nuclei.

\begin{figure}[ht!]
  \centering
     \includegraphics[width=16cm]{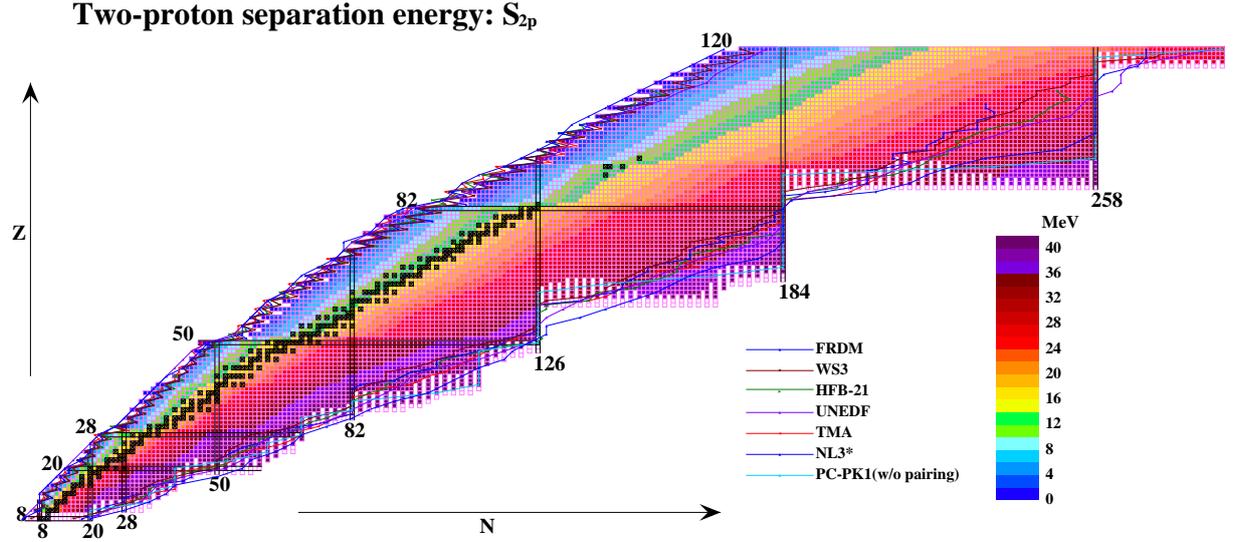}
   \caption{(Color online) Two-proton separation energies of bound nuclei in the RCHB calculations with PC-PK1 scaled by colors. Furthermore, the nucleon drip lines predicted by mass table TMA~\cite{Geng2005PTP} , HFB-21~\cite{Goriely2010PRC.82.035804}, FRDM~\cite{Moller1995ADNDT}, WS3~\cite{Wang2010PRC.82.044304}, UNEDF~\cite{Erler2012Nature}, NL3*~\cite{Afanasjev2013PLB,Agbemava2014PRC.89.054320}, and also PC-PK1 without pairing correlations are plotted for comparison.}\label{s2p}
\end{figure}

\begin{figure}[ht!]
  \centering
     \includegraphics[width=16cm]{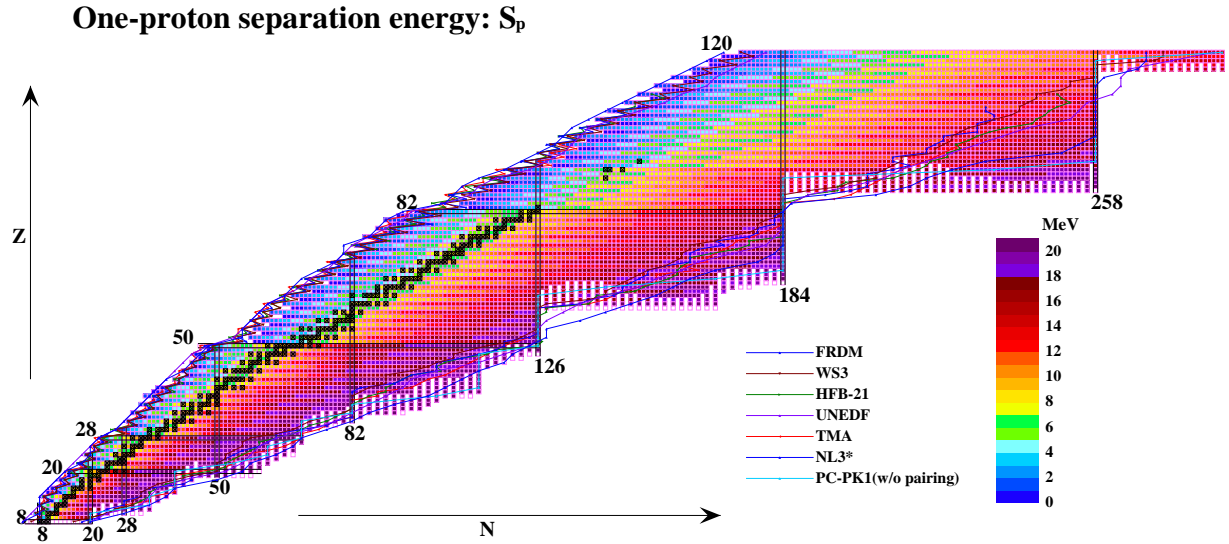}
   \caption{(Color online) One-proton separation energies of bound nuclei in the RCHB calculations with PC-PK1 scaled by colors. Furthermore, the nucleon drip lines predicted by the mass table TMA~\cite{Geng2005PTP} , HFB-21~\cite{Goriely2010PRC.82.035804}, FRDM~\cite{Moller1995ADNDT}, WS3~\cite{Wang2010PRC.82.044304}, UNEDF~\cite{Erler2012Nature}, NL3*~\cite{Afanasjev2013PLB,Agbemava2014PRC.89.054320}, and also PC-PK1 without pairing correlations are plotted for comparison.}\label{sp}
\end{figure}

In addition to the neutron separation energies $S_{2n}$ and $S_{n}$, the two-proton $S_{2p}$ and one-proton separation energies $S_{p}$ of the bound nuclei are scaled by colors in Figs.~\ref{s2p} and~\ref{sp}.
The proton separation energies increase with the neutron number for a given isotopic chain, while decrease with the proton number for a given isotonic chain.
In the present calculations, it is found that there are 520 nuclei with $S_{2p}\geqslant$ 40 MeV, 594 nuclei with $36\leqslant S_{2p} < 40$ MeV, 802 nuclei with $32\leqslant S_{2p} < 36$ MeV, 977 nuclei with $28\leqslant S_{2p} < 32$ MeV, 1150 nuclei with $24\leqslant S_{2p} < 28$ MeV, 1088 nuclei with $20\leqslant S_{2p} < 24$ MeV, 957 nuclei with $16\leqslant S_{2p} < 20$ MeV, 857 nuclei with $12\leqslant S_{2p} < 16$ MeV, 781 nuclei with $8\leqslant S_{2p} < 12$ MeV, 731 nuclei with $4\leqslant S_{2p} < 8$ MeV and 381 nuclei with $0\leqslant S_{2p} < 4$ MeV.
It is noted that the proton drip line predicted by the RCHB calculations is close to the  proton drip lines predicted by the other mass tables.
At the proton magic number 20, 28, 50 and 82, $S_{2p}$ changes abruptly. This indicates that the RCHB theory reproduces the traditional proton closed shells quite well.
Note that a sudden change of the $S_{2p}$ can be also found at $Z = 92$, while this has been considered as a pseudo shell in the previous relativistic mean field calculations~\cite{Geng2006CPL}.

In Fig.~\ref{sp}, there are 589 nuclei with $S_{p}\geqslant$ 20 MeV, 1348 nuclei with $16\leqslant S_{p} < 20$ MeV, 2083 nuclei with $12\leqslant S_{p} < 16$ MeV, 2075 nuclei with $8\leqslant S_{p} < 12$ MeV, 1648 nuclei with $4\leqslant S_{p} < 8$ MeV and 1227 nuclei with $0\leqslant S_{p} < 4$ MeV.
Due to the pairing correlations, $S_{p}$ is staggering between odd- and even-particle nuclei for a given isotonic chain.
It should be noted that the number of the nuclei with $S_{2p}\leqslant 1$ MeV is 176, which is much
less than the number of the nuclei with $S_{2n}\leqslant 1$ MeV.
This is consistent with the number of halo nuclei observed in the proton-rich side is much less than that in the neutron-rich side, which
should be attributed to the existence of the Coulomb barrier.
Moreover, although the proton Fermi surface is approaching  zero for the neutron-deficient nuclei, the contributions from the continuum are suppressed  due to the Coulomb barrier in comparison with that for the neutron-rich nuclei.
This also explains why the proton drip line obtained by the RCHB calculations is
very close to the ones predicted by the other mass models.

\subsubsection{Neutron drip line}
In Fig.~\ref{relativediff}, the nucleon drip lines predicted by the mass tables TMA~\cite{Geng2005PTP}, HFB-21~\cite{Goriely2010PRC.82.035804}, FRDM~\cite{Moller1995ADNDT}, WS3~\cite{Wang2010PRC.82.044304}, UNEDF~\cite{Erler2012Nature}, and NL3*~\cite{Afanasjev2013PLB, Agbemava2014PRC.89.054320} have been plotted for comparison, together with the drip line given by the PC-PK1 calculations
without taking pairing correlations into account.
At present, the experimental proton-rich border of the nuclear territory has been reached up to protactinium~\cite{http://www.nndc.bnl.gov/} ($Z=91$).
Due to the Coulomb repulsive interaction among protons, the proton drip line does not lie so far away from the valley of stability. Moreover, the proton continuum is effectively shifted up in energy as a result of the confining
effect of the Coulomb barrier. Therefore, the proton drip line obtained by the present calculations is very close to those from other mass tables, which are roughly the same with the experimental observations.

On the neutron-rich side, however, the neutron-rich boundary is known only up to oxygen ($Z=8$) experimentally.
In contrast to the proton drip line, the locations of neutron drip lines from the various mass tables obviously differ with each other, especially for the heavy mass region.
The differences of neutron drip lines increase with the proton number.
The neutron drip line predicted by the RCHB theory
is most extended compared with other models.
For example, the last bound neutron-rich nucleus of the Kr ($Z=36$) isotope is $^{118}$Kr in the calculations of FRDM, WS3, HFB-21, TMA and the PC-PK1 calculations without pairing correlations, $^{122}$Kr in UNDEF and $^{128}$Kr in NL3*, while
the present RCHB calculations predict $^{136}$Kr as the last bound nucleus of Kr isotopes in the neutron rich side.

\begin{table}[h]
\caption{The neutron number of the most neutron-rich even-even nuclei predicted to be bound in the RCHB theory, in comparison with the calculations without pairing correlations. }\label{table-dir}
\begin{center}
\footnotesize{{
\begin{tabular}{rcc|rcc|rcc}
\hline
\hline
\multicolumn{1}{c}{Element}
&\multicolumn{2}{c|}{Neutron number}
&\multicolumn{1}{c}{Element}
&\multicolumn{2}{c|}{Neutron number}
&\multicolumn{1}{c}{Element}
&\multicolumn{2}{c}{Neutron number}
\\
\multicolumn{1}{c}{($Z$)}
&\multicolumn{1}{c}{No pairing}
&\multicolumn{1}{c|}{RCHB}
&\multicolumn{1}{c}{($Z$)}
&\multicolumn{1}{c}{No pairing}
&\multicolumn{1}{c|}{RCHB}
&\multicolumn{1}{c}{($Z$)}
&\multicolumn{1}{c}{No pairing}
&\multicolumn{1}{c}{RCHB}
\\
\hline
O  ( 8)&	 20&	 20&    Pd (46) & 112	& 118 &  Po ( 84)  &  184  &	184\\
Ne (10)&	 20&	 32&    Cd (48) & 112	& 126 &  Rn ( 86)  &  184  &	184\\
Mg (12)&	 34&	 34&    Sn (50) & 126	& 126 &  Ra ( 88)  &  184  &	258\\
Si (14)&	 34&	 38&    Te (52) & 126	& 126 &  Th ( 90)  &  184  &	258\\
S  (16)&	 40&	 40&    Xe (54) & 126	& 126 &  U  ( 92)  &  234  &	258\\
Ar (18)&	 40&	 44&    Ba (56) & 126	& 126 &  Pu ( 94)  &  258  &	258\\
Ca (20)&	 40&	 60&    Ce (58) & 126	& 126 &  Cm ( 96)  &  258  &	258\\
Ti (22)&	 54&	 62&    Nd (60) & 126	& 168 &  Cf ( 98)  &  258  &	258\\
Cr (24)&	 62&	 68&    Sm (62) & 126	& 168 &  Fm (100)  &  258  &	258\\
Fe (26)&	 62&	 68&    Gd (64) & 142	& 176 &  No (102)  &  258  &	258\\
Ni (28)&	 70&	 70&    Dy (66) & 168	& 184 &  Rf (104)  &  258  &	258\\
Zn (30)&	 70&	 72&    Er (68) & 184	& 184 &  Sg (106)  &  258  &	258\\
Ge (32)&	 70&	 82&    Yb (70) & 184	& 184 &  Hs (108)  &  258  &	258\\
Se (34)&	 82&	 94&    Hf (72) & 184	& 184 &  Ds (110)  &  258  &	258\\
Kr (36)&	 82&	100&    W  (74) & 184	& 184 &  Cn (112)  &  258  &	258\\
Sr (38)&	102&	110&    Os (76) & 184	& 184 &  Fl (114)  &  258  &	258\\
Zr (40)&	112&	112&    Pt (78) & 184	& 184 &  Lv (116)  &  258  &	288\\
Mo (42)&	112&	112&    Hg (80) & 184	& 184 &  Og (118)  &  258  &	288\\
Ru (44)&	112&	112&    Pb (82) & 184	& 184 &  $Z=120$  &  288  &	288\\

\hline
\hline
\end{tabular}}}
\end{center}
\end{table}

\begin{table}[ht!]
\caption{Number of neutron-rich nuclei predicted to be bound by the RCHB, but unbound in the FRDM. }\label{table-frdm}
\begin{center}
\footnotesize{{
\begin{tabular}{rc|rc|rc|rc|rc}
\hline
\hline
\multicolumn{1}{c}{Element}
&\multicolumn{1}{c|}{$\Delta N$}
&\multicolumn{1}{c}{Element}
&\multicolumn{1}{c|}{$\Delta N$}
&\multicolumn{1}{c}{Element}
&\multicolumn{1}{c|}{$\Delta N$}
&\multicolumn{1}{c}{Element}
&\multicolumn{1}{c|}{$\Delta N$}
&\multicolumn{1}{c}{Element}
&\multicolumn{1}{c}{$\Delta N$}
\\
\multicolumn{1}{c}{($Z$)}
&\multicolumn{1}{c|}{ }
&\multicolumn{1}{c}{($Z$)}
&\multicolumn{1}{c|}{ }
&\multicolumn{1}{c}{($Z$)}
&\multicolumn{1}{c|}{ }
&\multicolumn{1}{c}{($Z$)}
&\multicolumn{1}{c|}{ }
&\multicolumn{1}{c}{($Z$)}
&\multicolumn{1}{c}{ }
\\
\hline
 O  (8)	&2	&  Ni (28)&	6	  &Rh (45)	&11	& Dy (66)	&25	&  Fr (87)&	29	\\
Ne (10)	&10	&  Cu (29)&	6	  &Pd (46)	&13	& Ho (67)	&23	&  Ra (88)&	28	\\
Na (11)	&10	&  Zn (30)&	1	  &Ag (47)	&9	& Er (68)	&24	&  Ac (89)&	28	\\
Mg (12)	&6	&  Ga (31)&	1	  &Cd (48)	&6	& Tm (69)	&26	&  Th (90)&	37	\\
Al (13)	&8	&  Ge (32)&	2	  &In (49)	&4	& Yb (70)	&20	&  Pa (91)&	43	\\
Si (14)	&6	&  As (33)&	1	  &Sn (50)	&6	& Lu (71)	&21	&   U (92)&	39	\\
 P (15)	&8	&  Se (34)&	6	  &Sb (51)	&8	& Hf (72)	&21	&  Np (93)&	38	\\
 S (16)	&6	&  Br (35)&	7	  &Te (52)	&2	& Ta (73)	&19	&  Pu (94)&	44	\\
 K (19)	&14	&  Kr (36)&	9	  &I (53)	 &1	& W (74)	&13	&  Am (95)&	45	\\
Ca (20)	&10	&  Rb (37)&	10	&Xe (54)	&1	& Re (75)	&16	&  Cm (96)&	38	\\
Sc (21)	&10	&  Sr (38)&	14	&Pr (59)	&20	& Os (76)	&6	&  Bk (97)&	38	\\
Ti (22)	&12	&   Y (39)&	21	&Nd (60)	&16	& Ir (77)	&6	&  Cf (98)&	37	\\
 V (23)	&10	&  Zr (40)&	19	&Pm (61)	&15	& Pt (78)	&6	&  Es (99)&	35	\\
Cr (24)	&10	&  Nb (41)&	12	&Sm (62)	&19	& Au (79)	&2	& Fm (100)&	36	\\
Mn (25)	&8	&  Mo (42)&	12	&Eu (63)	&22	& Hg (80)	&2	& Md (101)&	28	\\
Fe (26)	&5	&  Tc (43)&	13	&Gd (64)	&25	& Tl (81)	&2	& No (102)&	28	\\
Co (27)	&4	&  Ru (44)&	10	&Tb (65)	&27	& Rn (86)	&31	& Lr (103)& 28	\\

\hline
\hline
\end{tabular}}}
\end{center}
\end{table}

In order to explore where and how the continuum extends the neutron drip line, we compare the neutron drip lines obtained from the RCHB calculations with
those from the calculations without pairing correlation.
The most neutron-rich even-even nuclei predicted to be bound in the RCHB theory and the calculations without pairing is listed in Table~\ref{table-dir} for different isotopic chains.
Marked at sub-shell or shell closures $N=$40, 70, 82, 112, 126, 184 and 258, the extensions are dramastic
when the next sub-shell or shell is occupied.

A similar comparison is performed for the RCHB theory and the FRDM. The number of neutron-rich nuclei predicted to be bound by the RCHB, but unbound in the FRDM are listed in Table~\ref{table-frdm}.
It can be found that the neutron drip line of the RCHB results is much farther away from the $\beta$-stability line than that of the FRDM for almost all isotopic chains.
For example, from the isotopic chains Se ($Z=34$) to Zr ($Z=40$), the FRDM gives the drip line at $N=82$. The RCHB theory predicts 6 to 19 more bound neutron-rich nuclei from the isotopic chains Se ($Z=34$) to Zr ($Z=40$).

Comparing to the drip line of RHB calculations in harmonic oscillator (HO) basis with NL3*~\cite{Afanasjev2013PLB, Agbemava2014PRC.89.054320}, the present RCHB calculations show considerable difference from it.
The considerable difference of the drip line between RHB in HO calculations and present RCHB calculations are due to different functionals (PC-PK1 or NL3*), different pairing forces used, and the treatment of continuum in coordinate or HO space. Future investigations with same density functional and pairing force in both coordinate space and in harmonic oscillator basis are needed.

\subsubsection{Continuum effects}
\begin{figure}[t]
  \centering
     \includegraphics[width=14cm]{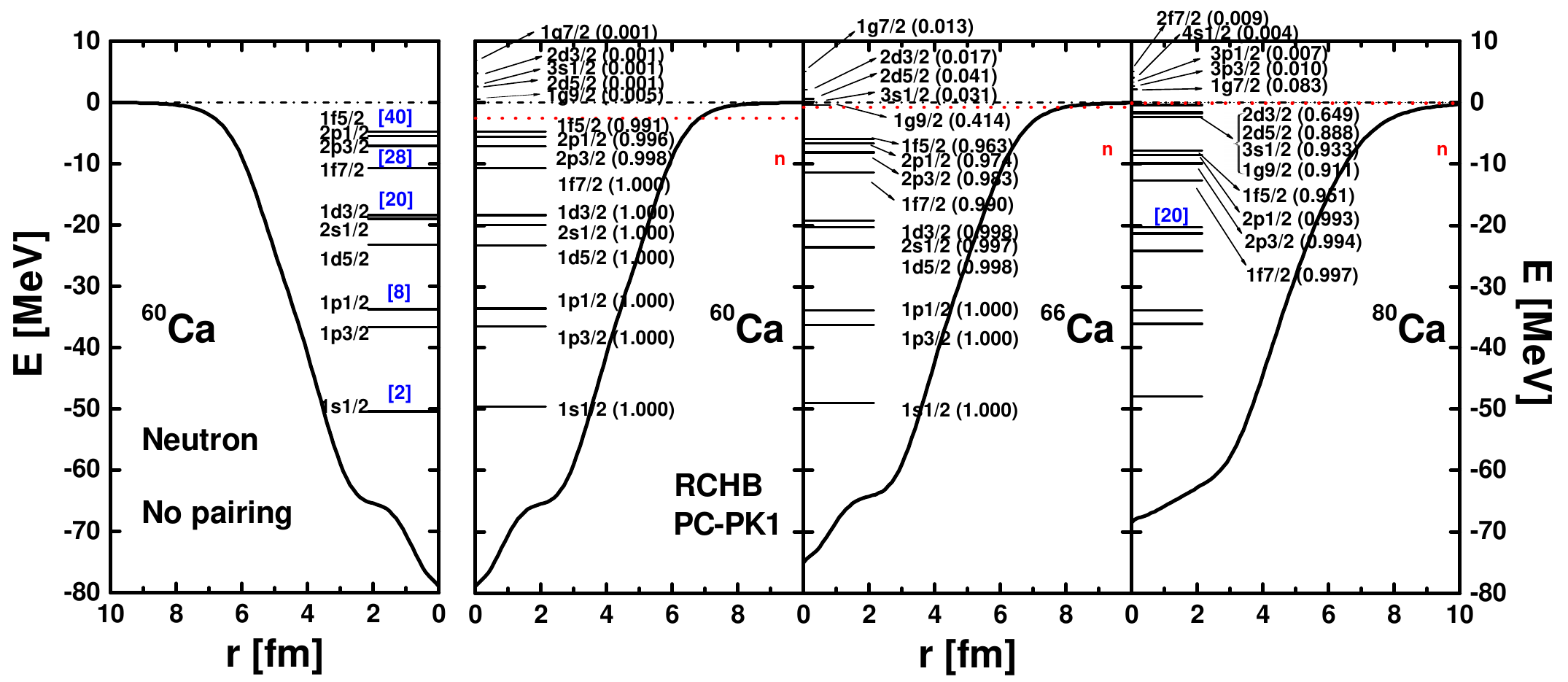}
   \caption{(Color online) The neutron single particle levels (occupation probability) for $^{60,66,80}$Ca in canonical basis from the RCHB calculations, in comparison with those for $^{60}$Ca from the calculations without pairing. The corresponding potentials $V+S$ are also shown.}\label{ca-pot-spl}
\end{figure}
\begin{figure}[ht!]
  \centering
     \includegraphics[width=14cm]{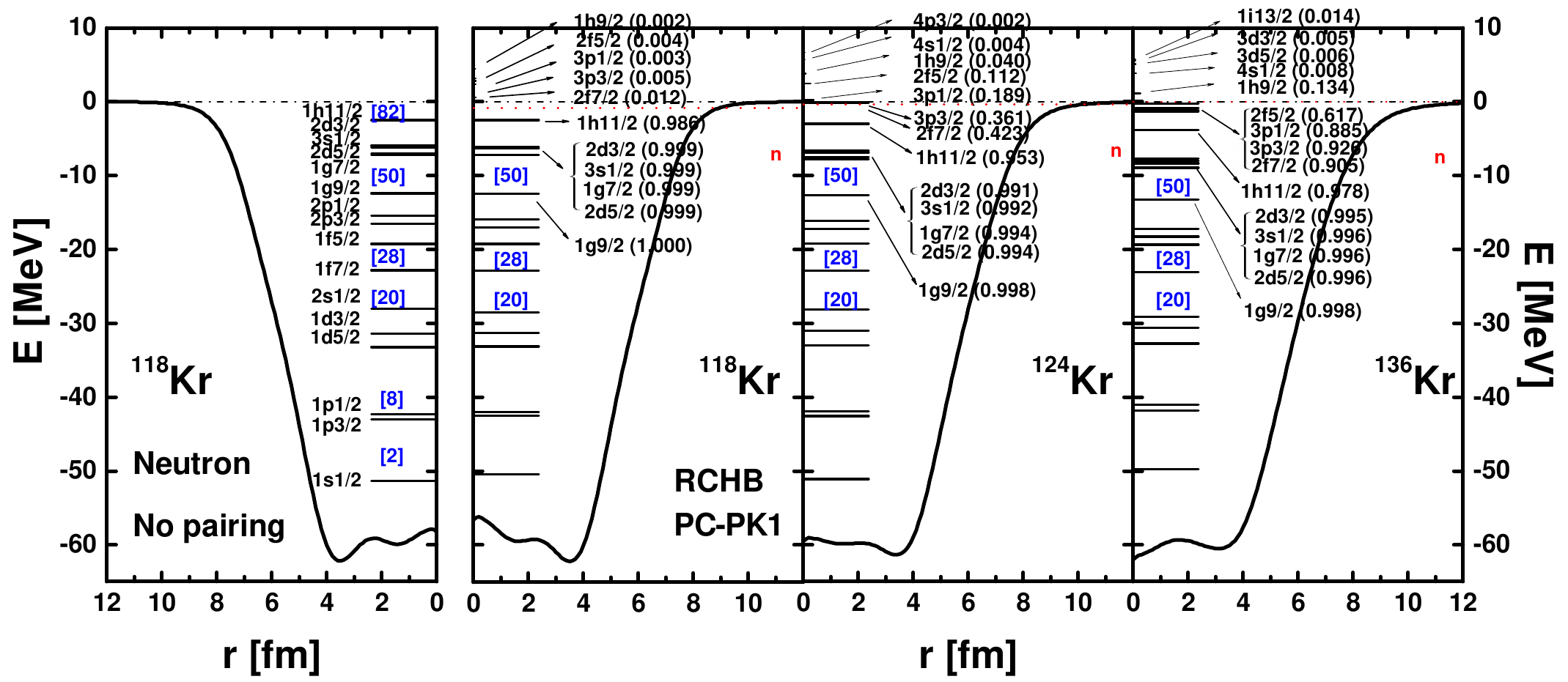}
   \caption{(Color online) Same as Fig.~\ref{ca-pot-spl}, but for $^{118,124,136}$Kr from the RCHB calculations and $^{118}$Kr from the calculations without pairing.}\label{kr-pot-spl}
\end{figure}
\begin{figure}[ht!]
  \centering
     \includegraphics[width=14cm]{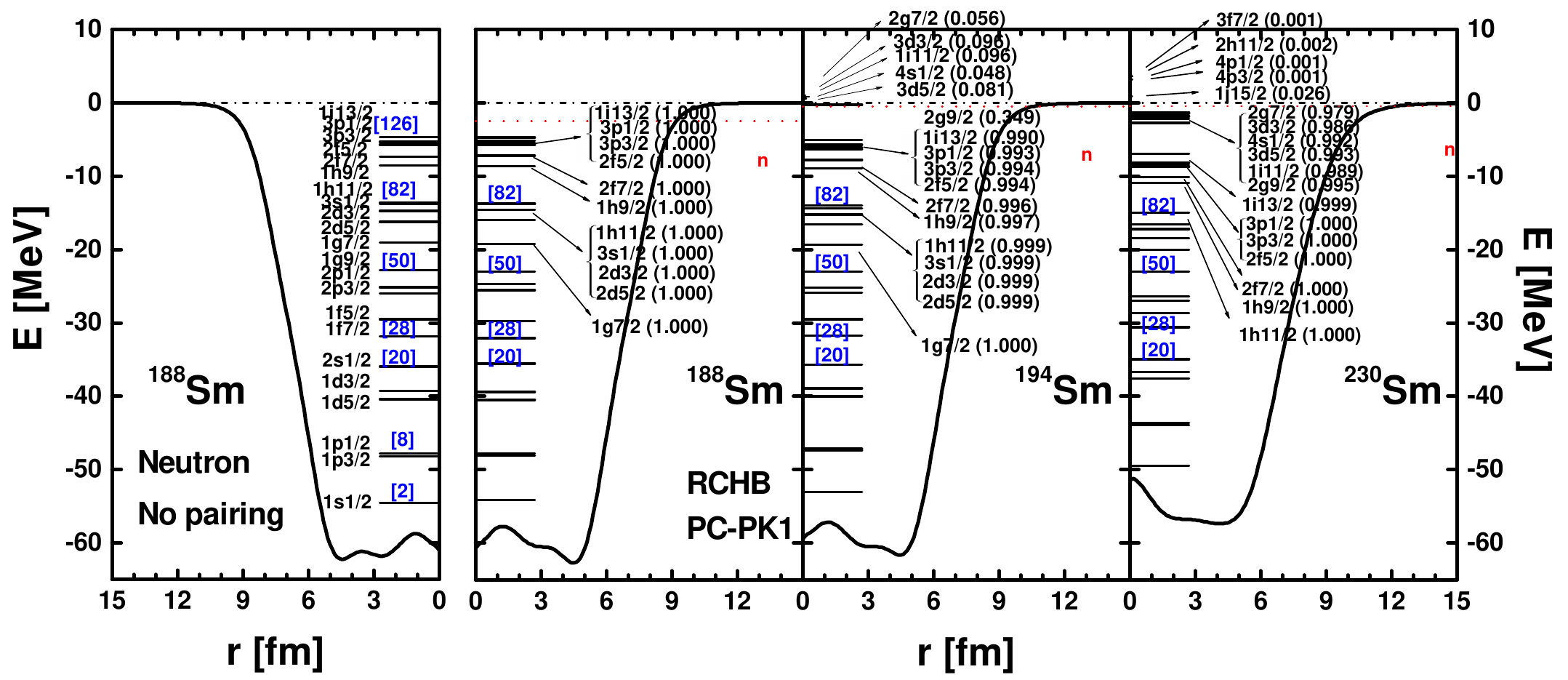}
   \caption{(Color online) Same as Fig.~\ref{ca-pot-spl}, but for $^{188,194,230}$Sm from the RCHB calculations and $^{188}$Sm from the calculations without pairing. }\label{sm-pot-spl}
\end{figure}

To illustrate the extension of the neutron drip line in RCHB theory, by taking $^{60,66,80}$Ca, $^{118,124,136}$Kr and $^{188,194,230}$Sm as examples,  we compare the neutron single particle levels in canonical basis from the RCHB calculations with those of $^{60}$Ca, $^{118}$Kr and $^{188}$Sm from the calculations without pairing, respectively in Figs.~\ref{ca-pot-spl}-\ref{sm-pot-spl}.

For Ca isotopes in Fig.~\ref{ca-pot-spl}, $^{60}$Ca is the neutron drip-line nucleus in the calculations without pairing, while with 20 more neutrons, $^{80}$Ca is the neutron drip-line nucleus in the RCHB calculations.
In  the calculations without pairing, the neutron Fermi surface  of $^{60}$Ca is $1f_{5/2}$, the last bound level.
The occupation of the next level $1g_{9/2}$ which has a positive single particle energy will lead to the unbound nucleus $^{62}$Ca.
In RCHB theory,  due to the pairing correlations, the neutron Fermi surface locates between $1f_{5/2}$ and $1g_{9/2}$, and the continuum in the canonical basis also have small occupation probabilities in $^{60}$Ca.
Adding more neutrons, see $^{66}$Ca in the figure, although the occupation probabilities of the continuum in the canonical basis become larger, the neutron Fermi surface remains negative. This means that the nucleus is still bound.
Moreover, with more neutrons added, the level $1g_{9/2}$ becomes negative and the neutron Fermi surface remains negative up to $^{80}$Ca, which is turned out to be the last bound nucleus in the RCHB calculations.

The mechanisms are similar for the Kr and Sm isotopic chains, as shown in Figs. \ref{kr-pot-spl} and \ref{sm-pot-spl}, respectively. Therefore, we conclude that the pairing correlations and the coupling
between continuum and bound states are responsible for the extension of the neutron drip line.

\subsubsection{Discussion on deformation effects}
It should be noted that the present survey is performed with the assumption of spherical symmetry.
A natural question is whether and how the deformation will influence the position of the neutron drip line.

To explore the deformation effects on the position of neutron drip line, taking argon isotopic chain as an example, the calculations with the deformed relativistic Hartree-Bogoliubov theory in continuum (DRHBc)~\cite{Zhou2010PhysRevC.82.011301,Li2012PhysRevC.85.024312}  are performed,
where the axial deformation, continuum and pairing correlation
are treated self-consistently.
In the DRHBc calculations, the relativistic density functional PC-PK1 is employed.
We use a box size $R_{\rm box} =$ 20 fm, a mesh size $\Delta r =$ 0.1 fm, a cutoff energy of Woods-Saxon basis $E^{+}_{\rm cut} =$ 160 MeV and a cutoff of angular moment $J_{\rm max}=25/2 \hbar$.
For the particle-particle channel, the pairing force and pairing parameters are the same as Ref.~\cite{Li2012PhysRevC.85.024312}.

\begin{figure}[t]
  \centering
     \includegraphics[width=8cm]{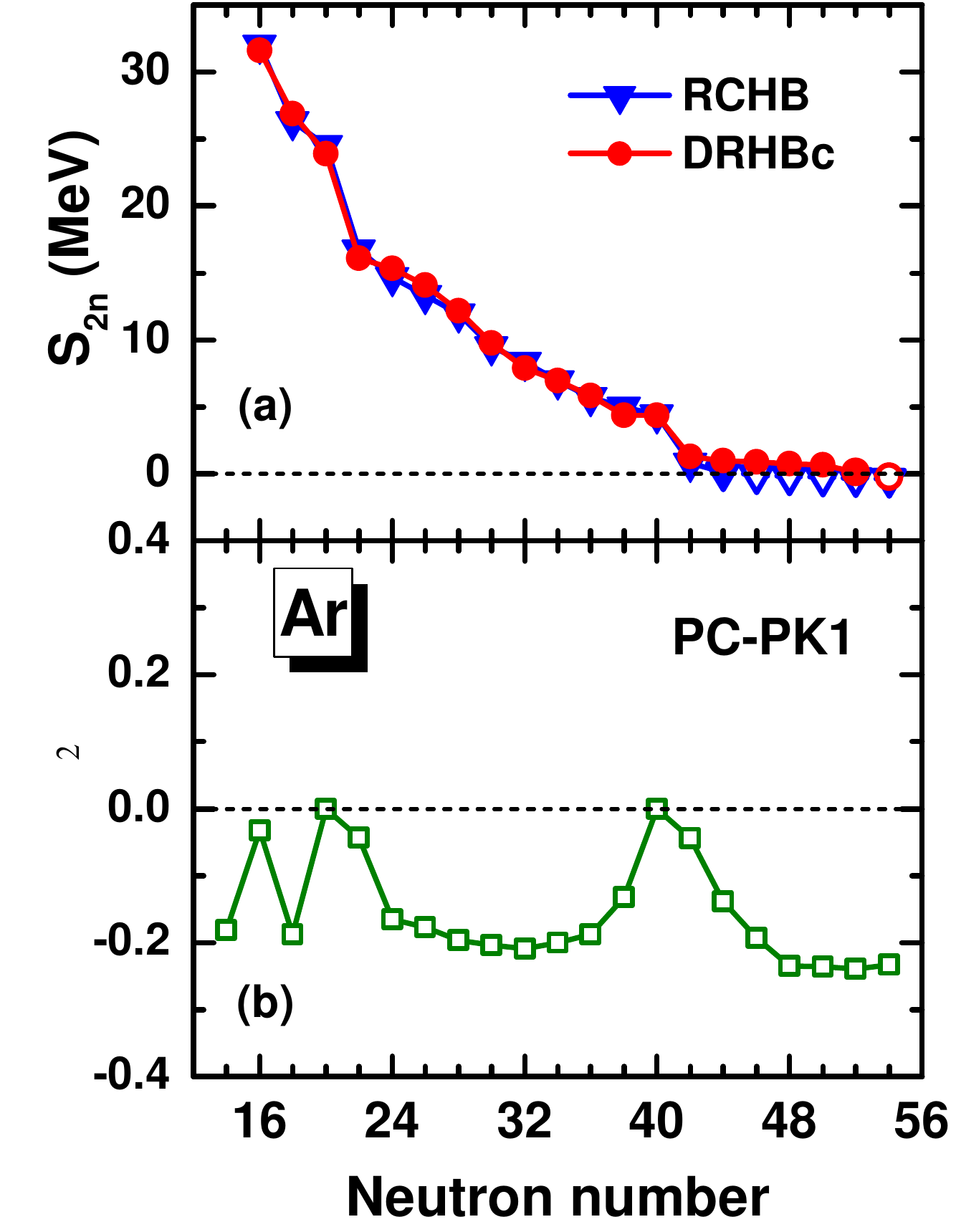}
   \caption{(Color online) (a) Two-neutron separation energy $S_{2n}$ from the RCHB calculations and DRHBc calculations for argon isotopes as a function of neutron number. (b) The ground-state quadrupole deformation $\beta_2$ for argon isotopes as a function of neutron number within the DRHBc calculations with PC-PK1. }\label{S2n-beta}
\end{figure}

In Fig.~\ref{S2n-beta}, the two-neutron separation energies from RCHB and DRHBc calculations for argon (Ar)
isotopes and the corresponding ground-state deformations from the DRHBc calculations are shown. It is can be seen that most argon isotopes are oblate in ground states, and the deformation parameter $\beta_2$ increases with the increase of neutron number for $N>40$. In Fig.~\ref{S2n-beta}(a),  $S_{2n}$ from the DRHBc calculations are close to those from the RCHB calculations globally. However, the deviation exists at $N>44$ where the
large deformation occurs. In the RCHB calculations, the $S_{2n}$ becomes negative after $N=44$ and still decreases with the increase of neutron number, while $S_{2n}$ from DRHBc calculations keeps positive for $44<N< 52$ and becomes negative for $N=54$, i.e. the neutron drip-line nuclei of argon isotopes are $^{62}$Ar in the RCHB and $^{70}$Ar in the DRHBc calculations, respectively.

\begin{figure}[t]
  \centering
     \includegraphics[width=8cm]{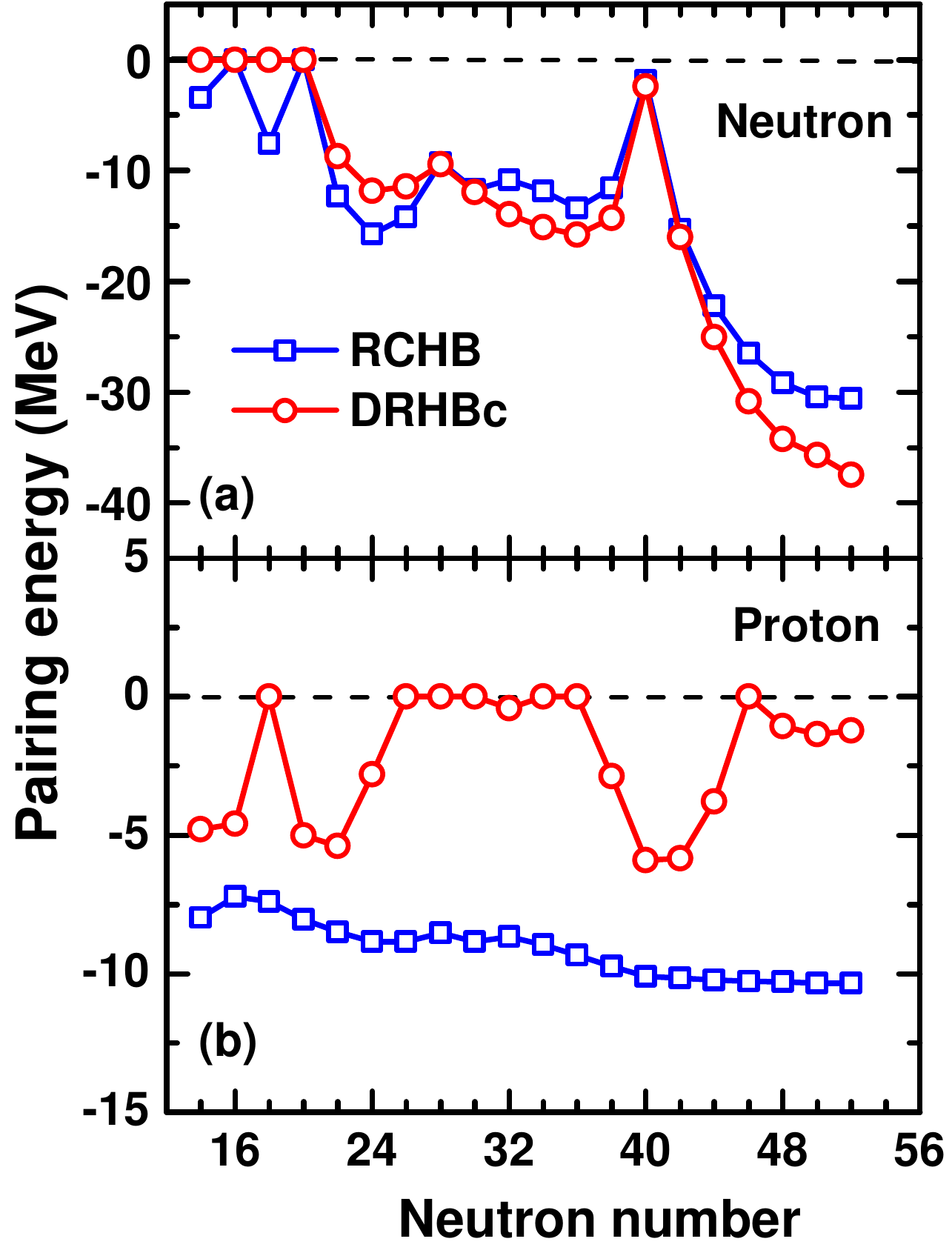}
   \caption{(Color online) Comparison of pairing energies between the DRHBc calculations and RCHB calculations for argon isotopic chain.}\label{epair-Ar}
\end{figure}
\begin{figure}[t]
  \centering
     \includegraphics[width=12cm]{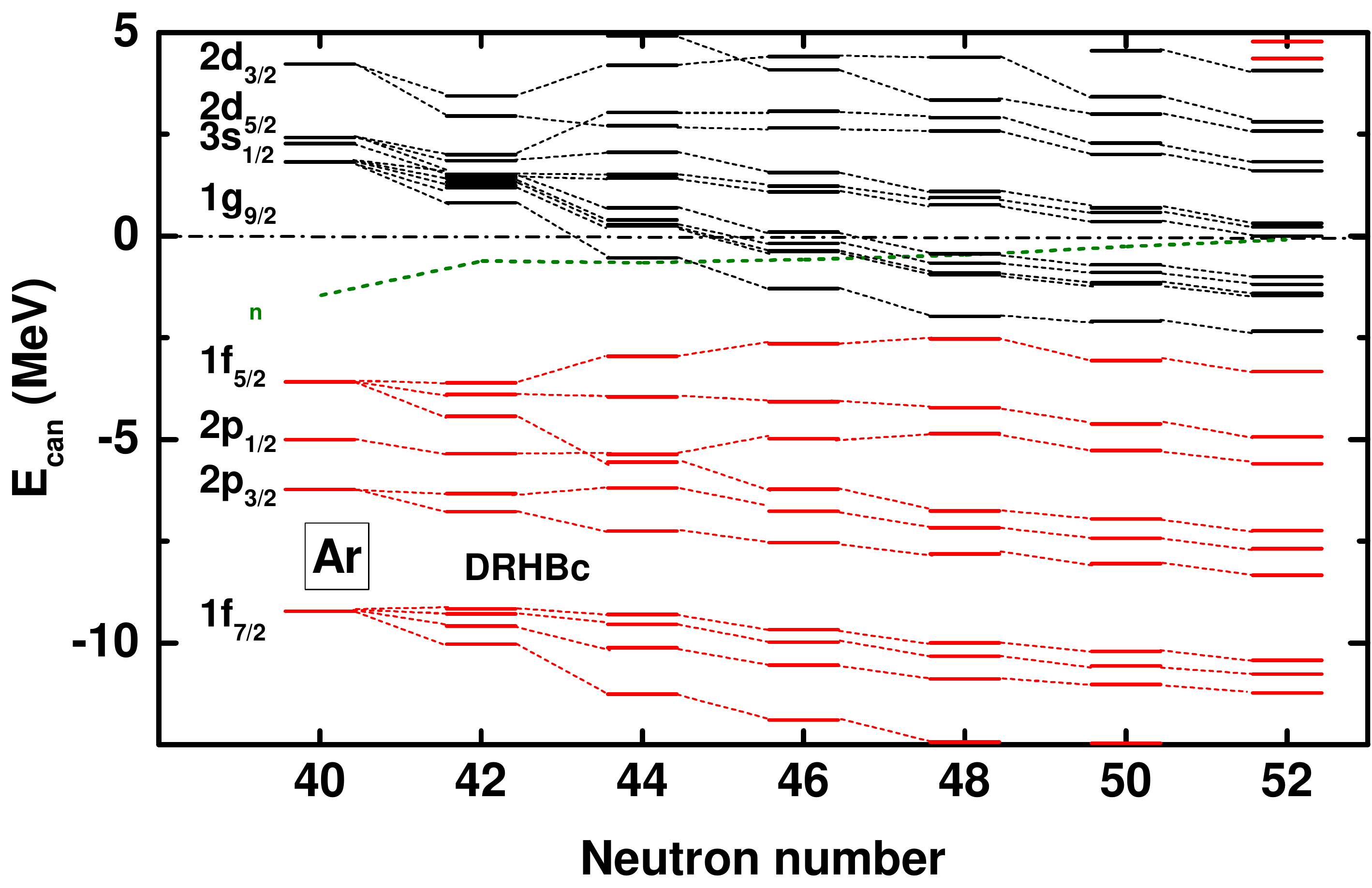}
   \caption{(Color online) The neutron single particle energy of argon isotopes as a function of neutron number within the DRHBc calculations with PC-PK1.}\label{Ar-spl}
\end{figure}

In Fig.~\ref{epair-Ar}, the pairing energies of argon isotopes from the DRHBc and RCHB calculations are shown. It can be seen that proton pairing energies of DRHBc calculations are close to zero and deviate from the ones of RCHB calculations for $^{36}$Ar,$^{44-56}$Ar and $^{64-70}$Ar which may be attributed to the occurrence of large deformation there. On the other hand, neutron pairing energies from these two calculations present similar evolution trend. For $N>44$, the pairing energy in RCHB calculations increases slowly with the  neutron number, while the pairing energy in DRHBc calculation increases rapidly.

Fig.~\ref{Ar-spl} shows the neutron single particle energy from DRHBc calculations for $^{58-70}$Ar. It can be seen that, with the evolution of neutron number, the single particle levels around the Fermi surface are getting denser and denser, and the pairing correlation is enhanced. The deformation, continuum and pairing correlation extend the neutron drip-line nucleus from $^{62}$Ar in the RCHB to $^{70}$Ar in the DRHBc calculations.

From above discussions, one can find that the deformation would affect the position of neutron drip line.
For argon isotopes,
self-consistent treatment of the deformation and continuum extends
the neutron drip line to more neutron-rich region.

In Fig.~\ref{relativediff}, the deformed RHB calculations with NL3* in harmonic oscillator basis~\cite{Afanasjev2013PLB} predicted less neutron-rich drip-line nuclei than the RCHB calculations for most of the isotopic chains. Although quite time-consuming and numerically challenging, considering the deformation and continuum effects simultaneously is essential to determine the drip line and should be investigated in the future.

\subsection{Radii of nucleon distributions}
\subsubsection{Charge radii}
\begin{figure}[t]
  \centering
     \includegraphics[width=12cm]{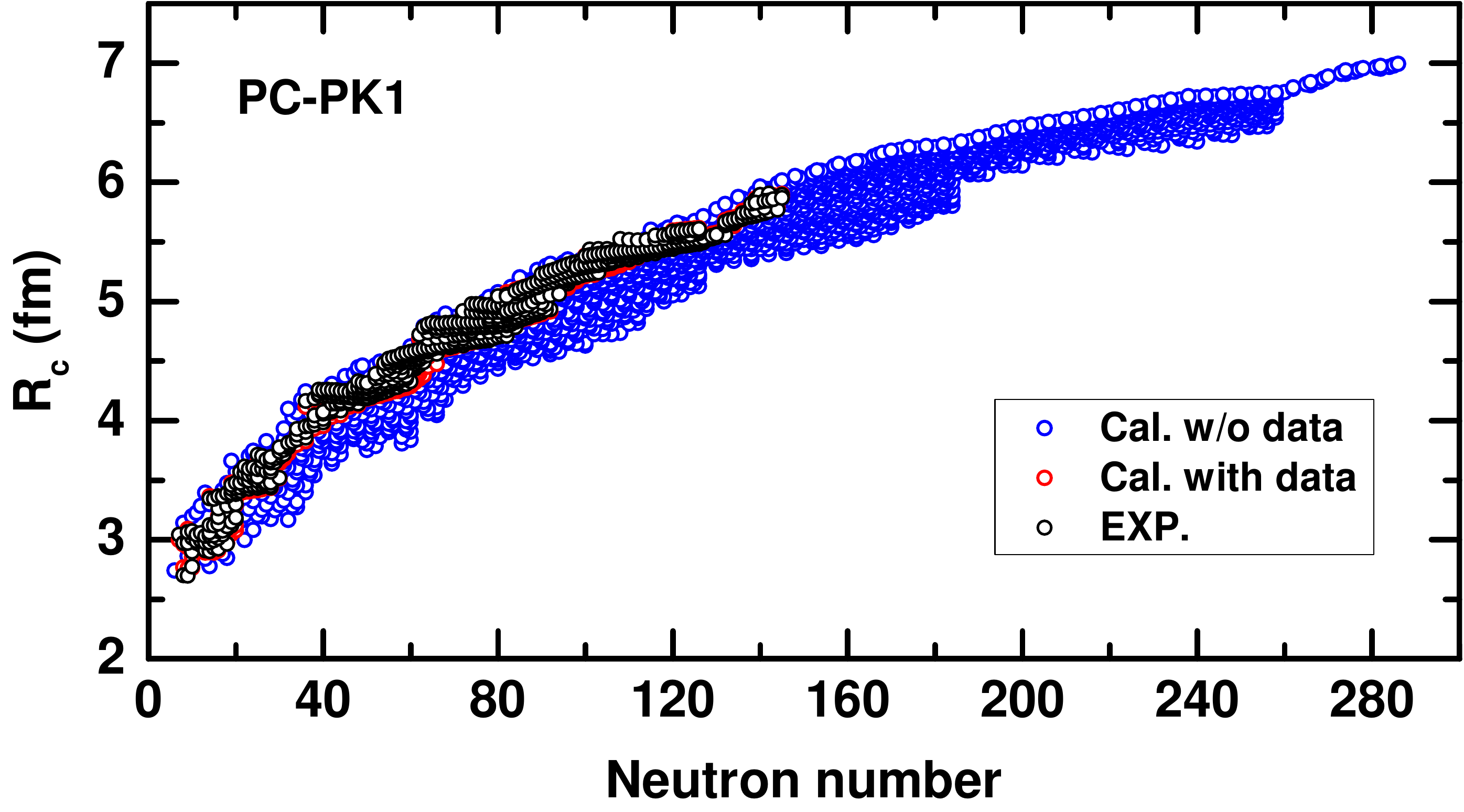}
   \caption{(Color online) The charge radii calculated by the RCHB theory with PC-PK1 as a function of
neutron number in comparison with the data available~\cite{Angeli2013ADNDT}. The black circles are
the experimental values, the red circles represent the calculated ones for the nuclei
with data available, and the blue circles are the predictions for the nuclei without
experimental data.}\label{chargeradii1}
\end{figure}
\begin{figure}[t]
  \centering
     \includegraphics[width=12cm]{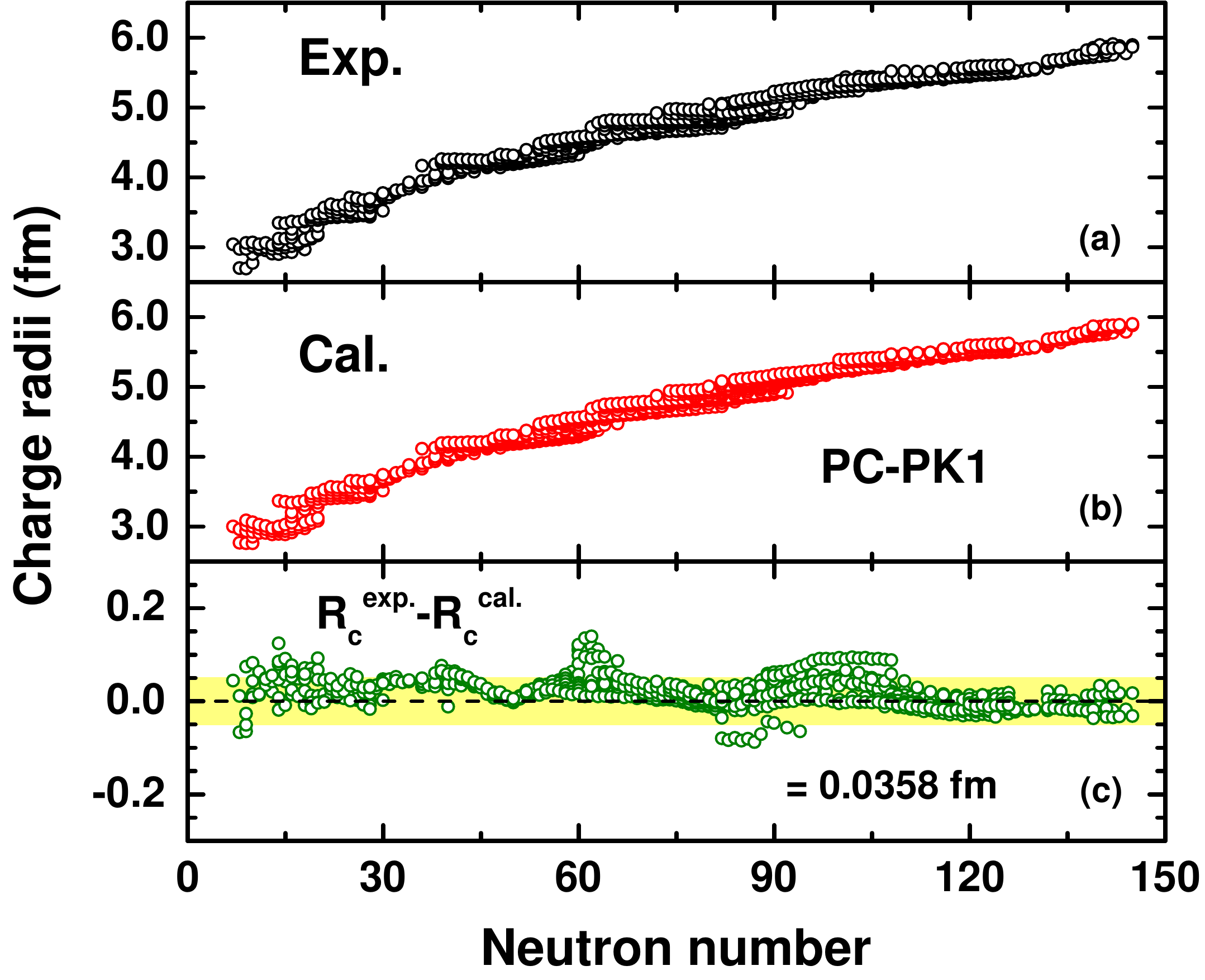}
   \caption{(Color online) The available experimental charge radii~\cite{Angeli2013ADNDT} (a), the RCHB calculated charge radii
with PC-PK1 (b), and the difference between the RCHB calculated charge radii and
the experimental data [90] (c) as a function of the neutron number. The rms deviation
is also shown.}\label{chargeradii}
\end{figure}

The nuclear charge radius is one of the most important nuclear observables that estimates the size of nuclei. In Fig.~\ref{chargeradii1}, the charge radii calculated by the RCHB theory with PC-PK1 are shown as a function of neutron number in comparison with the data available. The charge radii in the RCHB calculations reproduce the experiment quite well for the nuclei with data available~\cite{Angeli2013ADNDT}, and the charge radii of the nuclei without experimental data are predicted.

In Fig.~\ref{chargeradii}, the available experimental charge radii, the RCHB calculated charge radii
with PC-PK1, and the difference between the RCHB calculated charge radii and the experimental data~\cite{Angeli2013ADNDT} are shown.
One can see that most deviations are in the range of $-0.05 \sim 0.05$ fm, and the rms deviation $\sigma$ of the RCHB calculations  from the data is 0.0358 fm, which is in comparison with the results from the RHB calculations for NL3* $\sigma=0.0407$ fm, for DD-ME2 $\sigma=0.0376$ fm, for DD-ME$\delta$ $\sigma=0.0412$ fm, for DD-PC1 $\sigma=0.0402$ fm ~\cite{Agbemava2014PRC.89.054320}, the HFB calculations $\sigma=$0.026 fm~\cite{Goriely2013PhysRevC.88.024308}, and the formula based on the Weizsacker-Skyrme mass model $\sigma=$0.022 fm~\cite{WangN2013PhysRevC.88.011301}.

In Fig.~\ref{chargeradii}, larger deviations are located between $N=28$ and 50, $N=50$ and 82, as well as $N=82$ and 126. These deviations are expected to be reduced when the deformation effects are included.

\subsubsection{Neutron radii}
\begin{figure}[ht!]
  \centering
  \includegraphics[width=12cm]{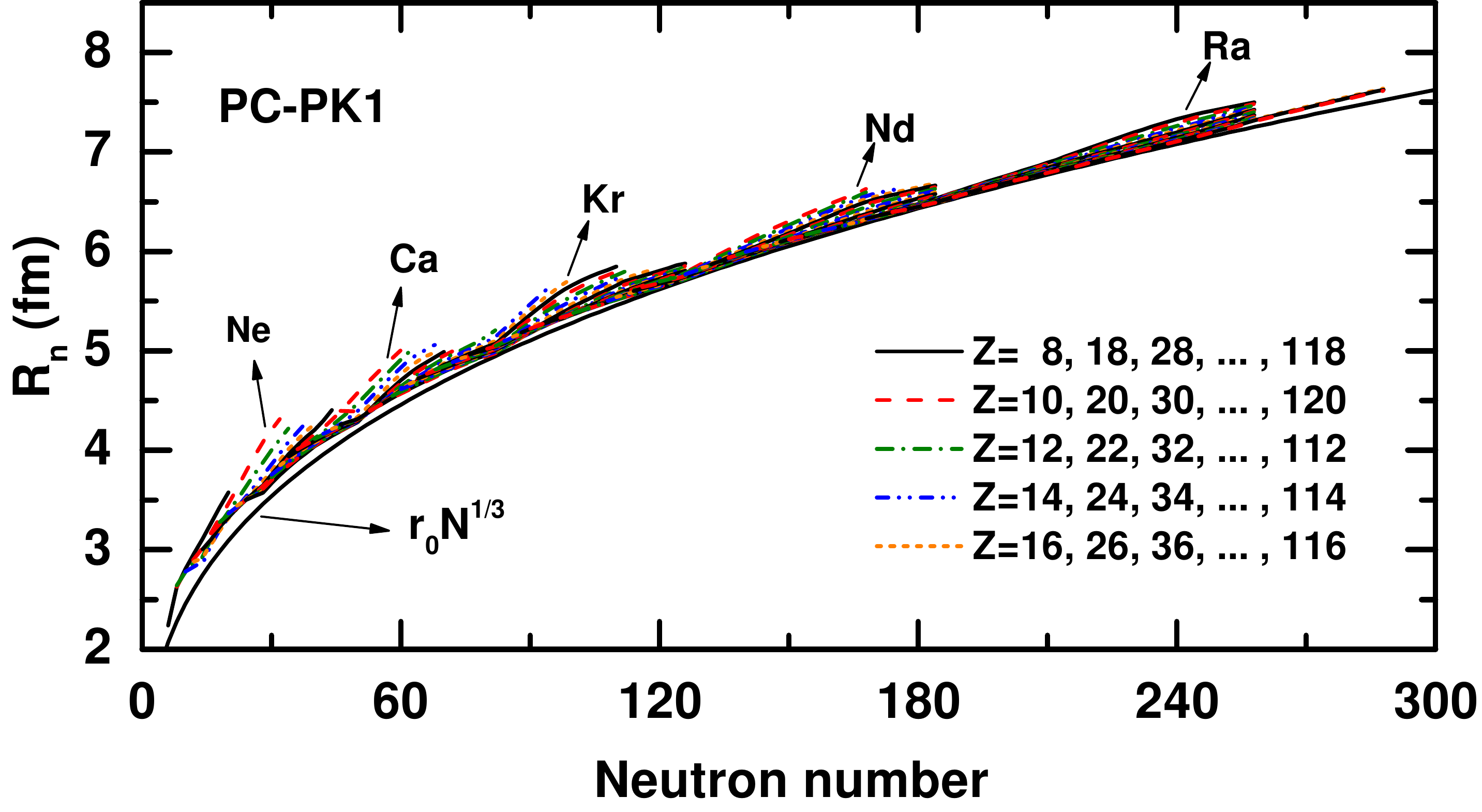}
   \caption{(Color online) The neutron rms radii for the even-even nuclei with $8\leqslant Z\leqslant 120$ from the RCHB calculations with PC-PK1 as a function of the neutron number. The empirical formula $r_0N^{1/3}$ with $r_0=1.140$ determined by $R_n$ of $^{208}$Pb is also plotted for guidance.}\label{neutronradii}
\end{figure}

In Fig.~\ref{neutronradii}, the calculated neutron rms radii of the neutron density distribution for even-even nuclei with $8\leqslant Z\leqslant 120$  are shown as a function of neutron number.
In addition, the empirical formula $R_n=r_0N^{1/3}$ is shown for guidance with $r_0 = 1.140$ fm determined by $R_n$ of $^{208}$Pb.
Except for extremely neutron-rich nuclei, the systematic trend of the neutron radii follows the simple empirical formula quite well.
Pronounced deviations of RCHB calculations from the empirical formula can be found in some extremely neutron-rich nuclei near the drip line.
Such deviations are regarded as signals of the so-called giant halo predicted in Ca~\cite{Meng2002PhysRevC.65.041302} and Zr~\cite{Meng1998PhysRevLett.80.460}.

\begin{figure}[ht!]
  \centering
    \includegraphics[width=12cm]{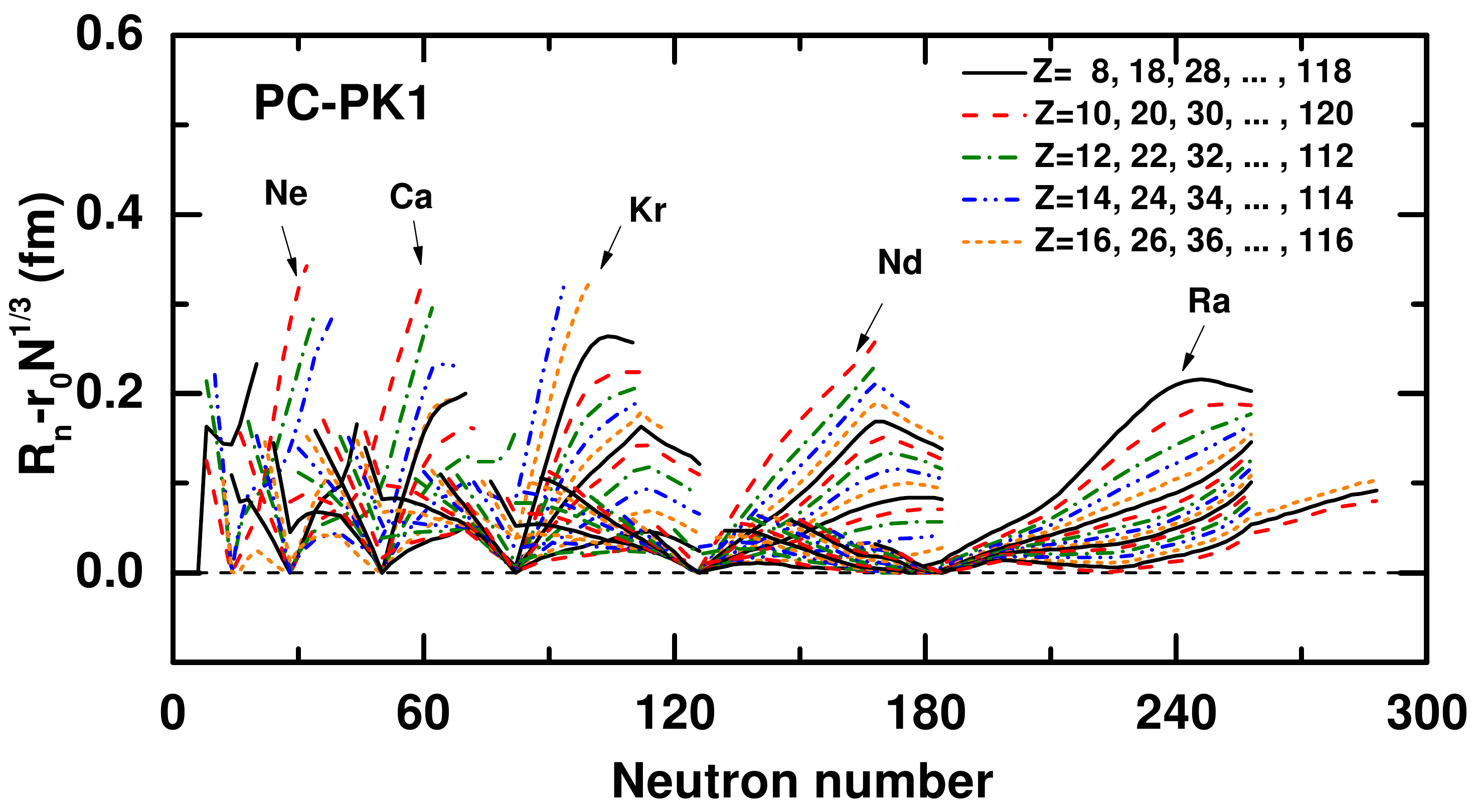}
   \caption{(Color online) The deviations of the neutron rms radii between the RCHB calculations with PC-PK1 and the empirical formula $r_0N^{1/3}$ for the even-even nuclei with $8\leqslant Z\leqslant 120$, in which the smallest ratio $R_{n}/N^{1/3}$ in each isotopic chain is chosen as $r_0$ to ensure non-negative differences.}\label{Rn-deviation}
\end{figure}

To give a deeper insight of the deviations from the empirical formula, the differences of neutron radii between the RCHB calculations and the empirical formula are shown in Fig.~\ref{Rn-deviation} for the even-even nuclei with $8\leqslant Z\leqslant 120$,
in which the smallest ratio $R_{n}/N^{1/3}$ in each isotopic chain is chosen as $r_0$ to ensure non-negative differences.

From Fig.~\ref{Rn-deviation}, one can see the pronounced deviations in some isotopic chains, such as Ne, Ca, Kr, Nd and Ra, which indicates the possible existence of the giant halos. The deviations show two different behaviors when approaching the neutron drip line. One increases with neutron number monotonically. The other increases with neutron number at the beginning then bends down at next stage. The different behavior is connected to the shell closure and the occupation of levels with high angular momentum.

\subsection{Neutron density distributions}
\begin{figure}[ht!]
  \centering
     \includegraphics[width=8cm]{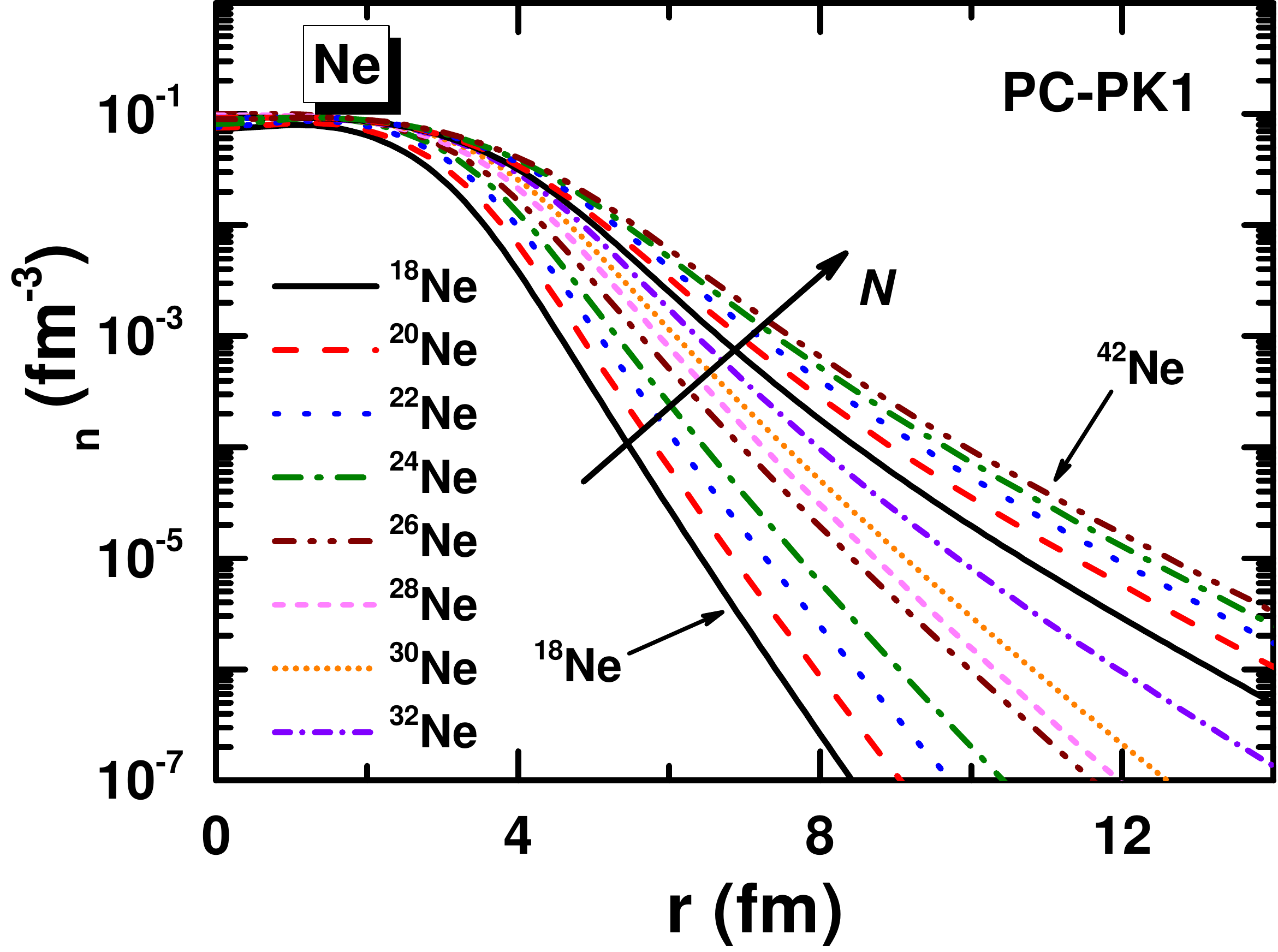}
     \includegraphics[width=8cm]{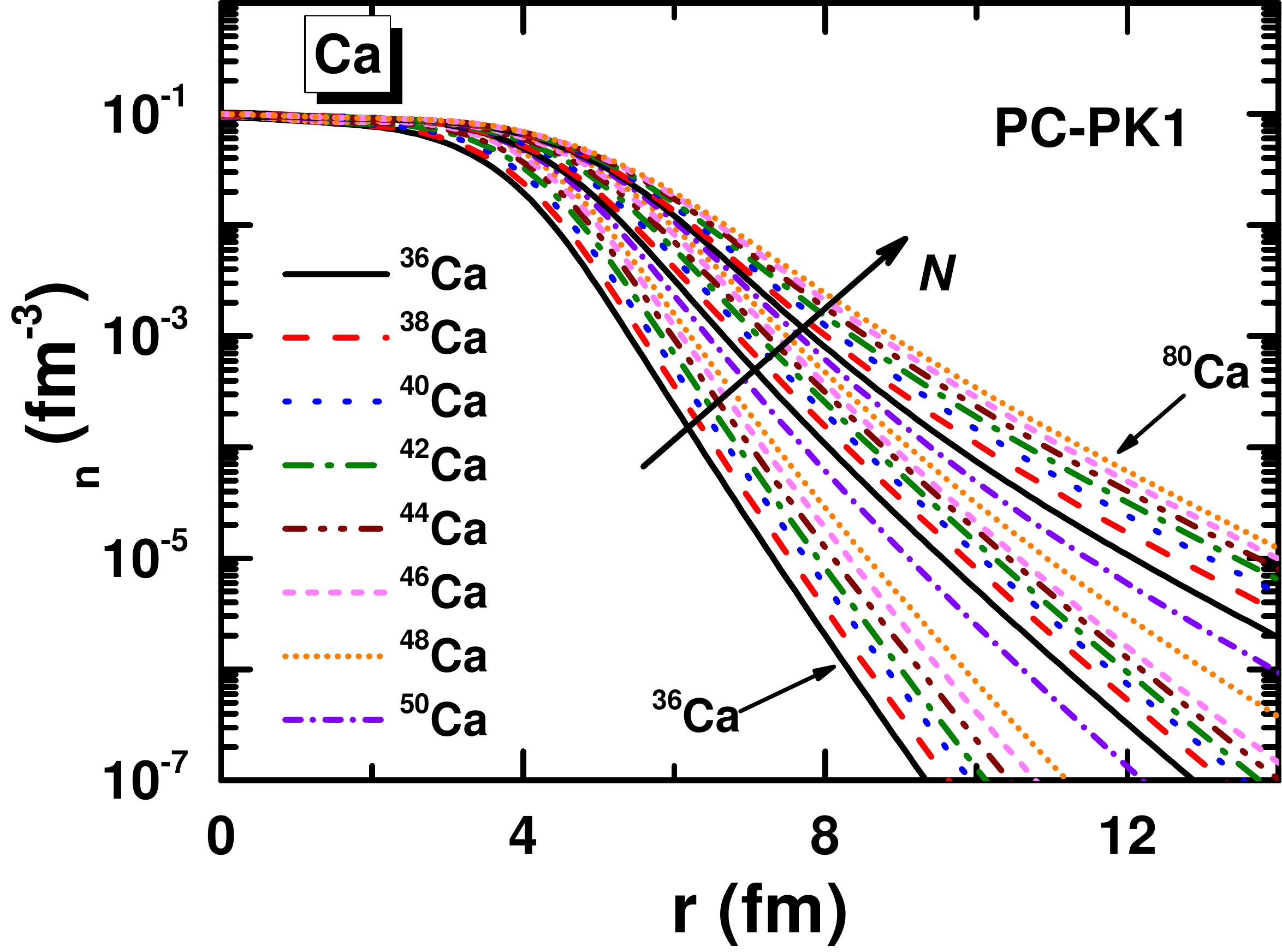}
     \includegraphics[width=8cm]{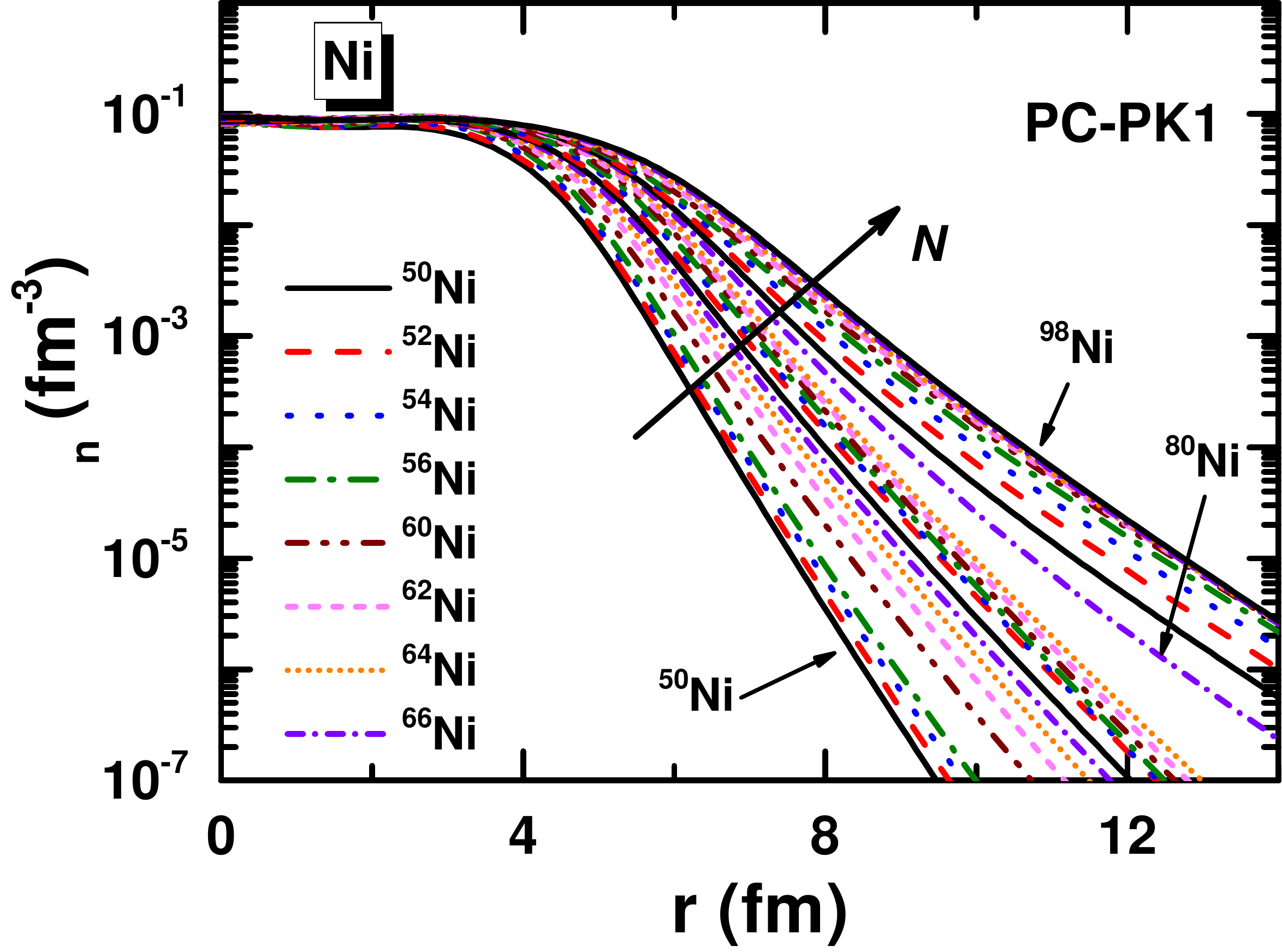}
     \includegraphics[width=8cm]{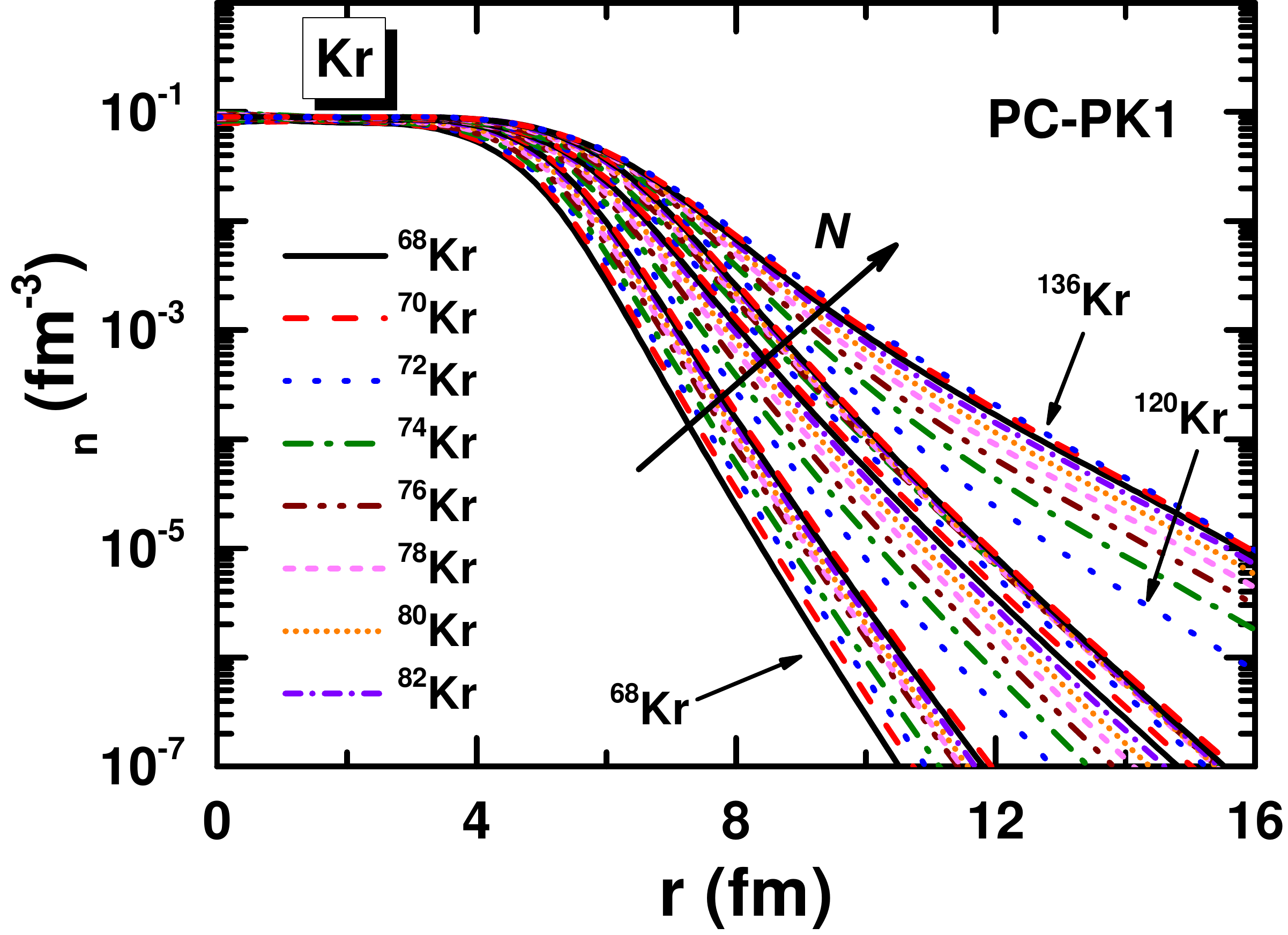}
     \includegraphics[width=8cm]{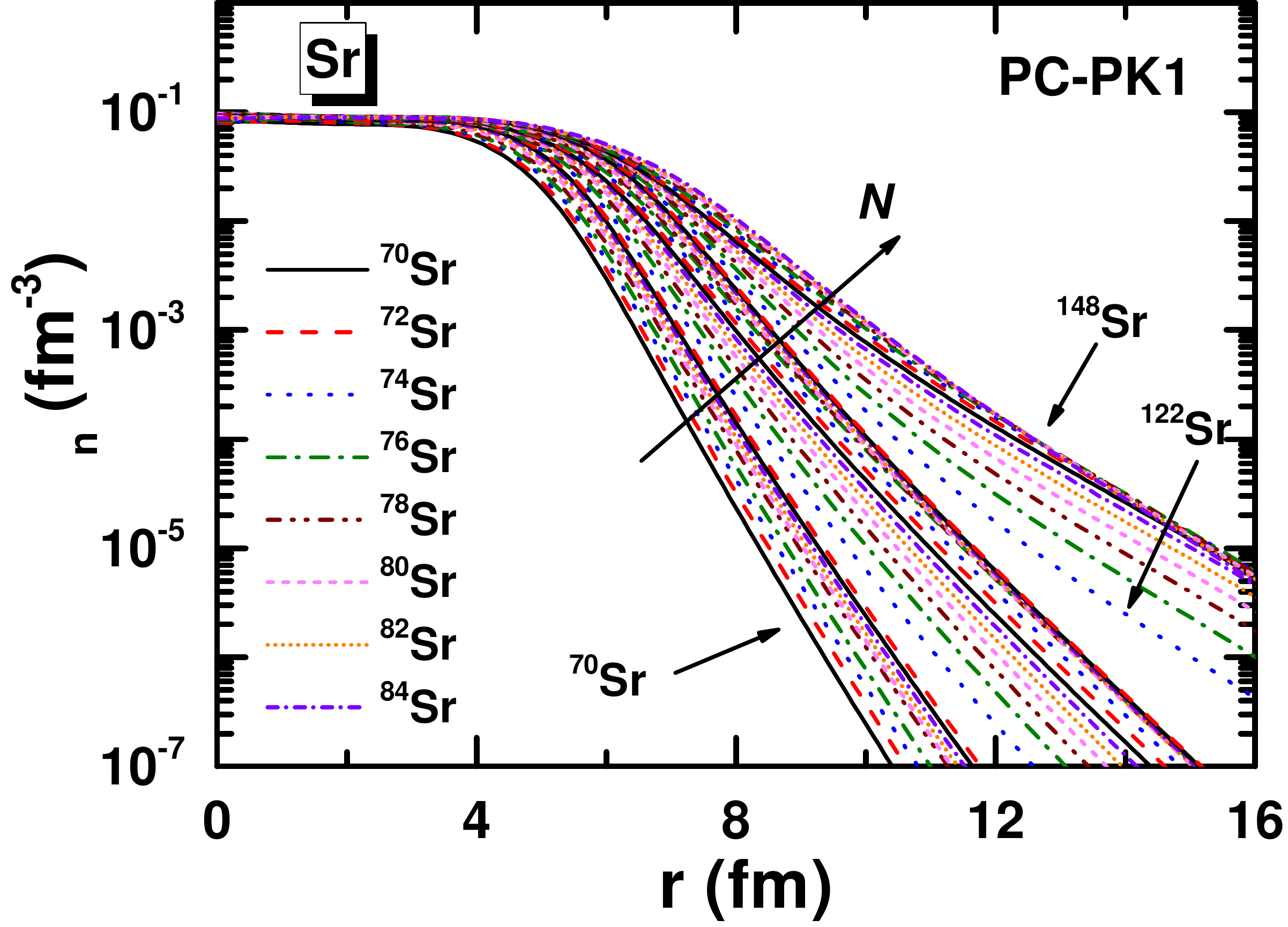}
     \includegraphics[width=8cm]{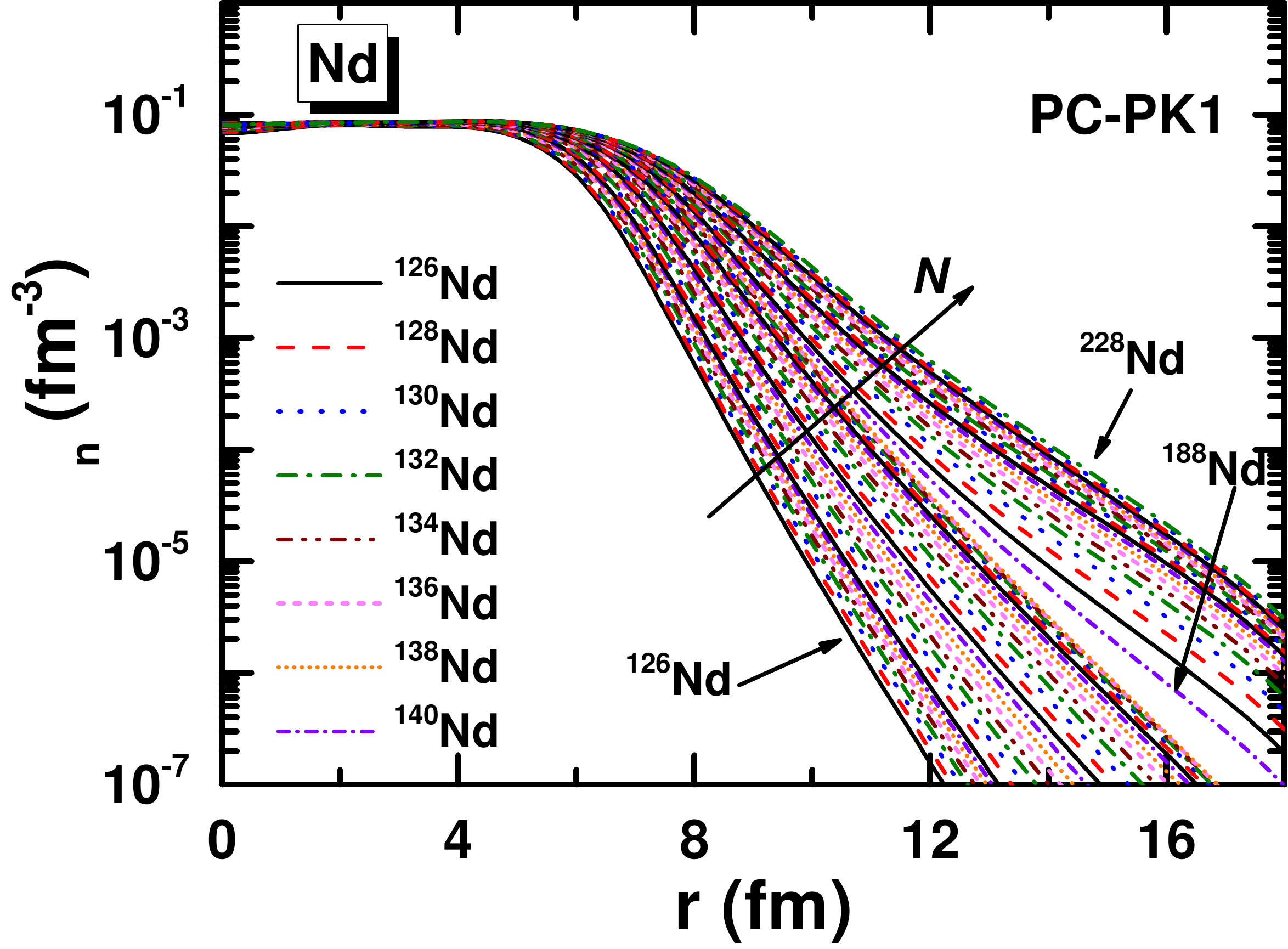}
   \caption{(Color online) The neutron density distributions of Ne, Ca, Ni, Kr, Sr and Nd isotopic chains obtained from the RCHB calculation with PC-PK1, where the thick arrows represent the evolution trend of density with neutron number.}\label{density}
\end{figure}

To investigate the evolution of the density distribution in each isotopic chain, in Fig.~\ref{density}, the neutron density distributions of even-even Ne, Ca, Ni, Kr, Sr and Nd isotopes are shown as examples, where the density distributions of proton  and neutron drip-line nuclei have been labeled, and the increase of the neutron number is marked.

There are several features to be noted in this figure.
Firstly, it is a global trend that the neutron density distributions are extended further with neutron number.
The surface expands outward rapidly, and the internal density distribution changes slightly.
Secondly, for the Ne, Ca and Kr isotopes, the tails of the density distributions extend with the neutron number monotonically until the neutron drip line.
For the Ni, Sr and Nd isotopes, however, the tails of the density distributions reach a maximum at certain neutron-rich nucleus, then saturate or even decrease with the neutron number till the neutron drip line.
Thirdly, the shell structures influence the density distribution significantly, which can be seen from the dramatic change at the tail of the
neutron density distribution when the neutron number $N$ in a nucleus crosses the magic numbers 20, 28, 50, 82 and 126.

\subsection{Neutron potential and diffuseness}
\begin{figure}[ht!]
  \centering
     \includegraphics[width=8cm]{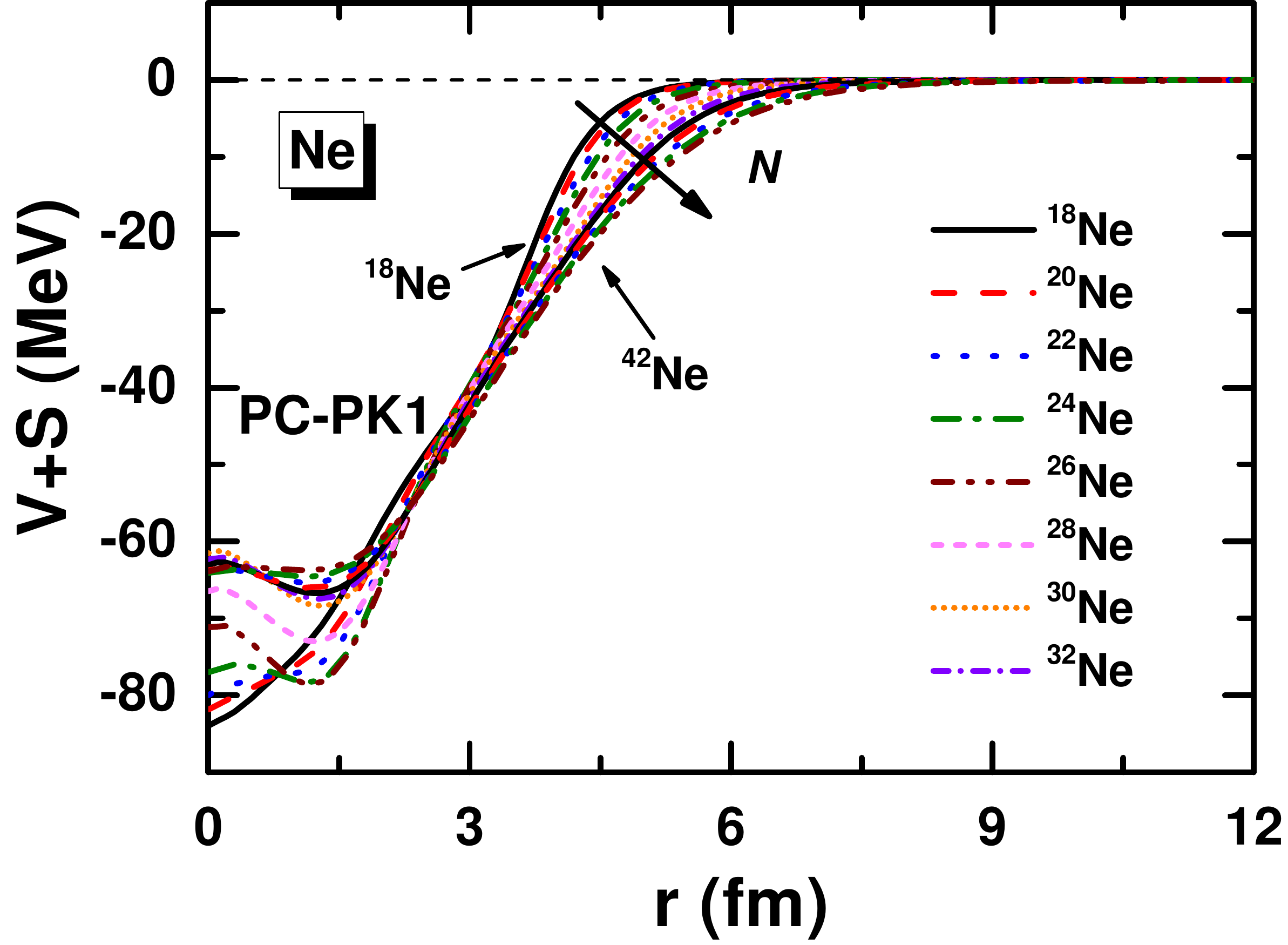}
     \includegraphics[width=8cm]{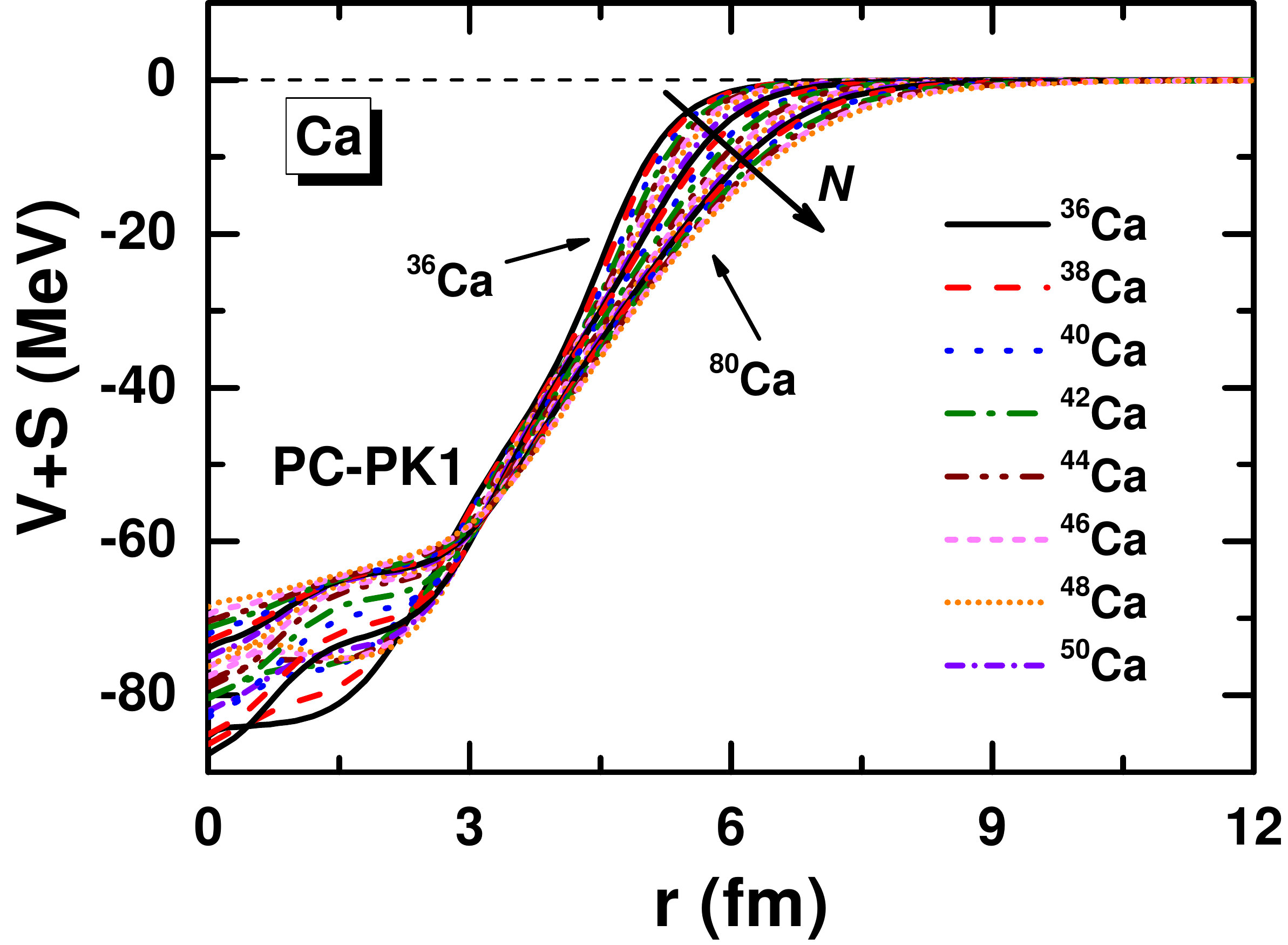}
     \includegraphics[width=8cm]{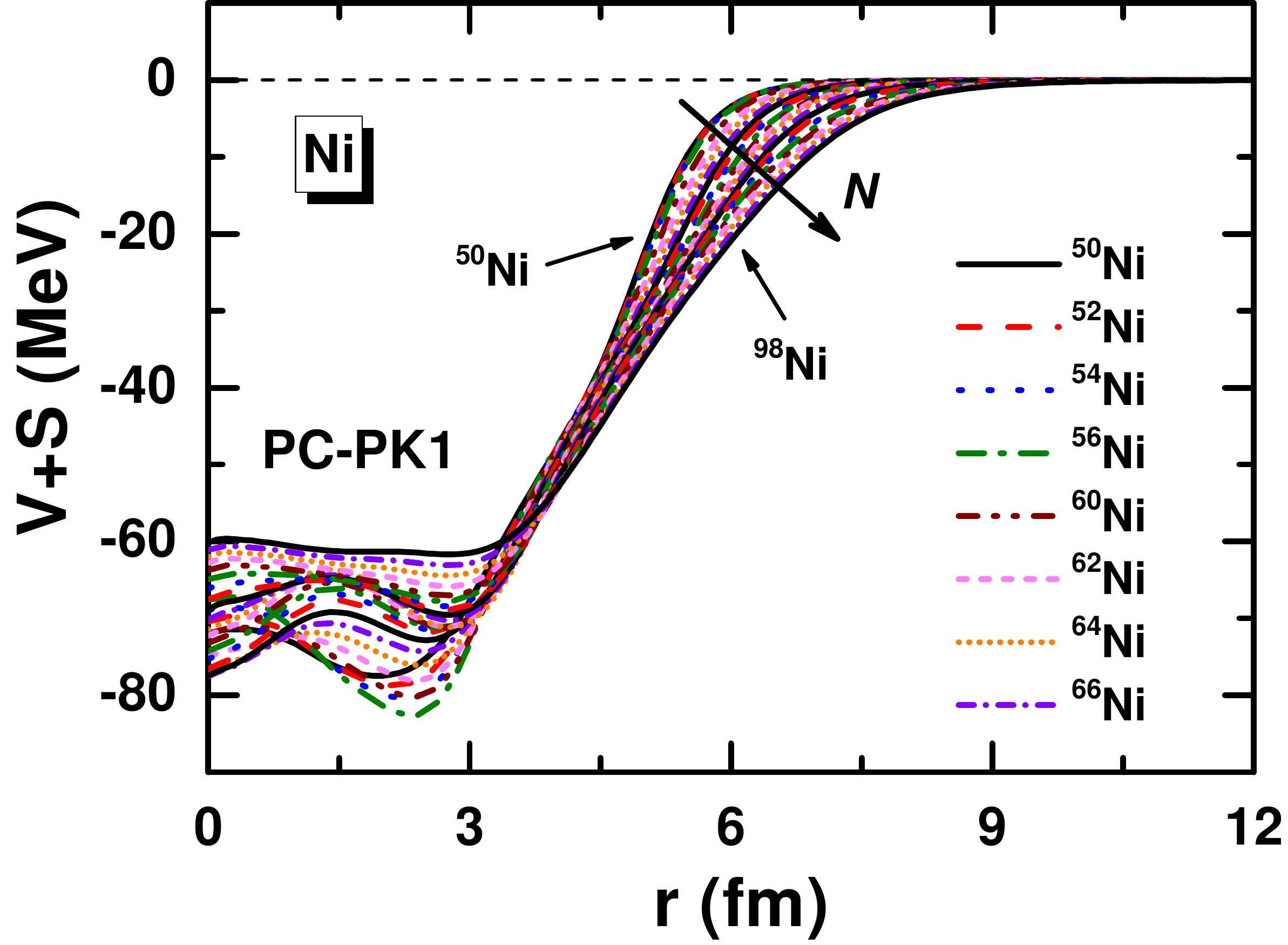}
     \includegraphics[width=8cm]{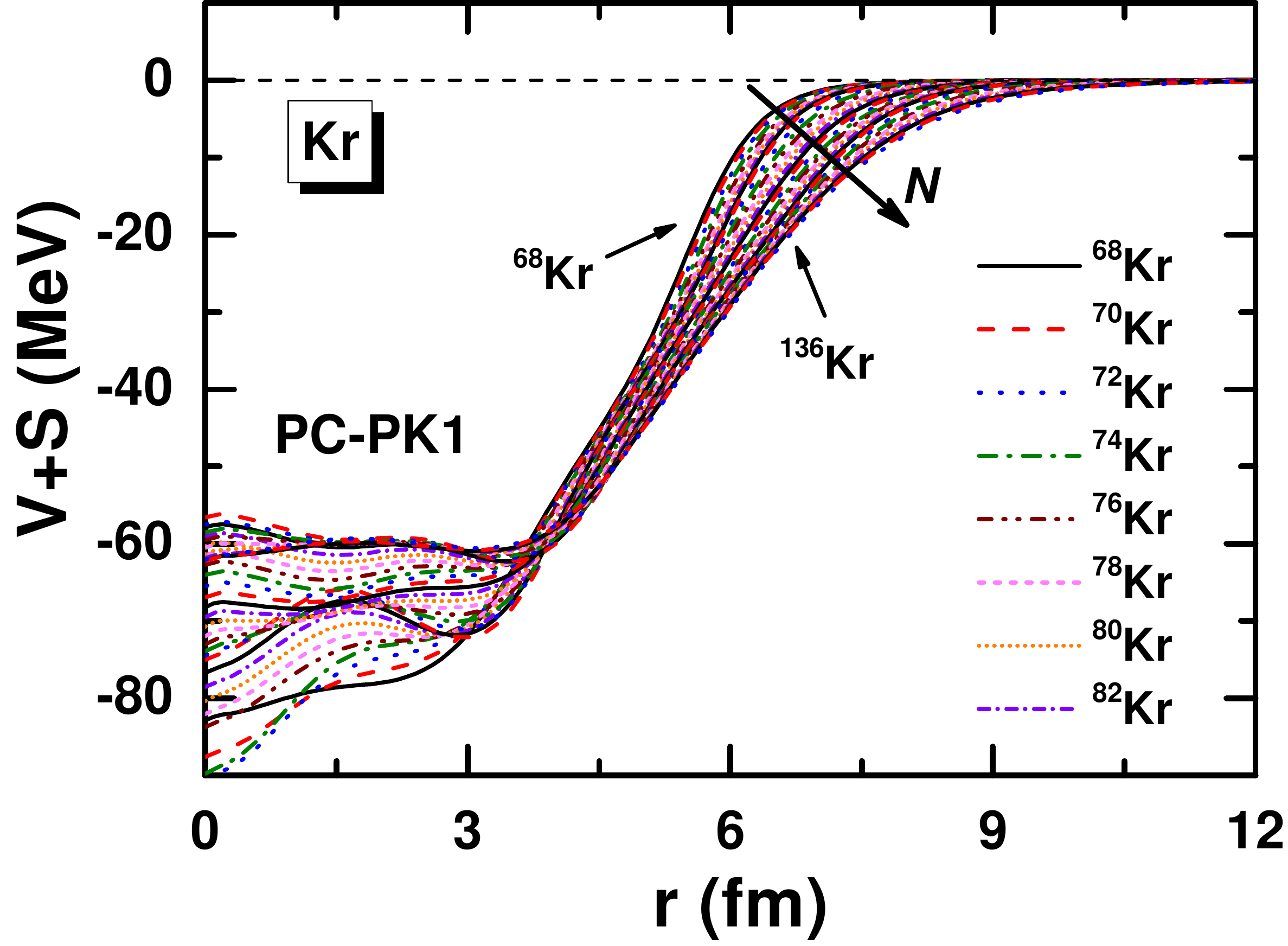}
     \includegraphics[width=8cm]{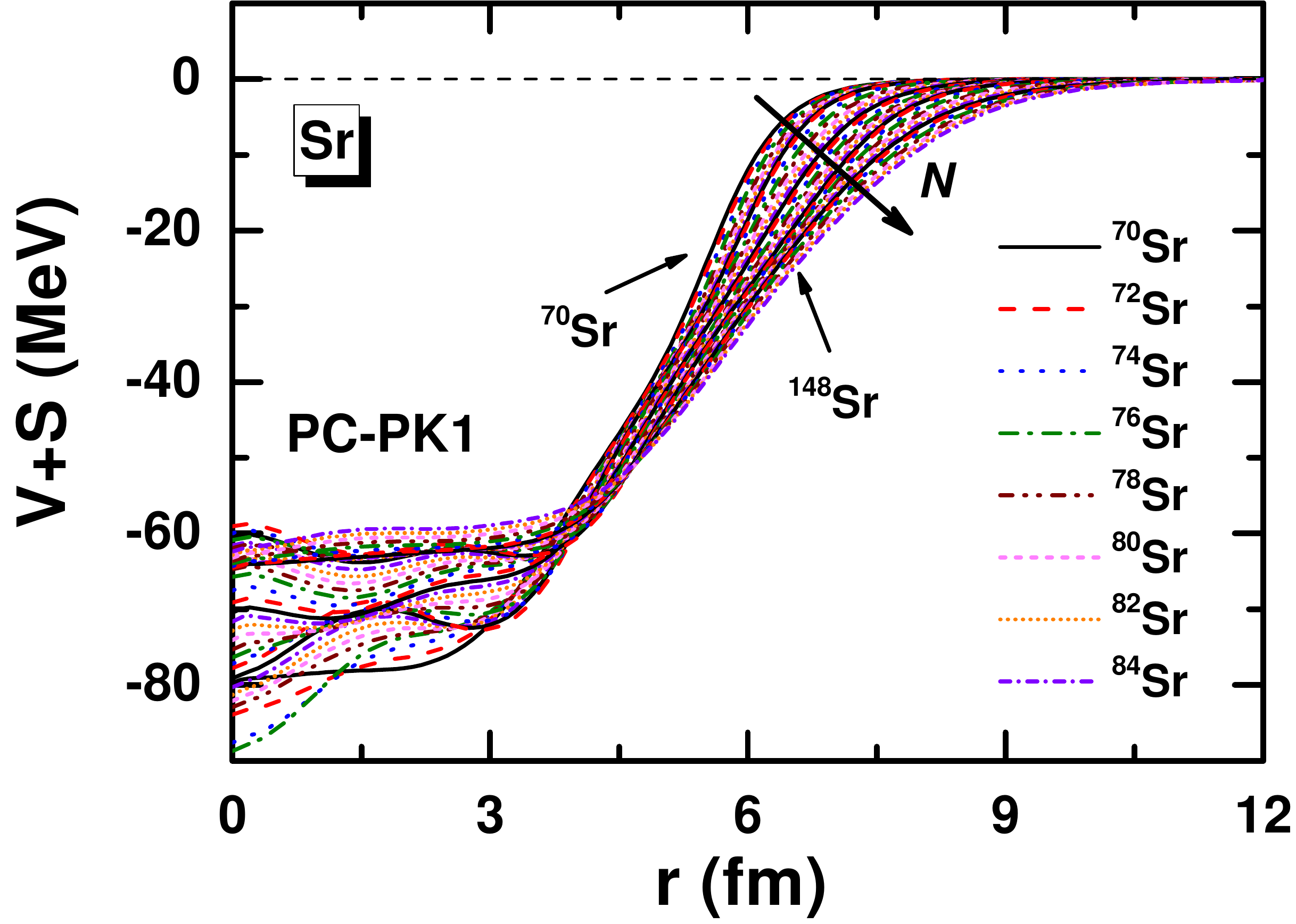}
     \includegraphics[width=8cm]{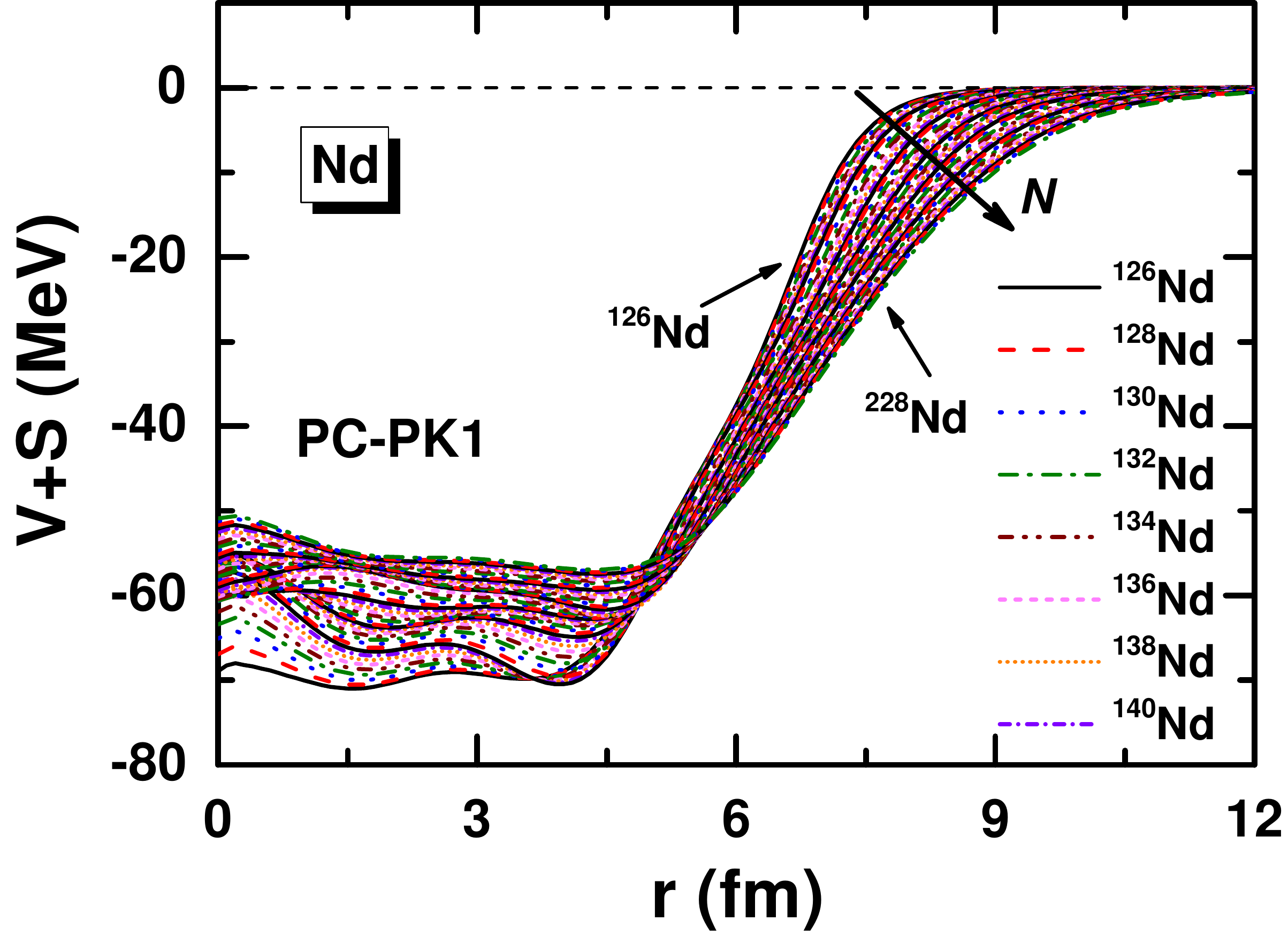}
   \caption{(Color online) Same with Fig.~\ref{density}, but for the neutron mean-field potentials $V+S$.}\label{potentialvps}
\end{figure}

To examine the isospin dependence of the mean-field potentials, we have investigated the neutron vector plus the scalar potentials $V(r)+S(r)$. The neutron potentials $V(r)+S(r)$ for even-even Ne, Ca, Ni, Kr, Sr and Nd isotopes from the proton drip-line nuclei to the neutron drip-line nuclei in the RCHB calculations are shown in Fig.~\ref{potentialvps}. The increase of the neutron number is marked.
Generally, the depths of the potentials rise with the neutron number, except for some fluctuation due to the shell structure.
At the surface, the potentials extend outward and the diffuseness increases significantly with the neutron number. As a result, the potentials for nuclei near the neutron drip line become highly diffused.
Such highly diffused potential will dramatically influence the low-$l$ orbits near the threshold, which may lead to the level crossing and the halos in drip-line nuclei.

\subsection{Pairing energies}
To examine the pairing correlation globally, we investigate the pairing energies $E_{\rm pair}$ for the even-even nuclei over the nuclear landscape.

\begin{figure}[t]
  \centering
  \includegraphics[width=12cm]{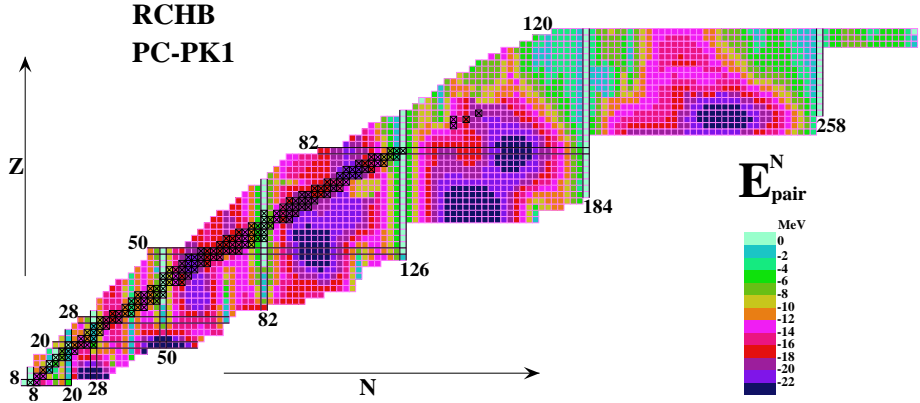}
\caption{(Color online) The neutron pairing energies of even-even nuclei with $8\leqslant Z\leqslant 120$ from the RCHB calculations with PC-PK1.}\label{epair-n}
\end{figure}

In the RCHB calculations, the pairing energy is defined as
\begin{eqnarray}
E_{\rm pair}=-\frac{1}{2}{\rm Tr} (\Delta\kappa).
\end{eqnarray}
In Fig.~\ref{epair-n}, the neutron pairing energies $E_{\rm pair}^{N}$ of even-even nuclei with $8\leqslant Z\leqslant 120$ from the RCHB calculations are shown.
One can see that the neutron pairing energies are approaching to zero or even vanish for the nuclei near the closed shells $N =$ 8, 20, 28, 50, 82 and 126, and they have maximum values for nuclei in the middle of the shells.
Furthermore, the neutron pairing energies vanish at $N =$ 184 and 258, which agrees with the predicted possible magic number $N =$ 184 in Ref.~\cite{Zhang2005NPA}.
For the isotopic chains with $Z$ around 10, 20 and 38, the pairing energies for nuclei near the drip line do not vanish, which suggest the disappearance of the neutron magic numbers 28, 50 and 82 in these nuclei.

\begin{figure}[t]
  \centering
  \includegraphics[width=12cm]{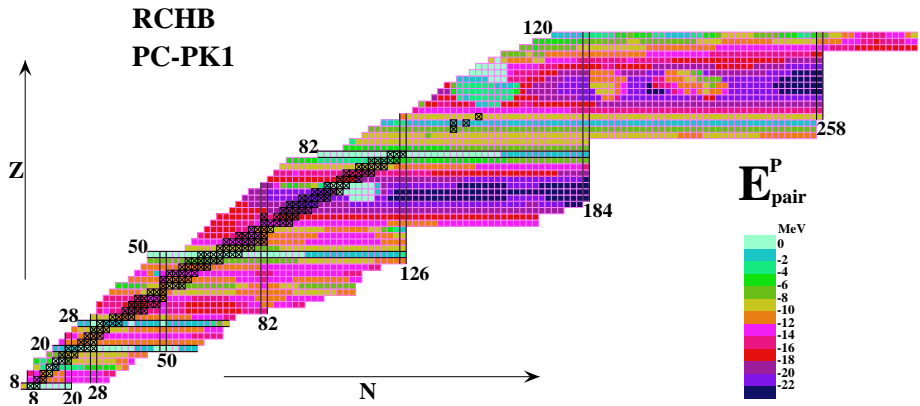}
\caption{(Color online) Same as Fig.~\ref{epair-n}, but for proton pairing energies.}\label{epair-p}
\end{figure}

\begin{figure}[t]
  \centering
  \includegraphics[width=12cm]{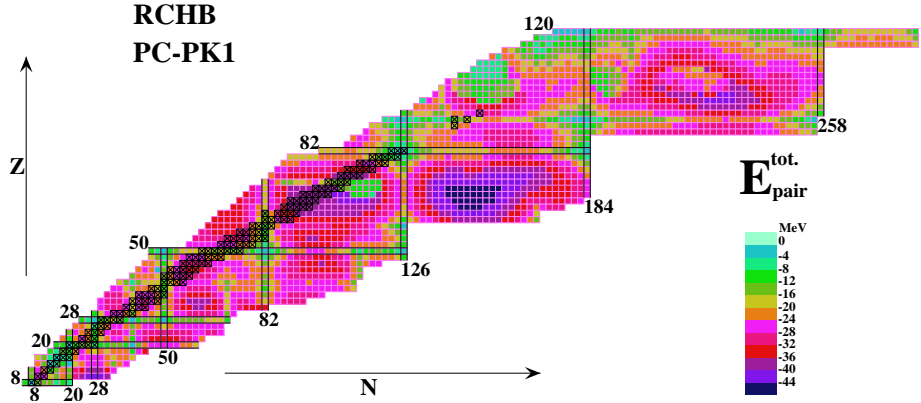}
   \caption{(Color online) Same as Fig.~\ref{epair-n}, but for total pairing energies.}\label{epair-t}
\end{figure}

In Fig.~\ref{epair-p}, the proton pairing energies $E_{\rm pair}^{P}$ of even-even nuclei with $8\leqslant Z\leqslant 120$ from the RCHB calculations  are shown. Similar to the neutron, the proton  pairing energies are also approaching to zero or even vanish at the closed shells $Z =$ 8, 20, 28, 50 and 82, and have the maximum value in the middle of the shells. In comparison with the neutron case, the proton pairing energies in each isotopic chain remain almost the same or change modestly with the neutron number.
Finally, the total pairing energies of even-even nuclei with $8\leqslant Z\leqslant 120$ from the RCHB calculations are shown in Fig.~\ref{epair-t}, where the shell structure effects appear again.


\subsection{Alpha decay energies}
\begin{figure}[ht!]
  \centering{}
  \includegraphics[width=12cm]{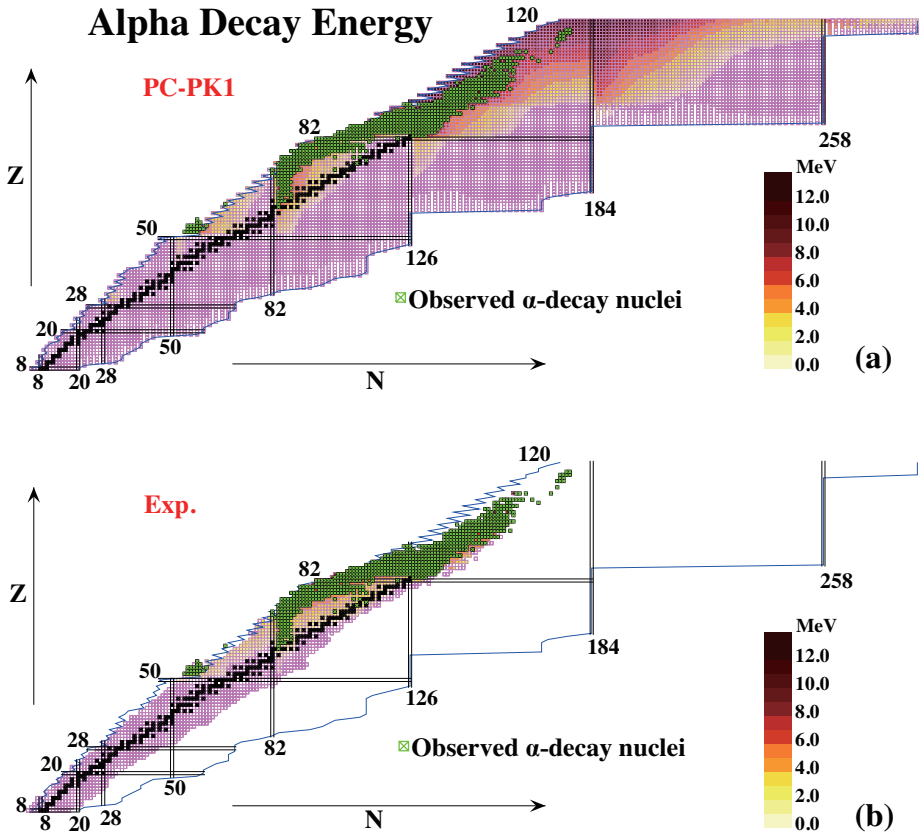}
   \caption{(Color online) $\alpha$-decay energies $Q_{\alpha}$ for nuclei with $10 \leqslant Z \leqslant 120$, provided by (a) RCHB theory with PC-PK1 and (b) available experimental values~\cite{Wang2012CPC}. Blue lines are proton and neutron drip lines predicted by the RCHB theory. The nuclei predicted to be bound in present work and observed experimentally are represented as the squares in panel (a) and (b), respectively. Furthermore, 719 nuclei observed experimentally with the radioactivity of $\alpha$-decay are marked with green crosses.}\label{alphadecay}
\end{figure}

By using the binding energies provided in the RCHB theory with PC-PK1, we can extract the $Q_\alpha$ values of the bound nuclei with $10 \leq Z \leq 120$ ~\cite{LFZhang2016CPC},
\begin{eqnarray}\label{eq1}
  Q_{\alpha}=E_B(Z-2,N-2)+E_B(2,2)-E_B(Z,N),
\end{eqnarray}
where $E_B(Z,N)$ is the binding energy for the nucleus with proton number $Z$ and neutron number $N$. In Fig.~\ref{alphadecay}, the $Q_\alpha$ obtained both from the RCHB calculations and the data in AME2012~\cite{Wang2012CPC} are shown, where the nuclei observed with $\alpha$-decay radioactivity experimentally~\cite{http://www.nndc.bnl.gov/} are marked with green crosses.

From the RCHB calculations, the $Q_{\alpha}$ values of 3703 nuclei are positive, and they are scaled by colors in Fig.~\ref{alphadecay}(a). Experimentally, as shown in Fig.~\ref{alphadecay}(b), the existing data give 1067 nuclei with their $Q_{\alpha}$ values positive.
Comparing the two panels of Fig.~\ref{alphadecay}, we see the following features.
Firstly, the positive $\alpha$-decay energies deduced from the RCHB results present a similar pattern with those from available experimental values, although spherical symmetry is assumed.
Secondly, in Fig.~\ref{alphadecay}(a), 719 nuclei observed experimentally with the radioactivity of $\alpha$-decay are marked with green crosses.
Most of these nuclei are  in the heavy and superheavy regions predicted to have $Q_{\alpha}$ values larger than 4 MeV.
Thirdly, in the nuclear region with $N > 184$ and $Z > 92$, most nuclei in the up-left corner have $Q_{\alpha}>$ 4 MeV. This indicates a possibility of $\alpha$-radioactivity for these nuclei.

\subsection{Proton emitters}
\begin{figure}[ht!]
  \centering
  \includegraphics[width=12cm]{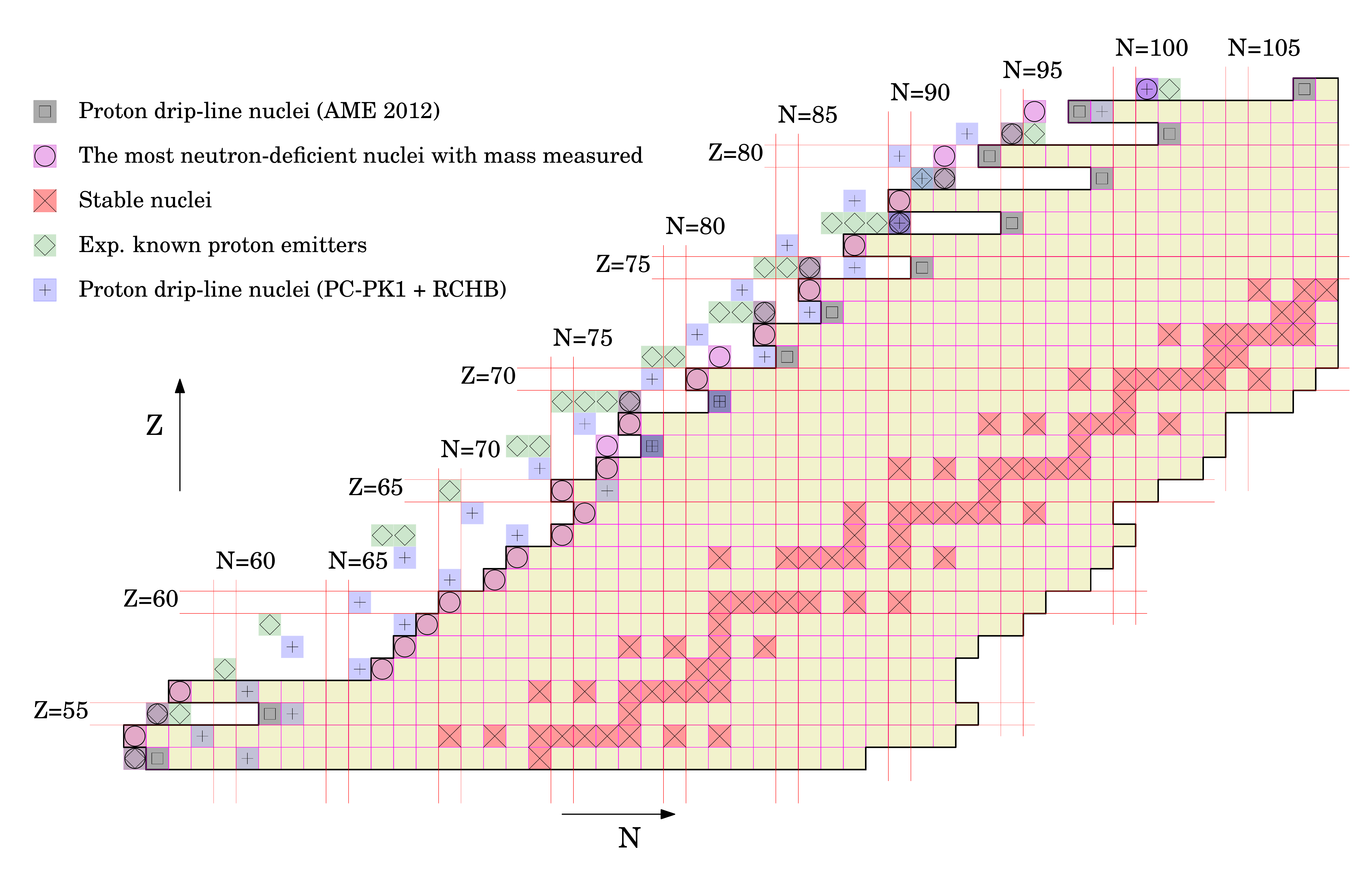}
   \caption{(Color online) Part of nuclear chart from I $(Z = 53)$ to Bi $(Z = 83)$ isotopes by the RCHB theory with PC-PK1. The experimentally known proton emitters are denoted with the symbol $\diamondsuit$, while the proton drip-line nuclei given by the spherical RCHB calculations are shown with the symbol $+$. The stable nuclei, the most neutron-deficient nuclei with mass measured and the proton drip-line nuclei in AME2012~\cite{Wang2012CPC} are also shown.}\label{protondecay}
\end{figure}
The proton radioactivity from neutron-deficient nuclei beyond the proton drip line provides not only spectroscopic information on nuclear structure but also valuable clues on the nuclear force beyond the proton drip line. These proton emitters  are also interesting open systems for quantum mechanical tunneling as the emitted proton goes through the barrier of Coulomb and centrifugal potentials.

Based on the RCHB mass table, the properties of (near) spherical proton emitters in the region from I ($Z = 53$) to Bi ($Z = 83$) isotopes are systematically studied~\cite{Lim2016PhysRevC.93.014314}. In Fig.~\ref{protondecay}, the nuclear chart from the I ($Z = 53$) to Bi ($Z = 83$) isotopes is shown, where the experimentally known proton emitters~\cite{Blank2008PPNP} are highlighted.

The  proton emitter candidates are located outside the proton drip line with positive $Q$ values. Accordingly, the proton emitter candidates in the calculations satisfy the Fermi surface $\lambda_{p} > 0$, and one-proton separation energy $S_{p} < 0$ or two-proton separation energy $S_{2p} < 0$.
The proton emitters observed in this region are well reproduced by the self-consistent RCHB calculations, except $^{170,171}$Au, $^{176,177}$Tl and $^{185} $Bi. The half-lives of the
proton emitters are calculated in Ref.~\cite{Lim2016PhysRevC.93.014314} by combining the RCHB calculations and Wentzel-Kramers-Brillouin (WKB) method, and good agreements with the data are obtained.

\section{Summary}{\label{sec.sum}}
In summary, we have performed systematic spherical calculations for all nuclei from $Z = 8$ to $Z = 120$ by using the RCHB theory with the relativistic density functional PC-PK1. The calculated binding energies, separation energies, neutron and proton Fermi surfaces, root-mean-square (rms) radii of neutron, proton, matter, and charge distributions, ground-state spins and parities are tabulated.

With the effects of the continuum included, there are totally 9035 nuclei predicted to be bound from $Z = 8$ to $Z = 120$. Comparing with the calculation without pairing, the RCHB theory predicts more bound nuclei in the neutron-rich side and extends the neutron drip line to more neutron-rich region.
For the nuclei close to the neutron drip line, as the neutron Fermi surface
is close to the continuum threshold, pairing correlations could scatter the nucleons
from bound states to the continuum, thus provide a significant coupling between the
continuum and bound states.
As a result, some unbound nuclei predicted without pairing
correlations can exist as bound ones.
Therefore, the RCHB theory which allows a proper treatment of
the continuum and the coupling to the bound states predicts a more extended
neutron drip line than the other models.
The considerable difference of the drip line between RHB in HO calculations with NL3* and present RCHB calculations are due to different functionals (PC-PK1 or NL3*), different pairing forces used, and the treatment of continuum in coordinate or HO space. Future investigations with same density functional and pairing force in both coordinate space and in harmonic oscillator basis are needed.
The separation energy evolutions for two-neutron $S_{2n}$, one-neutron $S_{n}$, two-proton $S_{2p}$ and one-proton $S_{p}$ are presented and discussed.

The deformation effects on the neutron drip line are investigated with the DRHBc theory by taking Ar isotopes as examples. It is found that
the deformation would affect the position of neutron drip line. For Ar isotopic chain, self-consistent treatment of the deformation
and continuum extends the neutron drip-line nucleus from $^{62}$Ar in the RCHB to $^{70}$Ar in the DRHBc calculations.

The RCHB calculated charge radii $R_{\rm c}$ with PC-PK1 are compared with the experimental values, the rms deviation $\sigma$ of the RCHB calculations from the data is 0.0358 fm. The comparison of the neutron rms radii of even-even nuclei with $8\leqslant Z\leqslant 120$ between the RCHB calculations and the empirical formula are performed. Except for extremely neutron-rich nuclei, the systematic trend of the neutron radii in the RCHB calculations follows the simple empirical formula quite well.

The neutron density distributions of even-even Ne, Ca, Ni, Kr, Sr and Nd isotopes are discussed.
For the Ne, Ca and Kr isotopes, the tails of the density distributions
extend with the neutron number monotonically until the neutron drip line. For the
Ni, Sr and Nd isotopes, the tails of the density distributions reach a maximum at certain neutron-rich nucleus, then saturate or even decrease with the neutron number till the neutron drip line.

The neutron potentials of even-even Ne, Ca, Ni, Kr, Sr and Nd isotopes are investigated. The depths of potentials generally rise
with the neutron number, except for some fluctuation due to the shell structure. At
the surface, the potentials extend outward and the diffuseness increases significantly
with the neutron number.

The pairing energies of the even-even nuclei over the nuclear landscape are discussed.
The pairing energies are approaching to zero or even vanish for the nuclei near the closed shells, and they have maximum values for nuclei in the middle of the shells. The possible magic numbers $N =$ 184 and 258 in superheavy are predicted.

In addition, the $\alpha$-decay energies and proton emitters based on the RCHB calculations are investigated.
The $\alpha$-decay energy pattern deduced from the RCHB calculations is similar to the one from available experimental values, and most of the proton emitters observed in the region from I ($Z = 53$) to Bi ($Z = 83$) isotopes are well reproduced by the self-consistent RCHB calculations.

The successful exploration of the nuclear chart by using the RCHB theory with the relativistic density functional
PC-PK1 demonstrated the important effects of the continuum on the nuclear landscape. It is quite encouraging to consider the deformation and continuum effects simultaneously in
determining the drip line  in the future although it is quite time-consuming and numerically challenging.

\ack

The authors would like to express gratitude to A. V. Afanasjev for his helpful discussions and careful reading of the manuscript.
We acknowledge the fruitful discussions with L. S. Geng, J. N. Hu, Z. M. Niu, P. Ring, I. J. Shin, B. H. Sun, N. Wang, Z.-H. Zhang,  and  S.-G. Zhou.
This work was partly supported by the National Natural
Science Foundation of China (Grants No. 11335002, No.
11375015, No. 11461141002, No. 11621131001 and No. 11605163), the Chinese Major
State 973 Program (Grant No. 2013CB834400), the Rare Isotope
Science Project of Institute for Basic Science funded by the Ministry of Science, ICT and Future Planning, the
National Research Foundation of Korea (2013M7A1A1075764), and  U.S. Department of Energy (DOE), Office of Science, Office of Nuclear Physics, under Contracts No. DE-AC02-06CH11357 (P.W.Z.).

\newpage


%

\newpage

\TableExplanation

\bigskip
\renewcommand{\arraystretch}{1.0}

\bigskip

\section*{Table 1.\label{tbl3te} Ground-state properties of nuclei calculated by RCHB theory with relativistic density functional PC-PK1}
\begin{tabular}{@{}p{0.0in}p{1.5in}p{4.5in}@{}}
& $Z$ & Proton number\\
& $N$ & Neutron number\\
& $A$ & Mass number\\
& $E^{\mathrm{Cal.}}_{\mathrm{b}}$ & Binding energy from RCHB calculations\\
& $E^{\mathrm{Exp.}}_{\mathrm{b}}$ & Binding energy from experimental data\\
& $E^{\mathrm{Cal.}}_{\mathrm{b}}/A$ & Per nucleon binding energy from RCHB calculations\\
& $E^{\mathrm{Exp.}}_{\mathrm{b}}/A$ & Per nucleon binding energy from experimental data\\
& $S_{2n}$ & Two-neutron separation energy\\
& $S_{2p}$ & Two-proton separation energy\\
& $S_{n}$ & One-neutron separation energy\\
& $S_{p}$ & One-proton separation energy\\
& $\lambda_{n}$ & Neutron Fermi surface\\
& $\lambda_{p}$ & Proton Fermi surface\\
& $R_{m}$ & Matter root-mean-square radius \\
& $R_{n}$ & Neutron root-mean-square radius \\
& $R_{p}$ & Proton root-mean-square radius \\
& $R^{\mathrm{Cal.}}_{\mathrm{c}}$ & Charge radius from RCHB calculations \\
& $R^{\mathrm{Exp.}}_{\mathrm{c}}$ & Charge radius from experimental data\\
& $j^{\pi}(P)$ & Ground state spin $j$ and parity $\pi$ of proton \\
& $j^{\pi}(N)$ & Ground state spin $j$ and parity $\pi$ of neutron \\
& $\sigma$ & rms deviations of binding energies and charge radii for each isotopic chain. \\
\\
\end{tabular}
\label{ExplTable6}

\newpage

\datatables 

\setlength{\LTleft}{0pt}
\setlength{\LTright}{0pt}


\setlength{\tabcolsep}{0.5\tabcolsep}

\renewcommand{\arraystretch}{1.0}

\footnotesize
\begin{longtable}{@{\extracolsep\fill}c c c c c c c c c c c c r r r r r r r@{}}
\caption{Ground-state properties of nuclei calculated by RCHB theory with PC-PK1, in comparison with the available data of masses and charge radii. In addition, the data labelled with underline means the nucleus is unbound.}\label{masstable}
$A$ & {$N$} & {$E^{\mathrm{Cal.}}_{\mathrm{b}}$} & {$E^{\mathrm{Exp.}}_{\mathrm{b}}$} & {$E^{\mathrm{Cal.}}_{\mathrm{b}}/A$} & {$E^{\mathrm{Exp.}}_{\mathrm{b}}/A$} & {$S_{2n}$} & {$S_{2p}$} & {$S_{n}$} & {$S_{p}$} & {$\lambda_n$}& {$\lambda_p$}& {$R_m$} & {$R_n$}& {$R_p$}& {$R^{\mathrm{Cal.}}_{\mathrm{c}}$}& {$R^{\mathrm{Exp.}}_{\mathrm{c}}$}& {$j^{\pi}(P)$}& {$j^{\pi}(N)$}\\
 & &{$(\rm MeV)$}&{$(\rm MeV)$}&{$(\rm MeV)$}&{$(\rm MeV)$}&{$(\rm MeV)$}&{$(\rm MeV)$}&{$(\rm MeV)$}&{$(\rm MeV)$}&{$(\rm MeV)$}&{$(\rm MeV)$}&{$(\rm fm)$}&{$(\rm fm)$}&{$(\rm fm)$}&{$(\rm fm)$}&{$(\rm fm)$}& & \\
\hline\\
\endfirsthead
\caption[]{(continued)}
$A$ & {$N$} & {$E^{\mathrm{Cal.}}_{\mathrm{b}}$} & {$E^{\mathrm{Exp.}}_{\mathrm{b}}$} & {$E^{\mathrm{Cal.}}_{\mathrm{b}}/A$} & {$E^{\mathrm{Exp.}}_{\mathrm{b}}/A$} & {$S_{2n}$} & {$S_{2p}$} & {$S_{n}$} & {$S_{p}$} & {$\lambda_n$}& {$\lambda_p$}& {$R_m$} & {$R_n$}& {$R_p$}& {$R^{\mathrm{Cal.}}_{\mathrm{c}}$}& {$R^{\mathrm{Exp.}}_{\mathrm{c}}$}& {$j^{\pi}(P)$}& {$j^{\pi}(N)$}\\
 & &{$(\rm MeV)$}&{$(\rm MeV)$}&{$(\rm MeV)$}&{$(\rm MeV)$}&{$(\rm MeV)$}&{$(\rm MeV)$}&{$(\rm MeV)$}&{$(\rm MeV)$}&{$(\rm MeV)$}&{$(\rm MeV)$}&{$(\rm fm)$}&{$(\rm fm)$}&{$(\rm fm)$}&{$(\rm fm)$}&{$(\rm fm)$}& & \\
\hline\\
\endhead

\multicolumn{8}{l}{$Z=8~~(\rm{O})$} &\\
12&  4&  60.44 & 58.68&5.04  &4.88&      &  &       &    &$-$19.45&\underline{0.44} & 2.738 & 2.335 & 2.919  & 3.027 && $0^+$   &$0^+$\\
13&  5&  77.81 & 75.56&5.99  &5.81&      &  &17.38  &    &$-$17.01&$-$0.99  & 2.619 & 2.373 & 2.762 & 2.876  &        & $0^+$   & $3/2^-$\\
14&  6& 101.86 & 98.73&7.28  &7.05&41.42 &  &24.05  &    &$-$19.96&$-$1.34  & 2.465 & 2.238 & 2.621 & 2.741  &        & $0^+$   & $0^+$\\
15&  7& 112.81 &111.96&7.52  &7.46&34.99 &  &10.95  &    &$-$15.48&$-$5.43  & 2.599 & 2.537 & 2.652 & 2.770  &        & $0^+$   & $1/2^-$\\
16&  8& 127.29 &127.62&7.96  &7.98&25.43 &  &14.48  &    &$-$11.64&$-$7.83  & 2.638 & 2.626 & 2.649 & 2.768  & 2.701  & $0^+$   & $0^+$\\
17&  9& 132.48 &131.76&7.79  &7.75&19.67 &  & 5.19  &    &$-$13.49&$-$9.79  & 2.690 & 2.733 & 2.642 & 2.760  & 2.695  & $0^+$   & $5/2^+$\\
18& 10& 141.63 &139.81&7.87  &7.77&14.34 &  & 9.15  &    &$-$6.94 &$-$11.79 & 2.736 & 2.807 & 2.644 & 2.763  & 2.775  & $0^+$   & $0^+$\\
\hline
\end{longtable}

\end{document}